\documentclass[12pt]{article}
\usepackage{graphicx}
\usepackage{longtable}
\usepackage{amsfonts}

\newcommand{\compl}{{\mathbb C}}

\begin{document}

\title{General Quantum Modeling of Combining Concepts: \vspace{0.3\baselineskip} \\
 \large A Quantum Field Model in Fock Space}
\author{Diederik Aerts\\
        \normalsize\itshape
        Center Leo Apostel for Interdisciplinary Studies \\
         \normalsize\itshape
         Department of Mathematics and Department of Psychology\\
        \normalsize\itshape
        Vrije Universiteit Brussel, 1160 Brussels, 
       Belgium \\
        \normalsize
        E-Mail: \textsf{diraerts@vub.ac.be}
		}
\date{}
\maketitle

\begin{abstract}
\noindent
We extend a quantum model in Hilbert space developed in Aerts (2007a) into a quantum field theoric model in Fock space for the modeling of the combination of concepts. Items and concepts are represented by vectors in Fock space and membership weights of items are modeled by quantum probabilities. We apply this theory to model the disjunction of concepts and show that the predictions of our theory for the membership weights of items regarding the disjunction of concepts match with great accuracy the complete set of results of an experiment conducted by Hampton (1988b). It are the quantum effects of interference and superposition of that are at the origin of the effects of overextension and underextension observed by Hampton as deviations from a classical use of the disjunction. It is essential for the perfect matches we obtain between the predictions of the quantum field model and Hampton's experimental data that items can be in superpositions of `different numbers states' which proves that the genuine structure of quantum field theory is needed to match predictions with experimental data.
\end{abstract} 

\medskip
\begin{quotation}
\noindent
Keywords: concept theories, concepts combinations, quantum field theory, disjunction, Fock space. 
\end{quotation}

\medskip
%%%%%%%%%%%%%%%%%%%%%%%%%%%%%%%%%%%%%%%%%%%%%%%%%%%%%%%%%%%
%%%%%%%%%%%%%%%%%%%%%%%%%%%%%%%%%%%%%%%%%%%%%%%%%%%%%%%%%%%
\section*{Introduction}
We propose a model for the description of the disjunction of concepts by using the mathematical structure of quantum field theory. The model predicts with very great accuracy the results of an experiment published in Hampton (1988b). The disjunction of concepts was studied by Hampton as part of the general problem of the combination of concepts within an approach where membership is considered to be a fuzzy notion (Rosch, 1973a,b, 1975, 1978, 1983; Smith \& Medin, 1981; Komatsu, 1992). The conjunction of concepts has been studied much more than the disjunction of the concepts (Chater, Lyon \& Meyers, 1990; Hampton, 1987, 1988a, 1996, 1997a; Osherson \& Smith, 1981, 1982; Rips, 1995; Storms, De Boeck, Van Mechelen \& Geeraerts, 1993; Storms, De Boeck, Van Mechelen \& Ruts, 1996; Storms, De Boeck, Hampton \& Van Mechelen, 1999). This is the reason why we apply in Aerts (2007b) the same quantum field model for the description of the conjunction of concepts, and show that the quantum field model predicts with equal accuracy the results of experiments on the conjunction of concepts.

The main problem of the modeling of the combination of concepts was originally put forward by Osherson and Smith (1981) using the example of the concepts {\it Pet} and {\it Fish} and their conjunction {\it Pet-Fish}. Osherson and Smith observed that the item {\it Guppy} is considered to be a very typical item for the conjunction {\it Pet-Fish}, while it is not very typical neither for the concept {\it Pet} and nor for the concepts {\it Fish}. Osherson and Smith showed that this effect, meanwhile refered to as the `guppy effect', cannot be modeled by the minimum rule of fuzzy set theory (Zadeh, 1965). The `guppy effect' was identified and studied in many other occasions of conjunctions of concepts, and the problem of the modeling of the conjunction of concepts taking into account the guppy effect is commonly referred to as the `pet-fish problem'. The pet-fish problem is considered to be a fundamental problem of the modeling of the combination of concepts, and has not been solved within the existing concept theories (Osherson \& Smith, 1981, 1982; Smith \& Medin, 1981; Smith \& Osherson, 1984; Hampton, 1987, 1988a, 1991, 1993, 1996, 1997a,b; Smith, Osherson, Rips \& Keane, 1988; Chater et al., 1990; Kunda, Miller \& Claire, 1990; Springer \& Murphy, 1992; Rips, 1995; Storms et al., 1993, 1996, 1999).

Hampton (1987, 1988a, 1991, 1993, 1996, 1997a) performed many experiments on the conjunction of concepts. Apart from typicality, and the related guppy effect as observed and studied by Osherson and Smith (1981), Hampton studied a similar effect for the membership weights of items with respect to the conjunction of concepts. He called the guppy effect for membership weights of items `overextension', i.e. an item is estimated with a greater than expected weight to be a member of the conjunction of two concepts if taken into account the membership weights of this item with respect to both concepts apart. As an example consider the concepts {\it Sports} and {\it Games} and the item {\it Pool}. In an experiment in Hampton (1988a) subjects estimate the membership weight of the item {\it Pool} for the concept {\it Sports} to be 0.50, and the membership weight of the item {\it Pool} for the concept {\it Games} to be 0.94. When asked for the membership weight of the item {\it Pool} for the concept {\it Sports and Games} the subjects estimate this membership to be 0.73. This means that subjects find {\it Pool} to be `more strongly a member of the conjunction {\it Sports and Games}' than they find it to be a member of the concept {\it Sports} on its own. If one thinks intuitively of the `logical' meaning of a conjunction, this is a strange effect. Indeed, someone who finds that {\it Pool} is a {\it Sport} and a {\it Game}, would be expected to agree at least equally with the statement that {\it Pool} is a {\it Sport} if the conjunction of concept behaved in a similar way as the conjunction of logical propositions behaves. The effect of overextension is abundant for the conjunction of concepts and has been studied intensively since (Osherson \& Smith, 1981, 1982; Smith \& Osherson, 1984, 1988; Hampton, 1988a,b, 1997a,b; Storms et al., 1993, 1996; Rips, 1995).

Hampton (1988b) studied the analogous effect for the disjunction of concepts, and identified a systematic underextension in this case, with however also substantial cases of overextension. Consider as example the concepts {\it House Furnishings} and {\it Furniture} and the item {\it Refrigerator}. In an experiment published in Hampton subjects estimated the membership weight of the item {\it Refrigerator} for the concept {\it House Furnishings} to be 0.9, and the membership weight for the item {\it Refrigerator} for the concept {\it Furniture} to be 0.7. However, when subjects where asked to estimate the membership weight of the item {\it Refrigerator} for the disjunction {\it House Furnishings or Furniture} of the two concepts, the result turned out to be 0.6. This means that subjects estimated {\it Refrigerator} to be less strongly a member of the disjunction of both concepts {\it House Furnishings} and {\it Furniture} than they estimated {\it Refrigerator} to be a member of either of the two concepts. If one thinks intuitively of the `logical' meaning of a disjunction, one would expect that someone who finds that {\it Refrigerator} is {\it House Furnishings}, would also find at least equally well that {\it Refrigerator} is {\it House Furnishing or Furniture}. Similarly for someone who finds that {\it Refrigerator} is {\it Furniture}.

The effect of overextension in the case of the conjunction of concepts as well as the effects of underextension and overextenstion in the case of the disjunction of concepts can be modeled in a very natural way within the quantum field model that we elaborate in the present article. It are the typical quantum effects of `interference' and `superposition' that give rise to the effects of overextension and underextension in case of the conjunction as well as in case of the disjunction of concepts.

The quantum field model elaborated in this article is an extension of the quantum model that we presented in Aerts (2007a). We take however an essential new step, which brings us to a quantum field model in Fock space for the disjunction and also the conjunction of concepts, rather than the Hilbert space quantum model that we presented in Aerts (2007a). We want to explain why this extension if necessary. After working out the quantum model in Hilbert space presented in Aerts (2007a), we contacted James Hampton asking him if he still had the original test results in all its details, including the tests on the items not explicitly mentioned in Hampton (1988b). We were eager to establish whether our Hilbert space model was also a good predictor of the data of the items not explicitly specified in Hampton's experiment. When we received the data, we started to apply the Hilbert space model to the items that had not been explicitly mentioned in Hampton. To our disappointement the data on some of the items did not match at all the predictions of our Hilbert space model. By considering the specifics of the data on items that did not match and the data on items that did match, we understood that a next step was necessary, namely the introduction of a quantum field model in Fock space.

Although initially reluctant to take this step, because of the mathematical and conceptual complexity of quantum field theory in Fock space as compared to quantum mechanics in Hilbert space, we soon recognized that this additional step introduced a wealth of new conceptual possibilities that make very much sense in relation with the cognitive phenomena we want to model. Indeed, we are now convinced that quantum field theory in Fock space is the proper theory to model the combination of concepts and the type of cognitive interactions that we envisaged when we introduced the quantum Hilbert space description. This wealth of new conceptual possibilities makes it possible to predict not only all the data of the disjunction experiments in Hampton (1988b) with perfect matches -- this is what we will work out in the present article --  but also all the data in conjunction experiments (Hampton, 1987, 1988a, 1991, 1993, 1996, 1997a,b; Storms et al., 1993, 1996, 1999) -- which is what we will work out in Aerts (2007b) -- and, more importantly, it introduces a conceptual framework for the modeling of cognitive interaction in an unexpected detailed way. In the next sections we introduce this quantum field model for cognitive interaction.

\section{One-Particle and Two-Particle Quantum Systems}

The items listed in Table 1 are the items of Hampton's (1988b) experiment that entail the biggest deviation from a classical fuzzy set model. Hence, the items not listed in Table 1 are the ones that in the tests showed a less big deviation from a classical fuzzy set model. As we have shown, the quantum model in Hilbert space introduced in Aerts (2007a) for the disjunction does extremely well for the items listed in Hampton, i.e. the items in Table 1, but, as we have said, it does less well for the items tested but not listed in Hampton. 

\scriptsize
\setlongtables 
\begin{longtable}{|l|l|l|l|l|l|l|l|} 
\caption{$\mu_{exp}(A)$, $\mu_{exp}(B)$ and $\mu _{\exp } (A{\rm{\ or\ }}B)$ are the membership weights of concepts $A$, $B$ and the disjunction $A$ or $B$, respectively, for the considered item, as measured in Hampton(1988b). $\beta - \alpha$ and $\beta' - \alpha'$ are the angles, which need to be chosen for the predicted quantum weights $\mu_{quant}(AB)$ to be equal to the experimental weights $\mu _{\exp } (A{\rm{\ or\ }}B)$.} \\
\hline 
 & $\mu_{exp}(A)$ & $\mu_{exp}(B)$ & $\mu _{\exp } (A{\rm{\ or\ }}B)$ & $\beta - \alpha$ & $\beta' - \alpha'$ & $\mu_{quant}(AB)$ & $|\mu_{exp} - \mu_{quant}| $ \\ 
\endfirsthead 
\hline 
 & $\mu_{exp}(A)$ & $\mu_{exp}(B)$ & $\mu _{\exp } (A{\rm{\ or\ }}B)$ & $\beta - \alpha$ & $\beta' - \alpha'$ & $\mu_{quant}(AB)$ & $|\mu_{exp} - \mu_{quant}| $ \\
\hline \hline
\endhead 
\hline \hline  
$A$ = {\it House Furnishings}  &  &  &  & & & & \\
$B$ = {\it Furniture} &  &  &  & & & & \\
\hline
{\it Ashtray} & 0.7 & 0.3 & 0.25 & 117.89811610$^\circ$ & 38.87869152$^\circ$ & 0.25 & 0 \\
{\it Waste-Paper Basket} & 1 & 0.5 & 0.25 & 0$^\circ$ & 122.0277601$^\circ$ & 0.6 & 0 \\
{\it Refrigerator} & 0.9 & 0.7 & 0.6 & 112.3118372$^\circ$ & 13.27346592$^\circ$ & 0.58 & 0 \\
{\it Sink Unit} & 0.9 & 0.6 & 0.6 & 96.18768564$^\circ$ & 9.605483409$^\circ$ & 0.6 & 0\\
{\it Park Bench} & 0 & 0.3 & 0.05 & 180$^\circ$ & 0$^\circ$ & 0.08166999 & 0.03166999 \\
\hline
$A$ = {\it Hobbies} & & & & & & & \\
$B$ = {\it Games} & & & & & & & \\
\hline
{\it Discus Throwing} & 1 & 0.78 & 0.7 & 141.0575587$^\circ$ & 0$^\circ$ & 0.7 & 0 \\
{\it Beer Drinking} & 0.78 & 0.2 & 0.58 & 86.30121954$^\circ$ & 66.43741408$^\circ$ & 0.58 & 0  \\
{\it Wrestling} & 0.9 & 0.6 & 0.63 & 92.82540729$^\circ$ & 26.91024902$^\circ$ & 0.63 & 0 \\
{\it Judo} & 1 & 0.7 & 0.8 & 110.9248324$^\circ$ & 0$^\circ$ & 0.8 & 0 \\
{\it Karate} & 1 & 0.7 & 0.8 & 110.9248324$^\circ$ & 0$^\circ$ & 0.8 & 0  \\
\hline
$A$ = {\it Pets} & & & & & & & \\
$B$ = {\it Farmyard Animals} & & & & & & & \\
\hline
{\it Camel} & 0.4 & 0 & 0.1 & 0$^\circ$ & 0$^\circ$ & 0.11270167 & 0.012770167 \\
{\it Monkey} & 0.5 & 0 & 0.25 & 160.5287794$^\circ$ & 0$^\circ$ & 0.25 & 0  \\
{\it Field Mouse} & 0.1 & 0.7 & 0.4 & 12.60438265$^\circ$ & 48.8103149$^\circ$ & 0.4 & 0 \\
{\it Rat} & 0.5 & 0.7 & 0.4 & 97.51275188$^\circ$ & 7.512751879$^\circ$ & 0.4 & 0  \\
{\it Spider} & 0.4 & 0.33 & 0.65 & 10.99366957$^\circ$ & 79.00633043$^\circ$ & 0.65 & 0  \\
{\it Guide Dog for the Blind} & 0.7 & 0 & 0.9 & 0$^\circ$ & 180$^\circ$ & 0.77386128 & 0.126139  \\
\hline
$A$ = {\it Spices} & & & & & & & \\
$B$ = {\it Herbs} & & & & & & & \\
\hline
{\it Vanilla} & 0.6 & 0 & 0.26 & 90.6568106$^\circ$ & 0$^\circ$ & 0.26 & 0 \\
{\it Horseradish} & 0.2 & 0.4 & 0.7 & 43.05027183$^\circ$ & 48.08071796$^\circ$ & 0.7 & 0  \\
{\it Sesame Seeds} & 0.33 & 0.4 & 0.63 & 20.40338955$^\circ$ & 66.8659453$^\circ$ & 0.63 & 0 \\
{\it Monosodium Glutamate} & 0.11 & 0.1 & 0.33 & 38.01837213$^\circ$ & 131.132395$^\circ$ & 0.33 & 0 \\
{\it Sugar} & 0 & 0 & 0.2 & 0$^\circ$ & 180$^\circ$ & 0 & 0.2 \\
\hline
$A$ = {\it Instruments} & & & & & & & \\
$B$ = {\it Tools} & & & & & & & \\
\hline
{\it Bicycle Pump} & 1 & 0.9 & 0.7 & 151.4512088$^\circ$ & 0$^\circ$ & 0.7 & 0  \\
{\it Pencil Eraser} & 0.4 & 0.7 & 0.45 & 151.4512088$^\circ$ & 123.0349258$^\circ$ & 0.45 & 0  \\
{\it Computer} & 0.6 & 0.8 & 0.6 & 97.18770632$^\circ$ & 67.36029726$^\circ$ & 0.6 & 0  \\
{\it Spoon} & 0.67 & 0.9 & 0.7 & 120.6923864$^\circ$ & 108.7619129$^\circ$ & 0.7 & 0  \\
\hline
$A$ = {\it Sportswear} & & & & & & & \\
$B$ = {\it Sports Equipment} & & & & & & & \\
\hline
{\it Sunglasses} & 0.4 & 0.2 & 0.1 & 141.4276456$^\circ$ & 89.18888998$^\circ$ & 0.1 & 0  \\
{\it Bathing Costume} & 1 & 0.8 & 0.8 & 123.9878436$^\circ$ & 0$^\circ$ & 0.8 & 0  \\
{\it Lineman's Flag} & 0.1 & 1 & 0.75 & 0$^\circ$ & 18$^\circ$ & 0.65811388 & 0.10188612  \\
\hline
$A$ = {\it Household Appliances} & & & & & & & \\
$B$ = {\it Kitchen Utensils} & & & & & & & \\
\hline
{\it Electric Toothbrush} & 0.8 & 0 & 0.55 & $97.13606039^\circ$ & $0^\circ$ & 0.55 & 0  \\
{\it Rubbish Bin} & 0.5 & 0.5 & 0.8 & $6.755014424^\circ $ & $120.1148836^\circ $ & 0.8 & 0  \\
{\it Cake Tin} & 0.4 & 0.7 & 0.95 & $34.37506977^\circ$ & $91.00980915^\circ$ & 0.95 &  0 \\
\hline
$A$ = {\it Fruits} & & & & & & & \\
$B$ = {\it Vegetables} & & & & & & & \\
\hline
{\it Elderberry } & 1 & 0 & 0.55 & 0$^\circ$ & 180$^\circ$ & 0.5 & 0.05  \\
{\it Mushroom} & 0 & 0.5 & 0.8 & 0$^\circ$ & 180$^\circ$ & 0.83355339 & 0.05355339 \\
{\it Yam} & 0.43 & 0.67 & 1 & $0^\circ$ & $180^\circ$ & 0.985227222 & 0.01477278 \\
{\it Coconut} & 0.7 & 0 & 1 & 0$^\circ$ & 180$^\circ$ & 0.77386128 & 0.22613872  \\
{\it Garlic} & 0.1 & 0.2 & 0.5 & $33.75015572^\circ $ & $65.84391343^\circ $ & 0.5 & 0 \\
{\it Olive} & 0.5 & 0.1 & 0.8 & $117.9054714^\circ $ & $69.71578529^\circ $ & 0.8 & 0 \\
{\it Tomato} & 0.7 & 0.7 & 1 & $106.6015485^\circ $ & $33.20309753^\circ $ & 1 & 0  \\
{\it Root Ginger} & 0 & 0.3 & 0.56 & $0^\circ $ & $142.8274109^\circ $ & 0.56 & 0  \\
{\it Almond} & 0.2 & 0.1 & 0.42 & $24.25296063^\circ $ & $75.34110853^\circ $ & 0.42 & 0 \\
{\it Parsley} & 0 & 0.2 & 0.45 & $0^\circ$ & $156.1709693^\circ $ & 0.45 & 0 
\\
{\it Broccoli } & 0 & 0.8 & 1 & $0^\circ$ & $180^\circ$ & 0.7236068 & 0.2763932 \\
{\it Green Pepper} & 0.3 & 0.6 & 0.8 & $87.0433271^\circ $ & $29.35485976^\circ $ & 0.8 & 0 \\
{\it Watercress} & 0 & 0.6 & 0.8 & $0^/circ$ & $37.76124391^\circ $ & 0.8 & 0
 \\
\hline
\hline 
\end{longtable}
\normalsize
One way to explain this would be to say that there are different types of behavior for different items with respect to the disjunction of concepts, one type of behavior that is more classical and that can therefore be modeled fairly well by means of a classical fuzzy set model, and another type of behavior that is more quantum and that can therefore be modeled well by the quantum model that we introduced in Aerts (2007a). After all, this is analogous to objects in the physical world, where microscopic objects behave differently than macroscopic objects and the former can be well modeled by quantum theory, while the latter cannot and require classical modeling. Let us put forward one of the cases not well modeled by the quantum model in Hilbert space developed in Aerts.

\subsection{The Case of Apple and Fruits or Vegetables}
First we write down the basic equation, namely equation (2.29) of Aerts (2007a), derived within the Hilbert space quantum model
\begin{equation} \label{eq1.1}
\mu (AB) = \frac{{\mu (A) + \mu (B) + 2\sqrt {\mu (A)\mu (B)} \cos (\beta  - \alpha )}}{{2 + 2\sqrt {\mu (A)\mu (B)} \cos (\beta  - \alpha ) + 2\sqrt {(1 - \mu (A))(1 - \mu (B))} \cos (\beta ' - \alpha ')}}
\end{equation}
where $\mu(AB)$ is the membership weight of the considered item with respect to the disjunction $A$ or $B$ of concept $A$ and concept $B$ predicted by the quantum model presented in Aerts (2007a), and $\mu(A)$ and $\mu(B)$ are the membership weights of the considered item with respect to concepts $A$ and concept $B$ as measured in the experiment presented in Hampton (1988b).

When Hampton (1988b) asked his subjects to estimate the membership weights for the item {\it Apple} with respect to the concept {\it Fruits} and with respect to the concept {\it Vegetables} and also with respect to the disjunction {\it Fruits or Vegetables}, the test gave 1.0 for {\it Fruits}, 0.0 for {\it Vegetables}, and 1.0 for {\it Fruits or Vegetables}. If we use equation (\ref{eq1.1}) to calculate the weight for the item {\it Apple} with respect to the disjunction {\it Fruits or Vegetables} as predicted by the quantum model in Aerts (2007a), we find
\begin{equation}
\mu_{quant} (Apple,Fruits\ or\ Vegetables) = 0.5
\end{equation}
This is very different from $\mu_{exp} (Apple,Fruits\ or\ Vegetables) = 1$, the value of the experimentally measured weight. And moreover, when one of the experimental weights $\mu_{exp}(X, A)$ or  $\mu_{exp}(X, B)$ equals zero and the other equals 1, the quantum interference terms disappear, which means that the quantum model developed in Aerts (2007a) predicts one fixed value, namely the value 0.5. More specifically, suppose that  $\mu_{exp}(X, A) = 1$ and  $\mu_{exp}(X, B) = 0$, which is the case for  $X$ = {\it Apple} and $A$ = {\it Fruits} and  $B$ = {\it Vegetables}, then equation (\ref{eq1.1}) becomes
\begin{equation}
\mu (AB) = 0.5
\end{equation}
In Hampton's (1988b) experiment, the situation of the item {\it Apple} with respect to the concept {\it Fruits} and the concept {\it Vegetables} and their disjunction {\it Fruits or Vegetables} is not the only one that gives rise to this type of data. Hampton tested 24 items for each one of the considered pair of concepts and their disjunction, and in Table 2 we have collected the items that cannot be modeled well by means of the quantum model that we introduced in Aerts (2007a). As we can see, the most dramatic situations are the ones similar to {\it Apple}, which we already mentioned explicitly. We have the items {\it Gardening}, {\it Theatre-Going} and {\it Guitar Playing} for the concepts {\it Hobbies} and {\it Games} and their disjunction {\it Hobbies or Games} and the item {\it Vacuum Cleaner} for the concepts {\it Household Appliances} and {\it Kitchen Utensils} and their disjunction {\it Household Appliances or Kitchen Utensils}, which behave like {\it Apple} with respect to the concepts {\it Fruits} and {\it Vegetables} and their disjunction {\it Fruits or Vegetables}.

\scriptsize
\setlongtables 
\begin{longtable}{|l|l|l|l|l|l|l|l|} 
\caption{The list of items that cannot be modeled well by the one-particle quantum model for the disjunction developed in Aerts (2007a). $\mu_{exp}(A)$, $\mu_{exp}(B)$ and $\mu _{\exp } (A{\rm{\ or\ }}B)$ are the membership weights of concepts $A$, $B$ and the disjunction $A$ or $B$, respectively, for the considered item, as measured in Hampton(1988b). $\beta - \alpha$ and $\beta' - \alpha'$ are the angles, which need to be chosen for the predicted quantum weights $\mu_{quant}(AB)$ to be equal to the experimental weights $\mu _{\exp } (A{\rm{\ or\ }}B)$.} \\
\hline 
 & $\mu_{exp}(A)$ & $\mu_{exp}(B)$ & $\mu _{\exp } (A{\rm{\ or\ }}B)$ & $\beta - \alpha$ & $\beta' - \alpha'$ & $\mu_{quant}(AB)$ & $|\mu_{exp} - \mu_{quant}| $ \\ 
\endfirsthead 
\hline 
 & $\mu_{exp}(A)$ & $\mu_{exp}(B)$ & $\mu _{\exp } (A{\rm{\ or\ }}B)$ & $\beta - \alpha$ & $\beta' - \alpha'$ & $\mu_{quant}(AB)$ & $|\mu_{exp} - \mu_{quant}| $ \\
\hline \hline
\endhead 
\hline \hline  
$A$ = {\it House Furnishings}  &  &  &  & & & & \\
$B$ = {\it Furniture} &  &  &  & & & & \\
\hline 
{\it Shelves} & 0.4 & 1 & 1 & 0$^\circ$ & 180$^\circ$ & 0.816227766 & 0.183772 \\
{\it Wall Hangings} & 0.4 & 0.9 & 0.95 & 0$^\circ$ & 180$^\circ$ & 0.922474487 & 0.027526 \\
{\it Wall Mirror} & 0.6 & 1 & 0.95 & 0$^\circ$ & 180$^\circ$ & 0.887298335 & 0.062702 \\
{\it Park Bench} & 0 & 0.3 & 0.05 & 180$^\circ$ & 0$^\circ$ & 0.08166999 & 0.03166999 \\
\hline
$A$ = {\it Hobbies} & & & & & & & \\
$B$ = {\it Games} & & & & & & & \\
\hline
{\it Gardening} & 1 & 0 & 1 & 0$^\circ$ & 180$^\circ$ & 0.5 & 0.5 \\
{\it Theatre-Going} & 1 & 0 & 1 & 0$^\circ$ & 180$^\circ$ & 0.5 & 0.5  \\
{\it Monopoly} & 0.7 & 1 & 1 & 0$^\circ$ & 180$^\circ$ & 0.918330013 & 0.081670 \\
{\it Fishing} & 1 & 0.6 & 1 & 0$^\circ$ & 180$^\circ$ & 0.887298335 & 0.112702 \\
{\it Camping} & 1 & 0.1 & 0.9 & 0$^\circ$ & 180$^\circ$ & 0.658113883 & 0.241886  \\
{\it Skating} & 1 & 0.5 & 0.9 & 0$^\circ$ & 180$^\circ$ & 0.853553391 & 0.046447  \\
{\it Guitar Playing} & 1 & 0 & 1 & 0$^\circ$ & 180$^\circ$ & 0.5 & 0.5  \\
{\it Autograph Hunting} & 1 & 0.2 & 0.9 & 0$^\circ$ & 180$^\circ$ & 0.723606798 & 0.176393  \\
{\it Jogging} & 1 & 0.4 & 0.9 & 0$^\circ$ & 180$^\circ$ & 0.816227766 & 0.083772  \\
{\it Keep Fit} & 1 & 0.3 & 0.95 & 0$^\circ$ & 180$^\circ$ & 0.773861279 & 0.176139  \\
{\it Noughts} & 0.5 & 1 & 0.9 & 0$^\circ$ & 180$^\circ$ & 0.853553391 & 0.046447  \\
{\it Rock Climbing} & 1 & 0.2 & 0.95 & 0$^\circ$ & 180$^\circ$ & 0.723606798 & 0.226393  \\
{\it Stamp Collecting} & 1 & 0.1 & 1 & 0$^\circ$ & 180$^\circ$ & 0.658113883 & 0.341886  \\
\hline
$A$ = {\it Pets} & & & & & & & \\
$B$ = {\it Farmyard Animals} & & & & & & & \\
\hline
{\it Goldfish} & 1 & 0 & 0.95 & 0$^\circ$ & 0$^\circ$ & 0.5 & 0.45 \\
{\it Collie Dog} & 1 & 0.7 & 1 & 0$^\circ$ & 180$^\circ$ & 0.918330013 & 0.081670  \\
{\it Camel} & 0.4 & 0 & 0.1 & 180$^\circ$ & 0$^\circ$ & 0.11270187 & 0.01270167 \\
{\it Guide Dog for the Blind} & 0.7 & 0 & 0.9 & 0$^\circ$ & 180$^\circ$ & 0.77386128 & 0.126139  \\
{\it Prize Bull} & 0.1 & 1 & 0.9 & 0$^\circ$ & 180$^\circ$ & 0.658113883 & 0.241886  \\
{\it Siamese Cat} & 1 & 0.1 & 0.95 & 0$^\circ$ & 180$^\circ$ & 0.658113883 & 0.291886  \\
{\it Ginger Tom-Cat} & 1 & 0.8 & 0.95 & 0$^\circ$ & 180$^\circ$ & 0.947213595 & 0.002786  \\
{\it Cart Horse} & 0.4 & 1 & 0.85 & 0$^\circ$ & 180$^\circ$ & 0.816227766 & 0.033772  \\
{\it Chicken} & 0.3 & 1 & 0.95 & 0$^\circ$ & 180$^\circ$ & 0.773861279 & 0.176139  \\
\hline
$A$ = {\it Spices} & & & & & & & \\
$B$ = {\it Herbs} & & & & & & & \\
\hline
{\it Chili Pepper} & 1 & 0.6 & 0.95 & 0$^\circ$ & 180$^\circ$ & 0.887298335 & 0.062702 \\
{\it Cinnamon} & 1 & 0.4 & 1 & 0$^\circ$ & 180$^\circ$ & 0.816227766 & 0.183772  \\
{\it Parsley} & 0.5 & 0.9 & 0.95 & 0$^\circ$ & 180$^\circ$ & 0.947213595 & 0.002786 \\
{\it Sugar} & 0 & 0 & 0.2 & 0$^\circ$ & 180$^\circ$ & 0 & 0.2 \\
{\it Chires} & 0.6 & 1 & 9.95 & 0$^\circ$ & 180$^\circ$ & 0.887298335 & 0.062702 \\
\hline
$A$ = {\it Instruments} & & & & & & & \\
$B$ = {\it Tools} & & & & & & & \\
\hline
{\it Magnetic Compass} & 0.9 & 0.5 & 1 & 0$^\circ$ & 180$^\circ$ & 0.947213595 & 0.052786  \\
{\it Tuning Fork} & 0.9 & 0.6 & 1 & 0$^\circ$ & 180$^\circ$ & 0.967423461 & 0.032577  \\
{\it Pen-Knife} & 0.65 & 1 & 0.95 & 0$^\circ$ & 180$^\circ$ & 0.903112887 & 0.046887  \\
{\it Skate Board} & 0.1 & 0 & 0 & 180$^\circ$ & 0$^\circ$ & 0.025658351 & 0.025658  \\
{\it Pliers} & 0.8 & 1 & 1 & 0$^\circ$ & 180$^\circ$ & 0.947213595 & 0.052786  \\
\hline
$A$ = {\it Sportswear} & & & & & & & \\
$B$ = {\it Sports Equipment} & & & & & & & \\
\hline
{\it Circus Clowns} & 0 & 0 & 0.1 & 0$^\circ$ & 180$^\circ$ & 0 & 0.1  \\
{\it Diving Mask} & 1 & 1 & 0.95 & 0$^\circ$ & 180$^\circ$ & 1 & 0.05  \\
{\it Frisbee} & 0.3 & 1 & 0.85 & 0$^\circ$ & 180$^\circ$ & 0.773861279 & 0.076139  \\
{\it Suntan Lotion} & 0 & 0 & 0.1 & 180$^\circ$ & 0$^\circ$ & 0 & 0.1  \\
{\it Gymnasium} & 0 & 0.9 & 0.825 & 180$^\circ$ & 0$^\circ$ & 0.341886117 & 0.483114  \\
{\it Wrist Sweat} & 1 & 1 & 0.95 & 0$^\circ$ & 180$^\circ$ & 1 & 0.05  \\
{\it Lineman's Flag} & 0.1 & 1 & 0.75 & 0$^\circ$ & 180$^\circ$ & 0.65811388 & 0.10188612  \\
\hline
$A$ = {\it Household Appliances} & & & & & & & \\
$B$ = {\it Kitchen Utensils} & & & & & & & \\
\hline
{\it Fork} & 0.7 & 1 & 0.95 & $0^\circ$ & $180^\circ$ & 0.918330013 & 0.031670  \\
{\it Freezer} & 1 & 0.6 & 0.95 & $0^\circ $ & $180^\circ $ & 0.887298335 & 0.062702  \\
{\it Extractor Fan} & 1 & 0.4 & 0.9 & $0^\circ$ & $180^\circ$ & 0.816227766 &  0.083772 \\
{\it Carving Knife} & 0.7 & 1 & 1 & $0^\circ$ & $180^\circ$ & 0.918330013 &  0.081670 \\
{\it Cooking Stove} & 1 & 0.5 & 1 & $0^\circ$ & $180^\circ$ & 0.853553391 &  0.146447 \\
{\it Iron} & 1 & 0.3 & 0.95 & $0^\circ$ & $180^\circ$ & 0.773861279 &  0.176139 \\
{\it Chopping Board} & 0.45 & 1 & 0.95 & $0^\circ$ & $180^\circ$ & 0.835410197 &  0.114590 \\
{\it Television} & 0.95 & 0 & 0.85 & $0^\circ$ & $180^\circ$ & 0.611803399 &  0.238197 \\
{\it Vacuum Cleaner} & 1 & 0 & 1 & $0^\circ$ & $180^\circ$ & 0.5 &  0.5 \\
{\it Rolling Pin} & 0.45 & 1 & 1 & $0^\circ$ & $180^\circ$ & 0.835410197 &  0.164590 \\
{\it Frying Pan} & 0.7 & 1 & 0.95 & $0^\circ$ & $180^\circ$ & 0.918330013 &  0.031670 \\
\hline
$A$ = {\it Fruits} & & & & & & & \\
$B$ = {\it Vegetables} & & & & & & & \\
\hline
{\it Apple} & 1 & 0 & 1 & 0$^\circ$ & 180$^\circ$ & 0.5 & 0.5  \\
{\it Broccoli} & 0 & 0.8 & 1 & 0$^\circ$ & 180$^\circ$ & 0.7236068 & 0.2763932 \\
{\it Raisin} & 1 & 0 & 0.9 & $0^\circ$ & $180^\circ$ & 0.5 & 0.4 \\
{\it Coconut} & 0.7 & 0 & 1 & 0$^\circ$ & 180$^\circ$ & 0.77386128 & 0.22613872  \\
{\it Mushroom} & 0 & 0.5 & 0.8 & 0$^\circ$ & 180$^\circ$ & 0.83355339 & 0.05355339 \\
{\it Yam} & 0.43 & 0.67 & 1 & $0^\circ $ & $180^\circ $ & 0.98522722 & 0.01477278 \\
{\it Elderberry} & 1 & 0 & 0.55 & $0^\circ $ & $180^\circ $ & 0.5 & 0.05 \\
\hline
\hline 
\end{longtable}
\normalsize
These items were not in the tables of Hampton (1988b) for a good reason. This behavior -- let us nickname it {\it Apple}-like' -- is a case of classical logic behavior. Indeed, subjects see {\it Apple} as a member of {\it Fruits}, assigning a weight equal to 1, which means `member in the classical sense'. They also see {\it Apple} as `non member' of {\it Vegetables}, assigning a weight equal to 0, which means `non member in the classical sense'. From classical logic it should then follow that {\it Apple} is also a member of the disjunction {\it Fruits or Vegetables} with a weight equal to 1, hence a `member in the classical sense'.

The second class of items, not behaving exactly similarly to {\it Apple}, but very {\it Apple}-like, are the item {\it Goldfish} for the concepts {\it Pets} and {\it Farmyard Animals} and their disjunction {\it Pets} or {\it Farmyard Animals} and the item {\it Raisin} for the concepts {\it Fruits} and {\it Vegetables} and their disjunction {\it Fruits or Vegetables}. These two examples already show some slight signs of a quantum effect, however. Although subjects take a classical view of both items with respect to concepts  $A$ and  $B$, assigning weights 0 and 1, they assign weights of less than 1 to the disjunction: 0.95 in the case of {\it Goldfish} and 0.9 in the case of {\it Raisin}. The third group of remaining items from Table 2 deviates more from the classical towards the quantum, but not enough to allow for a full description of the quantum model we have presented in Aerts (2007a).

If we take into account that the model we proposed in Aerts (2007a) is a pure quantum model, it should not really come as a surprise that it does not model this {\it Apple}-like behavior, since this concerns classical behavior. However, as we will show in the following of this article, even the {\it Apple}-like behavior appears to be part in a very natural way of a more extended and more general quantum model for the situation. This more extended quantum model is not only able to predict the data for all 24 of the items for all of the concepts and their disjunctions tested in Hampton (1988b), but it also contains the proposal of a dynamics that can be considered as a model for cognitive dynamics. We will use the mathematical and conceptual apparatus of quantum field theory to develop this extended quantum model. We need to make an important remark to be able to introduce this more extended quantum field theoretic model.

The quantum model we have introduced in Aerts (2007a) is a one-quantum particle model, more concretely this means that we consider a situation like the one in Figure 1, (which is a copy of Figure 4 in Aerts (2007a)) for the case of a quantum particle. Suppose now that we want to test a classical logical situation, for example the one of {\it Apple} being a member of the disjunction {\it Fruits or Vegetables}. We will then use `two identical copies' of {\it Apple}, the one to verify whether it is a member of {\it Fruits}, yes or no, and the other to verify whether it is a member of {\it Vegetables}, yes or no. In case we find one of the following combinations of results `yes, yes', `yes, no' and `no, yes', we conclude that {\it Apple} is a member of the disjunction {\it Fruits or Vegetables}.
\begin{figure}[h]
\centerline {\includegraphics[width=10cm]{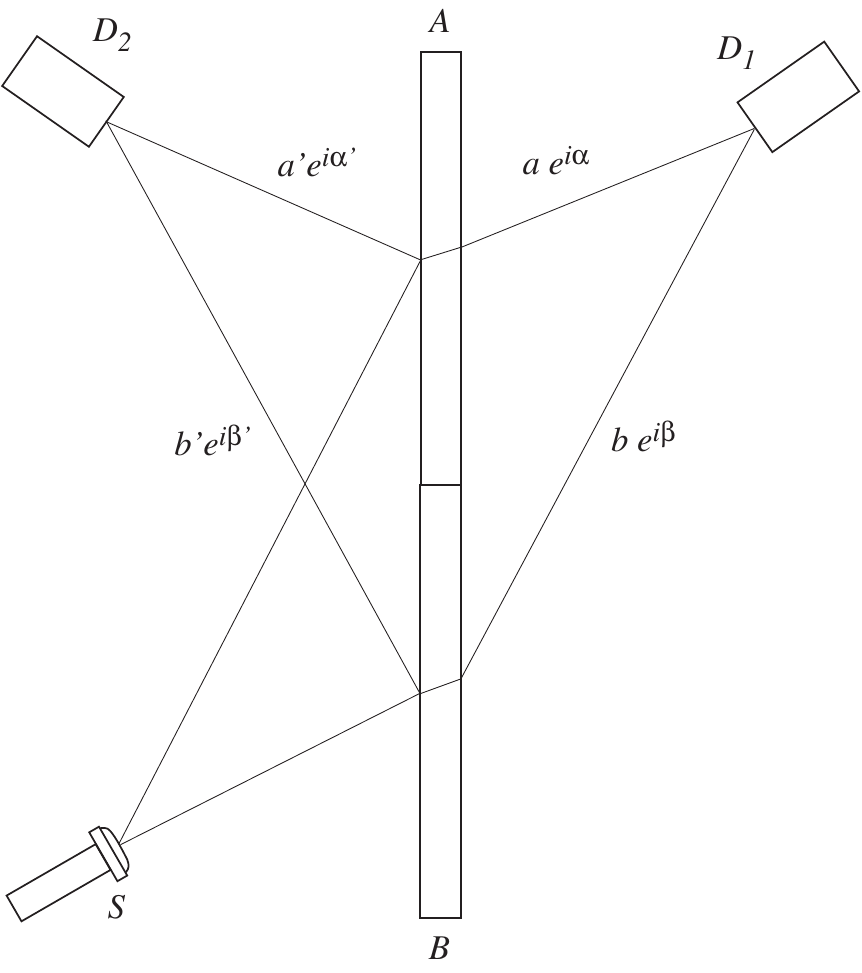}}
\caption{A quantum particle emitted from source  $S$ passes through a semi-transparent mirror  $A$ or through a semi-transparent mirror  $B$ and is detected by  $D_1$, or is reflected by mirror  $A$ or by mirror $B$ and detected by  $D_2$. Amplitudes  $ae^{i\alpha }  + be^{i\beta } $
 and  $a'e^{i\alpha '}  + b'e^{\beta '} $
 describe both processes.}
\end{figure}
This is a very different type of experimental situation from the one we considered in the quantum model presented in Aerts (2007a). Quantum mechanically speaking it is a `two-particle situation', instead of a `one-particle situation', like the one we considered in Aerts (2007a). Before we introduce our final quantum field theoretic model, we will develop this two-particle quantum model and show that it provides for a very good modeling of the Apple-like situations.

\subsection{The Two Identical Particle Quantum Model}
In Figure 2 we present a two identical particles quantum situation with respect to two semi-transparent mirrors  $A$ and  $B$. The difference with the `one-particle situations' considered in Aerts (2007a) and exposed in Figure 1 is that there are two sources of particles now, $S_1$ and  $S_2$, both of which can emit particles that will either pass through or be reflected by the semi-transparent mirrors $A$ and  $B$.
\begin{figure}[h]
\centerline {\includegraphics[width=10cm]{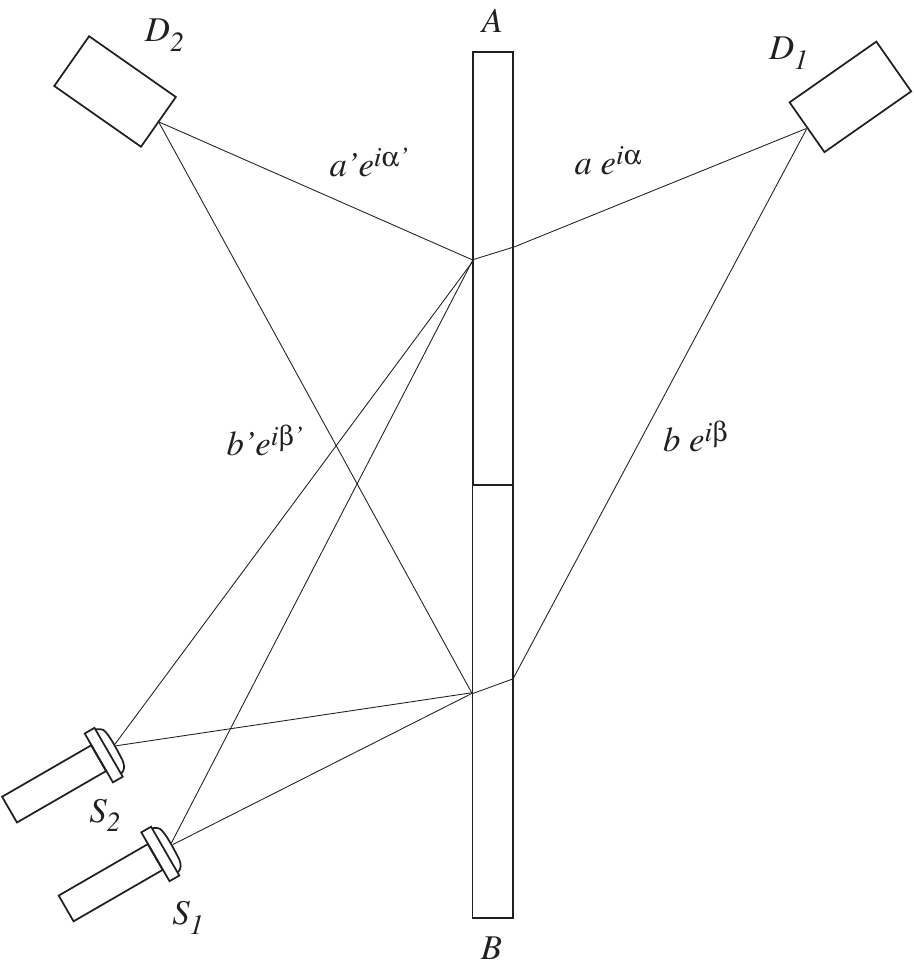}}
\caption{Two identical quantum particles emitted from sources $S_1$ and $S_2$ pass through or are reflected by two semi-transparent mirrors $A$ and  $B$. Amplitudes  $2abe^{i(\alpha  + \beta )} $,  $ab'e^{i(\alpha  + \beta ')}  + a'be^{i(\alpha ' + \beta )} $, and  $2a'b'e^{i(\alpha ' + \beta ')} $ describe the different processes that take place.}
\end{figure}
At this stage, we need to introduce a new quantum principle that we did not need to built the quantum model in Aerts (2007a), since there we only considered situations involving one quantum particle. Let us first state again the three quantum principles we introduced in Aerts (2007a).
\begin{quotation}
\noindent {\it Quantum Principle 1: A path which can be followed by a quantum particle is characterized by an amplitude, which is a complex number.}
\end{quotation}

\begin{quotation}
\noindent {\it Quantum Principle 2: The probability of detecting a quantum particle is proportional to the square of the magnitude of the complex number which characterizes the path leading to the detection apparatus.}
\end{quotation}
\begin{quotation}
\noindent
{\it Quantum Principle 3: When detection can occur in several alternative ways, the associated amplitude is the sum of the amplitudes for each way considered separately.}
\end{quotation}
The new and fourth quantum principle we introduce is the following:
\begin{quotation}
\noindent {\it Quantum Principle 4: When detection involves two events happening simultaneously, the amplitude connected to this detection is proportional to the product of the amplitudes of each one of the individual events.}
\end{quotation}
We refer to Aerts (2007a), and also to Feynman, Leighton \& Sands (1966) and Feynman (1985) for a detailed analysis of these quantum principles.

We are interested in events where two particles are detected having passed through different mirrors. In other words, we do not consider the case where both particles would pass through the same mirror, because such experiment would not constitute a test for their disjunction (or conjunction).

There are three possible events; (i) two particles are detected by $D_1$ such that the one passed through one of the mirrors and the other through the other mirror; (ii) one particle is detected by $D_1$ passing through one of the mirrors and one particle is detected by  $D_2$ passing through the other mirror; (iii) two particles are detected by $D_2$ such that the one passed through one of the mirrors and the other through the other mirror. Let us consider event (i) in greater detail. If two particles are detected by  $D_1$ such that the one passed through one of the mirrors and the other through the other mirror, there are two possibilities: the first particle passed through $A$ with amplitude $ae^{i\alpha}$ -- we explain in detail in Aerts (2007a) how amplitudes describe quantum dynamics and determine quantum weights, and also in Feynman et al. (1966) and Feynman (1985) this is explained in great length -- and the second particle passed through $B$ with amplitude $be^{i\beta}$ or the first particle passed through  $B$ with amplitude $be^{i\beta}$ and the second passed through $A$ with amplitude $ae^{i\alpha}$.

We now apply quantum principle 3 and quantum principle 4. Following quantum principle 4 we know that the amplitude of `the first particle passed through  $A$ and the second particle passed through $B$' is proportional to the product of the respective amplitudes, i.e. $ae^{i\alpha } be^{i\beta }  = abe^{i(\alpha  + \beta )} $. Again applying quantum principle 4, we have that the amplitude of `the first particle passed through  $B$ and the second particle passed through  $A$ equals $be^{i\beta } ae^{i\alpha }  = abe^{i(\alpha  + \beta )} $. These are the two alternative paths which lead to `two particles detected in  $D_1$' and hence we know from quantum principle 3 that we need to sum the amplitudes of these two alternatives to find the factor to which the total amplitude of the event `two particles detected in  $D_1$' is proportional. This means that amplitude of two particles detected by  $D_1$ is proportional to
\begin{equation} \label{eq1.4}
ae^{i\alpha } be^{i\beta }  + be^{i\beta } ae^{i\alpha }  = 2abe^{i(\alpha  + \beta )}
\end{equation}
In a completely analogous way we find that the amplitude of having one particle detected by $D_1$ and the other particle by $D_2$ is proportional to
\begin{equation} \label{eq1.5}
ae^{i\alpha } b'e^{i\beta '}  + be^{i\beta } a'e^{i\alpha '}  = ab'e^{i(\alpha  + \beta ')}  + a'be^{i(\alpha ' + \beta )}
\end{equation}
and the amplitude of having two particles detected in $D_2$ is proportional to
\begin{equation} \label{eq1.6}
a'e^{i\alpha '} b'e^{i\beta '}  + b'e^{i\beta '} a'e^{i\alpha '}  = 2a'b'e^{i(\alpha ' + \beta ')}
\end{equation}
Instead of working out in full the Feynman version (Feynman et al., 1966; Feynman, 1985) of the two identical particles situation, as we did in Aerts (2007a) for the case of the one particle situation, we prefer to switch immediately to its Hilbert space description, bearing in mind that the Feynman version readily gives us the kernels of the amplitudes as in (\ref{eq1.4}), (\ref{eq1.5}) and (\ref{eq1.6}). To be able to develop the final quantum field model for the cognitive situation, however, we need the mathematical details of the Hilbert space description. In this Hilbert space description we make use of the bra-ket notation introduced in Dirac (1958).

Each of the two quantum particles considered can be described in a two-dimensional Hilbert space over the complex numbers, since we only need to describe the following two possible events for each one of them: (i) the passing through the mirror and being detected by $D_1$ situation, which we describe in Hilbert space by means of the projection onto one base vector  $\left| {e_1 } \right\rangle $, and (ii) the being deflected by the mirror and being detected by $D_2$ situation, which we describe in Hilbert space by means of the projection onto a second base vector  $\left| {e_2 } \right\rangle $, orthogonal to  $\left| {e_1 } \right\rangle $. The two-dimensional complex Hilbert space is traditionally denoted by  $\compl^2$, and hence for each one of the particles we use the Hilbert space $\compl^2$ to describe the set of states.

We need one more quantum principle of Hilbert space quantum mechanics before we can proceed.

\begin{quotation}
\noindent {\it Quantum Principle 5: If one quantum entity is described in a Hilbert space ${\cal H}_1$ and a second quantum entity is described in a Hilbert space ${\cal H}_2$, then the joint quantum entity of these two quantum entities is described in the tensor product ${\cal H}_1 \otimes {\cal H}_2$ of both Hilbert spaces.}
\end{quotation}

Since we want the material of this article joined with the material of Aerts (2007a) to be self-contained, also from the point of view of the mathematics we use, we make a short digression to explain the tensor product of two Hilbert spaces. The tensor product ${\cal H}_1 \otimes {\cal H}_2$ is again a Hilbert space and it consists of the set of vectors
\begin{equation} \label{eq1.7}
\{\sum_i z_i \left|x\right\rangle_i \otimes \left|y\right\rangle_i\ \vert \left|x\right\rangle_i \in {\cal H}_1, \left|y\right\rangle_i \in {\cal H}_2, z_i \in \compl\}
\end{equation} 
where indeed $\otimes$ has the properties of a product, which means that for a complex number $z \in \compl$ and a vector $|x\rangle \otimes |y\rangle$ of the tensor product we have
\begin{equation} \label{eq1.8}
z(\left| x \right\rangle  \otimes \left| y \right\rangle ) = (z\left| x \right\rangle ) \otimes \left| y \right\rangle  = \left| x \right\rangle  \otimes (z\left| y \right\rangle )
\end{equation}
which is why we denote this expression as $z \left|x\right\rangle \otimes \left|y\right\rangle$ in (\ref{eq1.7}), not paying attention to the brackets ( and ) in (\ref{eq1.8}). Specifically for a Hilbert space, we have the following equality with respect to the tensor product and the bra and ket product. If $\left| x \right\rangle $, $\left| u \right\rangle \in {\cal H}_1$ and $\left| y \right\rangle $, $\left| v \right\rangle \in {\cal H}_2$, we have
\begin{equation}
(\left\langle x \right| \otimes \left\langle y \right|)(\left| u \right\rangle  \otimes \left| v \right\rangle ) = \left\langle {x}
 \mathrel{\left | {\vphantom {x u}}
 \right. \kern-\nulldelimiterspace}
 {u} \right\rangle \left\langle {y}
 \mathrel{\left | {\vphantom {y v}}
 \right. \kern-\nulldelimiterspace}
 {v} \right\rangle 
\end{equation}
It can be shown that, if  $\left\{ {\left| {e_1 } \right\rangle ,\dots,\left| {e_n } \right\rangle ,\dots} \right\}$ is a base of ${\cal H}_1$ and    $\left\{ {\left| f \right\rangle _1 ,\dots,\left| {f_m } \right\rangle ,\ldots} \right\}$ is a base of ${\cal H}_2$, then $\left\{ {\left| {e_1 } \right\rangle  \otimes \left| {f_1 } \right\rangle ,\ldots,\left| {e_1 } \right\rangle  \otimes \left| {f_m } \right\rangle ,\ldots,\left| {e_n } \right\rangle  \otimes \left| {f_1 } \right\rangle ,\ldots,\left| {e_n } \right\rangle  \otimes \left| {f_m } \right\rangle ,\ldots} \right\}$ is a base of the tensor product Hilbert space ${\cal H}_1 \otimes {\cal H}_2$. This means that if, for example, ${\cal H}_1$ has dimension $n$ and ${\cal H}_2$ has dimension $m$, then ${\cal H}_1 \otimes {\cal H}_2$ has dimension $n \times m$. We will not further elaborate on the tensor product of Hilbert spaces of arbitrary dimension and instead concentrate on the case where we have two Hilbert spaces of dimension 2, hence on the tensor product $\compl^2 \otimes \compl^2$, which is the tensor product we need for our specific problem of describing two identical quantum particles, each of them described in a two-dimensional Hilbert space $\compl^2$.

As said, we introduce two vectors $\left| {e_1 } \right\rangle ,\left| {e_2 } \right\rangle  \in \compl^2$ of unit length and orthogonal to each other, forming a base of  $\compl^2$. This means that  $\left\{ {\left| {e_1 } \right\rangle  \otimes \left| {e_1 } \right\rangle ,\left| {e_1 } \right\rangle  \otimes \left| {e_2 } \right\rangle ,\left| {e_2 } \right\rangle  \otimes \left| {e_1 } \right\rangle ,\left| {e_2 } \right\rangle  \otimes \left| {e_2 } \right\rangle } \right\}$ is a base of  $\compl^2 \otimes \compl^2$. A projection on $\left| {e_1 } \right\rangle  \otimes \left| {e_1 } \right\rangle $ stands for `both particles passing the mirrors and being detected by  $D_1$', and a projection on  $\left| {e_2 } \right\rangle  \otimes \left| {e_2 } \right\rangle $ for `both particles being deflected by the mirrors and being detected by  $D_2$'. What about the situation `one particle passing through one of the mirrors and being detected by $D_1$ and the other particle being deflected by the other mirror and being detected by  $D_2$'. Both vectors  $\left| {e_1 } \right\rangle  \otimes \left| {e_2 } \right\rangle $ and  $\left| {e_2 } \right\rangle  \otimes \left| {e_1 } \right\rangle $ seem to be able to represent this situation. Obviously, there is no preference, because the particles are identical so that we cannot distinguish between them. In this case, in quantum mechanics, one needs to make symmetric the vector in question, which means that instead of one of the two vectors  $\left| {e_1 } \right\rangle  \otimes \left| {e_2 } \right\rangle $ or  $\left| {e_2 } \right\rangle  \otimes \left| {e_1 } \right\rangle $ we need to consider the vector
\begin{equation} \label{eq1.10}
\frac{1}{{\sqrt 2 }}(\left| {e_1 } \right\rangle  \otimes \left| {e_2 } \right\rangle  + \left| {e_2 } \right\rangle  \otimes \left| {e_1 } \right\rangle )
\end{equation}
to represent the situation of `one particle passing through one of the mirrors and being detected by $D_1$ and the other particle being deflected by the other mirror and being detected by  $D_2$'. The factor $1/\sqrt{2}$  is a normalization coefficient to make the length of the symmetric vector equal to 1.

Hence, the projection on vector (\ref{eq1.10}) stands for the situation of `one particle passing through one of the mirrors and being detected by $D_1$ and the other particle being deflected by the other mirror and being detected by $D_2$'.

Let us proceed with our Hilbert space description of two identical quantum particles, and introduce the states of the particles. We denote the state of a particle interacting with the first mirror  $A$ by means of  $\left|u\right\rangle$ and the state of a particle interacting with the second mirror $B$ by means of  $\left|v\right\rangle$. This means that we have
\begin{eqnarray} \label{eq1.11}
\left| u \right\rangle  &=& ae^{i\alpha } \left| {e_1 } \right\rangle  + a'e^{i\alpha '} \left| {e_2 } \right\rangle \\ \label{eq1.12}
\left| v \right\rangle  &=& be^{i\beta } \left| {e_1 } \right\rangle  + b'e^{i\beta '} \left| {e_2 } \right\rangle 
\end{eqnarray}
The vector $\left| u \right\rangle  \otimes \left| v \right\rangle $
 of the tensor product Hilbert space $\compl^2 \otimes \compl^2$ represents a state of the two particles, one interacting with one of the mirrors and the other one interacting with the other mirror. However, this is also the case for the vector  $\left| v \right\rangle  \otimes \left| u \right\rangle $. Since the particles are identical, we again need to `make symmetric over the states', and hence consider the vector.
\begin{equation}
\left| {uv} \right\rangle  = \frac{1}{E}(\left| u \right\rangle  \otimes \left| v \right\rangle  + \left| v \right\rangle  \otimes \left| u \right\rangle )
\end{equation}
to represent the two particles. The factor $E$ is a normalization coefficient to make the vector of unit length.
To calculate  $E$, we need to solve the equation $\left\langle {{uv}}
 \mathrel{\left | {\vphantom {{uv} {uv}}}
 \right. \kern-\nulldelimiterspace}
 {{uv}} \right\rangle  = 1$. So 
\begin{eqnarray} \label{eq1.14}
 1 &=& \left\langle {{uv}}
 \mathrel{\left | {\vphantom {{uv} {uv}}}
 \right. \kern-\nulldelimiterspace}
 {{uv}} \right\rangle  \\ 
  &=& \frac{1}{{E^2 }}(\left\langle u \right| \otimes \left\langle v \right| + \left\langle v \right| \otimes \left\langle u \right|)(\left| u \right\rangle  \otimes \left| v \right\rangle  + \left| v \right\rangle  \otimes \left| u \right\rangle ) \\ 
  &=& \frac{1}{{E^2 }}(\left\langle {u}
 \mathrel{\left | {\vphantom {u u}}
 \right. \kern-\nulldelimiterspace}
 {u} \right\rangle \left\langle {v}
 \mathrel{\left | {\vphantom {v v}}
 \right. \kern-\nulldelimiterspace}
 {v} \right\rangle  + \left\langle {u}
 \mathrel{\left | {\vphantom {u v}}
 \right. \kern-\nulldelimiterspace}
 {v} \right\rangle \left\langle {v}
 \mathrel{\left | {\vphantom {v u}}
 \right. \kern-\nulldelimiterspace}
 {u} \right\rangle  + \left\langle {v}
 \mathrel{\left | {\vphantom {v u}}
 \right. \kern-\nulldelimiterspace}
 {u} \right\rangle \left\langle {u}
 \mathrel{\left | {\vphantom {u v}}
 \right. \kern-\nulldelimiterspace}
 {v} \right\rangle  + \left\langle {v}
 \mathrel{\left | {\vphantom {v v}}
 \right. \kern-\nulldelimiterspace}
 {v} \right\rangle \left\langle {u}
 \mathrel{\left | {\vphantom {u u}}
 \right. \kern-\nulldelimiterspace}
 {u} \right\rangle ) \\ 
  &=& \frac{1}{{E^2 }}(2 + 2\left| {\left\langle {u}
 \mathrel{\left | {\vphantom {u v}}
 \right. \kern-\nulldelimiterspace}
 {v} \right\rangle } \right|^2 ) 
\end{eqnarray}
where $\left| u \right\rangle $ and  $\left| v \right\rangle $ are vectors of a length equal to 1, i.e.  $\left\langle {u}
 \mathrel{\left | {\vphantom {u u}}
 \right. \kern-\nulldelimiterspace}
 {u} \right\rangle  = \left\langle {v}
 \mathrel{\left | {\vphantom {v v}}
 \right. \kern-\nulldelimiterspace}
 {v} \right\rangle  = 1$. From (\ref{eq1.14}) it follows that
\begin{equation} \label{eq1.15}
E^2  = 2 + 2\left| {\left\langle {u}
 \mathrel{\left | {\vphantom {u v}}
 \right. \kern-\nulldelimiterspace}
 {v} \right\rangle } \right|^2 
\end{equation}
Using (\ref{eq1.11}) and (\ref{eq1.12}) we have
\begin{eqnarray}
 \left\langle {u}
 \mathrel{\left | {\vphantom {u v}}
 \right. \kern-\nulldelimiterspace}
 {v} \right\rangle  &=& (ae^{ - i\alpha } \left\langle {e_1 } \right| + a'e^{ - i\alpha '} \left\langle {e_2 } \right|)(be^{i\beta } \left| {e_1 } \right\rangle  + b'e^{i\beta '} \left| {e_2 } \right\rangle ) \\ 
  &=& abe^{i(\beta  - \alpha )} \left\langle {{e_1 }}
 \mathrel{\left | {\vphantom {{e_1 } {e_1 }}}
 \right. \kern-\nulldelimiterspace}
 {{e_1 }} \right\rangle  + ab'e^{i(\beta ' - \alpha )} \left\langle {{e_1 }}
 \mathrel{\left | {\vphantom {{e_1 } {e_2 }}}
 \right. \kern-\nulldelimiterspace}
 {{e_2 }} \right\rangle  \nonumber \\
&& + a'be^{i(\beta  - \alpha ')} \left\langle {{e_2 }}
 \mathrel{\left | {\vphantom {{e_2 } {e_1 }}}
 \right. \kern-\nulldelimiterspace}
 {{e_1 }} \right\rangle  + a'b'e^{i(\beta ' - \alpha ')} \left\langle {{e_2 }}
 \mathrel{\left | {\vphantom {{e_2 } {e_2 }}}
 \right. \kern-\nulldelimiterspace}
 {{e_2 }} \right\rangle  \\ 
  &=& abe^{i(\beta  - \alpha )}  + a'b'e^{i(\beta ' - \alpha ')} 
\end{eqnarray}
This means that
\begin{eqnarray} \label{eq1.17}
 E^2  &=& 2 + 2(abe^{ - i(\beta  - \alpha )}  + a'b'e^{ - i(\beta ' - \alpha ')} )(abe^{i(\beta  - \alpha )}  + a'b'e^{i(\beta ' - \alpha ')} ) \\ 
  &=& 2 + 2a^2 b^2  + 2a'^2 b'^2  + 2aa'bb'e^{i(\beta ' - \alpha ' - \beta  + \alpha )}  + 2aa'bb'e^{ - i(\beta ' - \alpha ' - \beta  + \alpha )}  \\ 
  &=& 2 + 2a^2 b^2  + 2a'^2 b'^2  + 4aa'bb'\cos (\beta ' - \alpha ' - \beta  + \alpha ) \\ 
  &=& 2 + 2a^2 b^2  + 2(1 - a^2 )(1 - b^2 ) + 4aa'bb'\cos (\beta ' - \alpha ' - \beta  + \alpha ) \\ 
  &=& 4 + 4a^2 b^2  - 2a^2  - 2b^2  + 4aa'bb'\cos (\beta ' - \alpha ' - \beta  + \alpha )
\end{eqnarray}
We are interested in the weight of the disjunction, so we need to calculate the probability that at least one particle arrives at detector  $D_1$. Indeed, if at least one particle is counted in detector $D_1$, this means that the particle has passed at least through one of the mirrors. This probability is the sum of two other probabilities: (1) the probability that two particles pass through the mirrors and are detected by  $D_1$ plus (2) the probability that the one particle passes through one of the mirrors and is detected by $D_1$ and the other particle is reflected by the other mirror and is detected by $D_2$. Let us calculate these probabilities.

To calculate the probability that two particles are detected by  $D_1$, we first calculate the amplitude that carries this probability. This amplitude is given by
\begin{eqnarray} \label{eq1.18}
 (\left\langle {e_1 } \right| \otimes \left\langle {e_1 } \right|)(\left| {uv} \right\rangle ) &=& \frac{1}{E}(\left\langle {e_1 } \right| \otimes \left\langle {e_1 } \right|)(\left| u \right\rangle  \otimes \left| v \right\rangle  + \left| v \right\rangle  \otimes \left| u \right\rangle ) \\ 
  &=& \frac{1}{E}(\left\langle {{e_1 }}
 \mathrel{\left | {\vphantom {{e_1 } u}}
 \right. \kern-\nulldelimiterspace}
 {u} \right\rangle \left\langle {{e_1 }}
 \mathrel{\left | {\vphantom {{e_1 } v}}
 \right. \kern-\nulldelimiterspace}
 {v} \right\rangle  + \left\langle {{e_1 }}
 \mathrel{\left | {\vphantom {{e_1 } v}}
 \right. \kern-\nulldelimiterspace}
 {v} \right\rangle \left\langle {{e_1 }}
 \mathrel{\left | {\vphantom {{e_1 } u}}
 \right. \kern-\nulldelimiterspace}
 {u} \right\rangle ) \\ 
  &=& \frac{1}{E}(abe^{i(\alpha  + \beta )}  + abe^{i(\alpha  + \beta )} ) \\ 
  &=& \frac{2}{E}abe^{i(\alpha  + \beta )} 
\end{eqnarray}
The probability is the square of the modulus of this amplitude, so
\begin{eqnarray} \label{eq1.19}
 P(A \wedge B) &=& (\frac{{2ab}}{E}e^{ - i(\alpha  + \beta )} )(\frac{{2ab}}{E}e^{i(\alpha  + \beta )} ) \\ 
  &=& \frac{{4a^2 b^2 }}{{E^2 }} 
\end{eqnarray}
The amplitude corresponding to the situation of `one particle passing through one of the mirrors and detected by $D_1$ and the other particle reflected by the other mirror and detected by  $D_2$ is given by
\begin{eqnarray} \label{eq1.20}
&&\frac{1}{{\sqrt 2 }}(\left\langle {e_1 } \right| \otimes \left\langle {e_2 } \right| + \left\langle {e_2 } \right| \otimes \left\langle {e_1 } \right|)(\left| {uv} \right\rangle ) \\
&&= \frac{1}{{\sqrt 2 }}(\left\langle {e_1 } \right| \otimes \left\langle {e_2 } \right|  + \left\langle {e_2 } \right| \otimes \left\langle {e_1 } \right|)(\frac{1}{E}(\left| u \right\rangle  \otimes \left| v \right\rangle  + \left| v \right\rangle  \otimes \left| u \right\rangle )) \\ 
&&= \frac{1}{{\sqrt 2 E}}(\left\langle {{e_1 }}
 \mathrel{\left | {\vphantom {{e_1 } u}}
 \right. \kern-\nulldelimiterspace}
 {u} \right\rangle \left\langle {{e_2 }}
 \mathrel{\left | {\vphantom {{e_2 } v}}
 \right. \kern-\nulldelimiterspace}
 {v} \right\rangle  + \left\langle {{e_1 }}
 \mathrel{\left | {\vphantom {{e_1 } v}}
 \right. \kern-\nulldelimiterspace}
 {v} \right\rangle \left\langle {{e_2 }}
 \mathrel{\left | {\vphantom {{e_2 } u}}
 \right. \kern-\nulldelimiterspace}
 {u} \right\rangle + \left\langle {{e_2 }}
 \mathrel{\left | {\vphantom {{e_2 } u}}
 \right. \kern-\nulldelimiterspace}
 {u} \right\rangle \left\langle {{e_1 }}
 \mathrel{\left | {\vphantom {{e_1 } v}}
 \right. \kern-\nulldelimiterspace}
 {v} \right\rangle  + \left\langle {{e_2 }}
 \mathrel{\left | {\vphantom {{e_2 } v}}
 \right. \kern-\nulldelimiterspace}
 {v} \right\rangle \left\langle {{e_1 }}
 \mathrel{\left | {\vphantom {{e_1 } u}}
 \right. \kern-\nulldelimiterspace}
 {u} \right\rangle ) \\ 
&&= \frac{1}{{\sqrt 2 E}}(ae^{i\alpha } b'e^{i\beta '}  + be^{i\beta } a'e^{i\alpha '} + a'e^{i\alpha '} be^{i\beta }  + b'e^{i\beta '} ae^{i\alpha } ) \\ 
&&= \frac{2}{{\sqrt 2 E}}(ab'e^{i(\alpha  + \beta ')}  + a'be^{i(\alpha ' + \beta ')} )
\end{eqnarray}
The probability is given by
\begin{eqnarray} \label{eq1.21}
&&P((A \wedge B^C ) \vee (A^C  \wedge B)) \\
&&= \frac{4}{{2E^2 }}(ab'e^{ - i(\alpha  + \beta ')}  + a'be^{ - i(\alpha ' + \beta ')} )(ab'e^{i(\alpha  + \beta ')}  + a'be^{i(\alpha ' + \beta ')} ) \\ 
&&  = \frac{2}{{E^2 }}(a^2 b'^2  + a'^2 b^2  + aa'bb'e^{i(\alpha  + \beta ' - \alpha ' - \beta )}  + aa'bb'e^{ - i(\alpha  + \beta ' - \alpha ' - \beta )} ) \\ 
&&  = \frac{2}{{E^2 }}(a^2 b'^2  + a'^2 b^2  + 2aa'bb'\cos (\alpha  + \beta ' - \alpha ' - \beta ))
\end{eqnarray}
The probability for the disjunction, i.e. the event that at least one particle is detected by  $D_1$, is given by the sum of these two probabilities. This gives us
\begin{eqnarray}
 P(A \vee B) &=& P((A \wedge B^C ) \vee (A^C  \wedge B)) + P(A \wedge B) \\ 
  &=& \frac{1}{{E^2 }}(4a^2 b^2  + 2a^2 b'^2  + 2a'^2 b^2  + 4aa'bb'\cos (\alpha  + \beta ' - \alpha ' - \beta )) 
\end{eqnarray}
Substituting (\ref{eq1.17}) we get
\begin{eqnarray}
 P(A \vee B) &=& \frac{{4a^2 b^2  + 2a^2 b'^2  + 2a'^2 b^2  + 4aa'bb'\cos (\alpha  + \beta ' - \alpha ' - \beta )}}{{4 + 4a^2 b^2  - 2a^2  - 2b^2  + 4aa'bb'\cos (\beta ' - \alpha ' - \beta  + \alpha )}} \\ 
  &=& \frac{{4a^2 b^2  + 2a^2 (1 - b^2 ) + 2(1 - a^2 )b^2  + 4aa'bb'\cos (\alpha  + \beta ' - \alpha ' - \beta )}}{{4 + 4a^2 b^2  - 2a^2  - 2b^2  + 4aa'bb'\cos (\beta ' - \alpha ' - \beta  + \alpha )}} \\ 
  &=& \frac{{2a^2  + 2b^2  + 4aa'bb'\cos (\alpha  + \beta ' - \alpha ' - \beta )}}{{4 + 4a^2 b^2  - 2a^2  - 2b^2  + 4aa'bb'\cos (\beta ' - \alpha ' - \beta  + \alpha )}} \\ 
  &=& \frac{{a^2  + b^2  + 2aa'bb'\cos (\alpha  + \beta ' - \alpha ' - \beta )}}{{2 + 2a^2 b^2  - a^2  - b^2  + 2aa'bb'\cos (\beta ' - \alpha ' - \beta  + \alpha )}} 
\end{eqnarray}
Taking into account
\begin{eqnarray}
& a^2  = P(A) \quad {\rm{   }} \quad a'^2  = 1 - P(A) \quad {\rm{   }} \quad b^2  = P(B) \quad {\rm{  }} \quad b'^2  = 1 - P(B) \\ 
& a = \sqrt {P(A)} \quad  {\rm{   }} \quad a' = \sqrt {1 - P(A)} \quad {\rm{   }} \quad b = \sqrt {P(B)} \quad {\rm{   }} \quad b' = \sqrt {1 - P(B)}
\end{eqnarray}
we can write the formula which gives the probability for the disjunction in function of the component probabilities:
\begin{equation}
P(A \vee B) = \frac{{P(A) \! + \! P(B) \! + \! 2\sqrt {P(A)(1 \! - \! P(A))P(B)(1 \! - \! P(B))} \cos (\alpha \!  + \! \beta ' \! - \! \alpha ' \! - \! \beta )}}{{2 \! + \! 2P(A)P(B) \! - \! P(A) \! - \! P(B) + \! 2\sqrt {P(A)(1 \! - \! P(A))P(B)(1 \! - \! P(B))} \cos (\alpha \! + \! \beta ' \! - \! \alpha ' \! - \! \beta )}}
\end{equation}
\normalsize
In a very analogous way as we did with the one-particle quantum model in Aerts (2007a) we apply the two-particle quantum model to the situation of the combination of concepts.

\subsection{The Two Identical Items Quantum Model}
Hence we consider two concepts  $A$ and  $B$ and an item  $X$. We consider the following possibilities: (i) item  $X$ is a member of concept $A$ or the item is not a member of concept  $A$; (ii) item  $X$ is a member of concept  $B$ or item  $X$ is not a member of concept  $B$. These situations are described by vectors  $\left| U \right\rangle $ and  $\left| V \right\rangle $ of a two-dimensional vector space $\compl^2$ with orthonormal base $\left\{ {e_1 ,e_2 } \right\}$, such that
\begin{eqnarray} \label{eq1.26}
\left| U \right\rangle  &=& ae^{i\alpha } \left| {e_1 } \right\rangle  + a'e^{i\alpha '} \left| {e_2 } \right\rangle \\ \label{eq1.27}
\left| V \right\rangle  &=& be^{i\beta } \left| {e_1 } \right\rangle  + b'e^{i\beta '} \left| {e_2 } \right\rangle
\end{eqnarray}
and hence membership weights $\mu(A)$ and $\mu(B)$ of item  $X$ with respect to concept  $A$ and concept  $B$, respectively, are given by
\begin{eqnarray}
\mu (A) &=& \left| {\left\langle {{e_1 }}
 \mathrel{\left | {\vphantom {{e_1 } U}}
 \right. \kern-\nulldelimiterspace}
 {U} \right\rangle } \right|^2  = a^2 \\
\mu (B) &=& \left| {\left\langle {{e_1 }}
 \mathrel{\left | {\vphantom {{e_1 } V}}
 \right. \kern-\nulldelimiterspace}
 {V} \right\rangle } \right|^2  = b^2 
\end{eqnarray}
Next we consider the situation where item  $X$ is a member or not a member of concept  $A$ and simultaneously also a member or not a member of concept  $B$. We analyze this situation as if two identical items  $X$ are considered, and one of them is a member or not a member of concept $A$ while the other one is a member or not a member of concept  $A$. Inspired by how quantum mechanics describes the situation of two identical quantum entities, we introduce the tensor product Hilbert space  $\compl^2 \otimes \compl^2$ to describe the state of concept  $A$ and the state of concept $B$ simultaneously and the state of the two identical items  $X$. The state of the two identical items $X$ is described by the vector
\begin{equation}
\left| {UV} \right\rangle  = \frac{1}{E}(\left| U \right\rangle  \otimes \left| V \right\rangle  + \left| V \right\rangle  \otimes \left| U \right\rangle )
\end{equation}
element of  $\compl^2 \otimes \compl^2$, where, see (\ref{eq1.15}) and (\ref{eq1.17}), we have
\begin{equation} \label{eq1.31}
E^2  = 2 + 2\left| {\left\langle {U}
 \mathrel{\left | {\vphantom {U V}}
 \right. \kern-\nulldelimiterspace}
 {V} \right\rangle } \right|^2  = 4 + 4a^2 b^2  - 2a^2  - 2b^2  + 4aa'bb'\cos (\beta ' - \alpha ' - \beta  + \alpha )
\end{equation}
The state of concept  $A$ and concept  $B$ is described by the vector $\left| {e_1 } \right\rangle  \otimes \left| {e_1 } \right\rangle $. Hence the weight that one of the identical items  $X$ is a member of concept $A$ and the other item  $X$ is a member of concept  $B$ is given by
\begin{equation}
\mu (A \wedge B) = \left| {\left\langle {{e_1  \otimes e_1 }}
 \mathrel{\left | {\vphantom {{e_1  \otimes e_1 } {UV}}}
 \right. \kern-\nulldelimiterspace}
 {{UV}} \right\rangle } \right|^2 
\end{equation}
while the weight that one of the items $X$ is not a member of concept $A$ and the other item  $X$ is not a member of concept  $B$ is given by
\begin{equation}
\mu (A' \wedge B') = \left| {\left\langle {{e_2  \otimes e_2 }}
 \mathrel{\left | {\vphantom {{e_2  \otimes e_2 } {UV}}}
 \right. \kern-\nulldelimiterspace}
 {{UV}} \right\rangle } \right|^2 
\end{equation}
Further, the weight that one of the items  $X$ is a member of concept $A$ and the other item  $X$ is not a member of concept  $B$ or one of the items  $X$ is a member of concept  $B$ while the other one is not a member of concept $A$ is given by
\begin{equation}
\mu ((A \wedge B') \vee (A' \wedge B)) = \left| {\left\langle {{\frac{1}{{\sqrt 2 }}(e_1  \otimes e_2  + e_2  \otimes e_1 )}}
 \mathrel{\left | {\vphantom {{\frac{1}{{\sqrt 2 }}(e_1  \otimes e_2  + e_2  \otimes e_1 )} {UV}}}
 \right. \kern-\nulldelimiterspace}
 {{UV}} \right\rangle } \right|^2 
\end{equation}
A calculation analogous to the one that led to (\ref{eq1.18}) gives us
\begin{eqnarray}
 (\left\langle {e_1 } \right| \otimes \left\langle {e_1 } \right|)(\left| {UV} \right\rangle ) &=& \frac{1}{E}(\left\langle {e_1 } \right| \otimes \left\langle {e_1 } \right|)(\left| U \right\rangle  \otimes \left| V \right\rangle  + \left| V \right\rangle  \otimes \left| U \right\rangle ) \\ 
  &=& \frac{1}{E}(\left\langle {{e_1 }}
 \mathrel{\left | {\vphantom {{e_1 } U}}
 \right. \kern-\nulldelimiterspace}
 {U} \right\rangle \left\langle {{e_1 }}
 \mathrel{\left | {\vphantom {{e_1 } V}}
 \right. \kern-\nulldelimiterspace}
 {V} \right\rangle  + \left\langle {{e_1 }}
 \mathrel{\left | {\vphantom {{e_1 } V}}
 \right. \kern-\nulldelimiterspace}
 {V} \right\rangle \left\langle {{e_1 }}
 \mathrel{\left | {\vphantom {{e_1 } U}}
 \right. \kern-\nulldelimiterspace}
 {U} \right\rangle ) \\ 
  &=& \frac{1}{E}(abe^{i(\alpha  + \beta )}  + abe^{i(\alpha  + \beta )} ) \\ 
  &=& \frac{2}{E}abe^{i(\alpha  + \beta )}
\end{eqnarray}
In a similar way, we get
\begin{eqnarray}
 (\left\langle {e_2 } \right| \otimes \left\langle {e_2 } \right|)(\left| {UV} \right\rangle ) &=& \frac{1}{E}(\left\langle {e_2 } \right| \otimes \left\langle {e_2 } \right|)(\left| U \right\rangle  \otimes \left| V \right\rangle  + \left| V \right\rangle  \otimes \left| U \right\rangle ) \\ 
  &=& \frac{1}{E}(\left\langle {{e_2 }}
 \mathrel{\left | {\vphantom {{e_2 } U}}
 \right. \kern-\nulldelimiterspace}
 {U} \right\rangle \left\langle {{e_2 }}
 \mathrel{\left | {\vphantom {{e_2 } V}}
 \right. \kern-\nulldelimiterspace}
 {V} \right\rangle  + \left\langle {{e_2 }}
 \mathrel{\left | {\vphantom {{e_2 } V}}
 \right. \kern-\nulldelimiterspace}
 {V} \right\rangle \left\langle {{e_2 }}
 \mathrel{\left | {\vphantom {{e_2 } U}}
 \right. \kern-\nulldelimiterspace}
 {U} \right\rangle ) \\ 
  &=& \frac{1}{E}(a'b'e^{i(\alpha ' + \beta ')}  + a'b'e^{i(\alpha ' + \beta ')} ) \\ 
  &=& \frac{2}{E}a'b'e^{i(\alpha ' + \beta ')}
\end{eqnarray}
and making a calculation analogous to the one leading to (\ref{eq1.20}), we get
\begin{eqnarray}
&&\frac{1}{{\sqrt 2 }}(\left\langle {e_1 } \right| \otimes \left\langle {e_2 } \right| + \left\langle {e_2 } \right| \otimes \left\langle {e_1 } \right|)(\left| {UV} \right\rangle ) \\
&& = \frac{1}{{\sqrt 2 }}(\left\langle {e_1 } \right| \otimes \left\langle {e_2 } \right| + \left\langle {e_2 } \right| \otimes \left\langle {e_1 } \right|)(\frac{1}{E}(\left| U \right\rangle  \otimes \left| V \right\rangle  + \left| V \right\rangle  \otimes \left| U \right\rangle )) \\ 
&& = \frac{1}{{\sqrt 2 E}}(\left\langle {{e_1 }}
 \mathrel{\left | {\vphantom {{e_1 } U}}
 \right. \kern-\nulldelimiterspace}
 {U} \right\rangle \! \left\langle {{e_2 }}
 \mathrel{\left | {\vphantom {{e_2 } V}}
 \right. \kern-\nulldelimiterspace}
 {V} \right\rangle \! + \! \left\langle {{e_1 }}
 \mathrel{\left | {\vphantom {{e_1 } V}}
 \right. \kern-\nulldelimiterspace}
 {V} \right\rangle \! \left\langle {{e_2 }}
 \mathrel{\left | {\vphantom {{e_2 } U}}
 \right. \kern-\nulldelimiterspace}
 {U} \right\rangle \! + \! \left\langle {{e_2 }}
 \mathrel{\left | {\vphantom {{e_2 } U}}
 \right. \kern-\nulldelimiterspace}
 {U} \right\rangle \! \left\langle {{e_1 }}
 \mathrel{\left | {\vphantom {{e_1 } V}}
 \right. \kern-\nulldelimiterspace}
 {V} \right\rangle \! + \! \left\langle {{e_2 }}
 \mathrel{\left | {\vphantom {{e_2 } V}}
 \right. \kern-\nulldelimiterspace}
 {V} \right\rangle \! \left\langle {{e_1 }}
 \mathrel{\left | {\vphantom {{e_1 } U}}
 \right. \kern-\nulldelimiterspace}
 {U} \right\rangle ) \\ 
&&  = \frac{1}{{\sqrt 2 E}}(ae^{i\alpha } b'e^{i\beta '}  + be^{i\beta } a'e^{i\alpha '}  + a'e^{i\alpha '} be^{i\beta }  + b'e^{i\beta '} ae^{i\alpha } ) \\ 
&&  = \frac{2}{{\sqrt 2 E}}(ab'e^{i(\alpha  + \beta ')}  + a'be^{i(\alpha ' + \beta ')} ) 
\end{eqnarray}
This makes it possible to calculate all relevant weights in a way analogous to the calculation leading to the quantum probabilities (\ref{eq1.19}) and (\ref{eq1.21}). Hence we have
\begin{eqnarray}
 \mu (A \wedge B) &=& \left| {\left\langle {{e_1  \otimes e_1 }}
 \mathrel{\left | {\vphantom {{e_1  \otimes e_1 } {UV}}}
 \right. \kern-\nulldelimiterspace}
 {{UV}} \right\rangle } \right|^2  \\ 
  &=& \left\langle {{e_1  \otimes e_1 }}
 \mathrel{\left | {\vphantom {{e_1  \otimes e_1 } {UV}}}
 \right. \kern-\nulldelimiterspace}
 {{UV}} \right\rangle^*\left\langle {{e_1  \otimes e_1 }}
 \mathrel{\left | {\vphantom {{e_1  \otimes e_1 } {UV}}}
 \right. \kern-\nulldelimiterspace}
 {{UV}} \right\rangle  \\ 
  &=& (\frac{{2ab}}{E}e^{ - i(\alpha  + \beta )} )(\frac{{2ab}}{E}e^{i(\alpha  + \beta )} ) \\ 
  &=& \frac{{4a^2 b^2 }}{{E^2 }}
\end{eqnarray}
for the weight of the conjunction of concepts $A$ and  $B$, the conjunction in the sense of two identical items interacting with the concepts. We have
\begin{eqnarray}
 \mu (A^C  \wedge B^C ) &=& \left| {\left\langle {{e_2  \otimes e_2 }}
 \mathrel{\left | {\vphantom {{e_2  \otimes e_2 } {UV}}}
 \right. \kern-\nulldelimiterspace}
 {{UV}} \right\rangle } \right|^2  \\ 
  &=& \left\langle {{e_2  \otimes e_2 }}
 \mathrel{\left | {\vphantom {{e_2  \otimes e_2 } {UV}}}
 \right. \kern-\nulldelimiterspace}
 {{UV}} \right\rangle^*\left\langle {{e_2  \otimes e_2 }}
 \mathrel{\left | {\vphantom {{e_2  \otimes e_2 } {UV}}}
 \right. \kern-\nulldelimiterspace}
 {{UV}} \right\rangle  \\ 
  &=& (\frac{{2a'b'}}{E}e^{ - i(\alpha ' + \beta ')} )(\frac{{2a'b'}}{E}e^{i(\alpha ' + \beta ')} ) \\ 
  &=& \frac{{4a'^2 b'^2 }}{{E^2 }}
\end{eqnarray}
for the weight of the item  $X$ that is neither a member of concept  $A$ nor of concept  $B$. We can also calculate the weight of the event consisting of the item  $X$ that is a member of one of the concepts but not a member of the other concept. This gives
\begin{eqnarray}
&& \mu ((A \wedge B') \vee (A' \wedge B)) \\
&& = \left| {\left\langle {{\frac{1}{{\sqrt 2 }}(e_1  \otimes e_2  + e_2  \otimes e_1 )}}
 \mathrel{\left | {\vphantom {{\frac{1}{{\sqrt 2 }}(e_1  \otimes e_2  + e_2  \otimes e_1 )} {UV}}}
 \right. \kern-\nulldelimiterspace}
 {{UV}} \right\rangle } \right|^2  \\ 
&&  = \left\langle {{\frac{1}{{\sqrt 2 }}(e_1  \otimes e_2  + e_2  \otimes e_1 )}}
 \mathrel{\left | {\vphantom {{\frac{1}{{\sqrt 2 }}(e_1  \otimes e_2  + e_2  \otimes e_1 )} {UV}}}
 \right. \kern-\nulldelimiterspace}
 {{UV}} \right\rangle^*\left\langle {{\frac{1}{{\sqrt 2 }}(e_1  \otimes e_2  + e_2  \otimes e_1 )}}
 \mathrel{\left | {\vphantom {{\frac{1}{{\sqrt 2 }}(e_1  \otimes e_2  + e_2  \otimes e_1 )} {UV}}}
 \right. \kern-\nulldelimiterspace}
 {{UV}} \right\rangle  \\ 
&&  = \frac{4}{{2E^2 }}(ab'e^{ - i(\alpha  + \beta ')}  + a'be^{ - i(\alpha ' + \beta ')} )(ab'e^{i(\alpha  + \beta ')}  + a'be^{i(\alpha ' + \beta ')} ) \\ 
&&  = \frac{2}{{E^2 }}(a^2 b'^2  + a'^2 b^2  + aa'bb'e^{i(\alpha  + \beta ' - \alpha ' - \beta )}  + aa'bb'e^{ - i(\alpha  + \beta ' - \alpha ' - \beta )} ) \\ 
&&  = \frac{2}{{E^2 }}(a^2 b'^2  + a'^2 b^2  + 2aa'bb'\cos (\alpha  + \beta ' - \alpha ' - \beta ))
\end{eqnarray}
The weight of the event consisting of the item  $X$ that is a member of one of the concepts, i.e. $\mu(A \vee B)$, is given by
\begin{eqnarray}
 \mu (A \vee B) &=& \mu (A \wedge B) + \mu ((A \wedge B') \vee (A' \wedge B)) \\ 
  &=& \frac{{4a^2 b^2 }}{{E^2 }} + \frac{2}{{E^2 }}(a^2 b'^2  + a'^2 b^2  + 2aa'bb'\cos (\alpha  + \beta ' - \alpha ' - \beta )) \\ 
  &=& \frac{2}{{E^2 }}(2a^2 b^2  + a^2 b'^2  + a'^2 b^2  + 2aa'bb'\cos (\alpha  + \beta ' - \alpha ' - \beta )) \\ 
  &=& \frac{2}{{E^2 }}(a^2 b^2  + a^2 b'^2  + a^2 b^2  + a'^2 b^2  + 2\sqrt {a^2 a'^2 b^2 b'^2 } \cos (\alpha  + \beta ' - \alpha ' - \beta )) \\ 
  &=& \!\!\frac{2}{{E^2 }}(a^2 (b^2 \! + \! b'^2 ) \! + \! (a^2 \! + \! a'^2 )b^2 \! + \! 2\sqrt {\!a^2 (1 \! \! -  \! a^2 )b^2 (1 \! \! -  \! b^2 )} \cos (\alpha \! + \!\! \beta ' \! \! - \! \! \alpha ' \! \! - \! \! \beta )) \\ 
  &=& \frac{2}{{E^2 }}(a^2  + b^2  + 2\sqrt {(a^2  - a^4 )(b^2  - b^4 )} \cos (\alpha  + \beta ' - \alpha ' - \beta )) \\ 
  &=& \frac{{2a^2  + 2b^2  + 4\sqrt {(a^2  - a^4 )(b^2  - b^4 )} \cos (\alpha  + \beta ' - \alpha ' - \beta )}}{{4 + 4a^2 b^2  - 2a^2  - 2b^2  + 4aa'bb'\cos (\beta ' - \alpha ' - \beta  + \alpha )}} \\ 
  &=& \frac{{a^2  + b^2  + 2\sqrt {(a^2  - a^4 )(b^2  - b^4 )} \cos (\alpha  + \beta ' - \alpha ' - \beta )}}{{2 + 2a^2 b^2  - a^2  - b^2  + 2\sqrt {(a^2  - a^4 )(b^2  - b^4 )} \cos (\beta ' - \alpha ' - \beta  + \alpha )}}
\end{eqnarray}
and, since we have  $a^2 = \mu(A)$, $b^2 = \mu(B)$, we get
\begin{equation} \label{eq1.42}
\mu (A \vee B) = \frac{{\mu (A) + \mu (B) + 2\sqrt {(\mu (A) - \mu (A)^2 )(\mu (B) - \mu (B)^2 )} \cos (\alpha  + \beta ' - \alpha ' - \beta )}}{{2 \! + \! 2\mu (A)\mu (B) \!\! - \!\! \mu (A) \!\! - \!\! \mu (B) \! + \! 2\sqrt {(\mu (A) \!\! - \!\! \mu (A)^2 )(\mu (B) \!\! - \!\! \mu (B)^2 )} \cos (\alpha \! + \! \beta ' \! - \! \alpha ' \! - \! \beta )}}
\end{equation}
This gives us the equation to calculate the quantum weight for the disjunction modeled on a two-identical particle quantum model. Let us see how it applies to those items that are badly modeled with the one-particle quantum model of the first sections of this article. 
\subsection{Modeling Hampton's Data with the Two Identical Items Quantum Model}
Consider again the item {\it Apple} with respect to the concepts {\it Fruits} and {\it Vegetables} and {\it Fruits or Vegetables}. We have  $\mu_{exp}(A) = 1$, $\mu_{exp}(B) = 0$ and $\mu_{exp}(A\ {\rm or}\ B) = 1$ for {\it Apple} and if we insert these values of $\mu(A)$ and  $\mu(B)$ into (\ref{eq1.42}) we get  $\mu(A \vee B) = 1$. This means that we obtain a correspondence with the experimentally measured value. Moreover, if one of the values of $\mu(A)$ or  $\mu(B)$ equals 1, then we get (this case assumes $\mu(B) = 1$, but the same result is obtained with  $\mu(B) = 1$)
\begin{equation}
\mu (A \vee B) = \frac{{\mu (A) + 1}}{{2 + 2\mu (A) - \mu (A) - 1}} = \frac{{\mu (A) + 1}}{{\mu (A) + 1}} = 1
\end{equation}
This means that the two identical particles quantum model for concept disjunction predicts a classical logic like outcome for the case where the weight of an item with respect to one of the concepts equals 1.

Another interesting observation is that the quantum interference term in (\ref{eq1.42}) disappears if $\mu(A) = \mu(A)^2$ or if $\mu(B) = \mu(B)^2$. Since  $\mu(A)$ and  $\mu(B)$ are numbers between 0 and 1, this is only the case if  $\mu(A) = 1$ or  $\mu(A) = 0$ or  $\mu(B) = 1$ or $\mu(B) = 0$. Hence classical logic behavior of the two identical quantum particle model only occurs in the extreme cases of the weights with respect to the individual concepts being 1 or 0. In all other cases, also the two identical particle quantum model intrinsically contains quantum interference terms.

Table 2 contains the 64 items of a totality of 176 items considered in the Hampton (1988b) experiment that are not modeled with an exact match by means of the one particle quantum model for concepts developed in Aerts (2007a). In Table 3 we consider the set of items of Table 2 and see how these items can be modeled by means of the two identical particles model. Of these 64 items, 18 can now be modeled by means of the two identical particles model such that there is an exact match between the predicted weight  $\mu_{quant}(A \vee B)$ and the experimental weight $\mu_{exp}(A\ {\rm or}\ B)$.

More specifically, the items modeled with a perfect match by means of the two identical quantum particles model are: {\it Shelves} with respect to the concepts {\it House Furnishings} and {\it Furniture}; {\it Gardening}, {\it Theatre-Going}, {\it Monopoly}, {\it Fishing}, {\it Guitar Playing} and {\it Stamp Collecting} with respect to the concepts {\it Hobbies} and {\it Games}; {\it Collie Dog} with respect to the concepts {\it Pets} and {\it Farmyard Animals}; {\it Cinnamon} with respect to the concepts {\it Spices} and {\it Herbs}; {\it Pliers} with respect to the concepts {\it Instruments} and {\it Tools}; {\it Baseball Bat}, {\it Sailing Life Jacket} and {\it Tennis Racket} with respect to the concepts {\it Sportswear} and {\it Sports Equipment}; {\it Carving Knife} and {\it Cooking Stove} with respect to the concepts {\it Household Appliances} and {\it Kitchen Utensils}; and {\it Apple} with respect to the concepts {\it Fruits} and {\it Vegetables}.

\scriptsize
\setlongtables 
\begin{longtable}{|l|l|l|l|l|l|l|} 
\caption{The list of items that cannot be modeled well by the one-particle quantum model for the disjunction developed in Aerts (2007a). $\mu_{exp}(A)$, $\mu_{exp}(B)$ and $\mu _{\exp } (A{\rm{\ or\ }}B)$  are the membership weights for concept  $A$ and concept $B$, as measured in Hampton (1988b). $\alpha + \beta' - \alpha' - \beta$ are the quantum angles, which need to be chosen for the predicted quantum weights $\mu_{quant}(A \vee B)$ within the two identical items model to be equal to the experimental weights $\mu _{\exp } (A{\rm{\ or\ }}B)$.} \\
\hline 
 & $\mu_{exp}(A)$ & $\mu_{exp}(B)$ & $\mu _{\exp } (A{\rm{\ or\ }}B)$ & $\alpha + \beta' - \alpha' - \beta$ & $\mu_{quant}(A \vee B)$ & $|\mu_{exp} - \mu_{quant}| $ \\ 
\endfirsthead 
\hline 
 & $\mu_{exp}(A)$ & $\mu_{exp}(B)$ & $\mu _{\exp } (A{\rm{\ or\ }}B)$ & $\alpha + \beta' - \alpha' - \beta$ & $\mu_{quant}(A \vee B)$ & $|\mu_{exp} - \mu_{quant}| $ \\
\hline \hline
\endhead 
\hline \hline  
$A$ = {\it House Furnishings}  &  &  &  & & &  \\
$B$ = {\it Furniture} &  &  &  & & &  \\
\hline
{\it Shelves} & 0.4 & 1 & 1 & 0$^\circ$ & 1 & 0 \\
{\it Wall Hangings} & 0.4 & 0.9 & 0.95 & 0$^\circ$ & 0.92998583 & 0.020014 \\
{\it Wall Mirror} & 0.6 & 1 & 0.95 & 0$^\circ$ & 1 & 0.05 \\
{\it Park Bench} & 0 & 0.3 & 0.05 & 180$^\circ$ & 0.176470588 & 0.126471 \\
\hline
$A$ = {\it Hobbies} & & & & & & \\
$B$ = {\it Games} & & & & & & \\
\hline
{\it Gardening} & 1 & 0 & 1 & 0$^\circ$ & 1 & 0 \\
{\it Theatre-Going} & 1 & 0 & 1 & 0$^\circ$ & 1 & 0  \\
{\it Monopoly} & 0.7 & 1 & 1 & 0$^\circ$ & 1 & 0 \\
{\it Fishing} & 1 & 0.6 & 1 & 0$^\circ$ & 1 & 0 \\
{\it Camping} & 1 & 0.1 & 0.9 & 0$^\circ$ & 1 & 0.1  \\
{\it Skating} & 1 & 0.5 & 0.9 & 0$^\circ$ & 1 & 0.1  \\
{\it Guitar Playing} & 1 & 0 & 1 & 0$^\circ$ & 1 & 0  \\
{\it Autograph Hunting} & 1 & 0.2 & 0.9 & 0$^\circ$ &  1 & 0.1  \\
{\it Jogging} & 1 & 0.4 & 0.9 & 0$^\circ$ & 1 & 0.1  \\
{\it Keep Fit} & 1 & 0.3 & 0.95 & 0$^\circ$ &  1 & 0.05  \\
{\it Noughts} & 0.5 & 1 & 0.9 & 0$^\circ$ &  1 & 0.1  \\
{\it Rock Climbing} & 1 & 0.2 & 0.95 & 0$^\circ$ & 1 & 0.05  \\
{\it Stamp Collecting} & 1 & 0.1 & 1 & 0$^\circ$ & 1 & 0  \\
\hline
$A$ = {\it Pets} & & & & & & \\
$B$ = {\it Farmyard Animals} & & & & & & \\
\hline
{\it Goldfish} & 1 & 0 & 0.95 & 0$^\circ$  & 1 & 0.05 \\
{\it Collie Dog} & 1 & 0.7 & 1 & 0$^\circ$  & 1 & 0  \\
{\it Camel} & 0.4 & 0 & 0.1 & 180$^\circ$ & 0.25 & 0.15 \\
{\it Guide Dog for the Blind} & 0.7 & 0 & 0.9 & 0$^\circ$  & 0.538461538 & 0.361538  \\
{\it Prize Bull} & 0.1 & 1 & 0.9 & 0$^\circ$ & 1 & 0.1  \\
{\it Siamese Cat} & 1 & 0.1 & 0.95 & 0$^\circ$ & 1 & 0.05  \\
{\it Ginger Tom-Cat} & 1 & 0.8 & 0.95 & 0$^\circ$ & 1 & 0.05  \\
{\it Cart Horse} & 0.4 & 1 & 0.85 & 0$^\circ$  & 1 & 0.15  \\
{\it Chicken} & 0.3 & 1 & 0.95 & 0$^\circ$  & 1 & 0.05  \\
\hline
$A$ = {\it Spices} & & & & & & \\
$B$ = {\it Herbs} & & & & & & \\
\hline
{\it Chili Pepper} & 1 & 0.6 & 0.95 & 0$^\circ$  & 1 & 0.05 \\
{\it Cinnamon} & 1 & 0.4 & 1 & 0$^\circ$  & 1 & 0  \\
{\it Parsley} & 0.5 & 0.9 & 0.95 & 0$^\circ$  & 0.944444444 & 0.226393 \\
{\it Sugar} & 0 & 0 & 0.2 & 0$^\circ$  & 0 & 0.2 \\
{\it Chires} & 0.6 & 1 & 9.95 & 0$^\circ$  & 1 & 0.05 \\
\hline
$A$ = {\it Instruments} & & & & & & \\
$B$ = {\it Tools} & & & & & & \\
\hline
{\it Magnetic Compass} & 0.9 & 0.5 & 1 & 0$^\circ$  & 0.944444444 & 0.055556  \\
{\it Tuning Fork} & 0.9 & 0.6 & 1 & 0$^\circ$  & 0.957309171 & 0.042691  \\
{\it Pen-Knife} & 0.65 & 1 & 0.95 & 0$^\circ$  & 1 & 0.05  \\
{\it Skate Board} & 0.1 & 0 & 0 & 180$^\circ$  & 0.052631579 & 0.052632  \\
{\it Pliers} & 0.8 & 1 & 1 & 0$^\circ$ & 1 & 0  \\
\hline
$A$ = {\it Sportswear} & & & & & & \\
$B$ = {\it Sports Equipment} & & & & & & \\
\hline
{\it Circus Clowns} & 0 & 0 & 0.1 & 180$^\circ$ & 0 & 0.1  \\
{\it Diving Mask} & 1 & 1 & 0.95 & 0$^\circ$ & 1 & 0.05  \\
{\it Frisbee} & 0.3 & 1 & 0.85 & 0$^\circ$ & 1 & 0.15  \\
{\it Suntan Lotion} & 0 & 0 & 0.1 & 180$^\circ$ & 0 & 0.1  \\
{\it Gymnasium} & 0 & 0.9 & 0.825 & 180$^\circ$ & 0.818181818 & 0.006818  \\
{\it Wrist Sweat} & 1 & 1 & 0.95 & 0$^\circ$ & 1 & 0.05  \\
{\it Lineman's Flag} & 0.1 & 1 & 0.75 & 0$^\circ$ & 1 & 0.25  \\
\hline
$A$ = {\it Household Appliances} & & & & & & \\
$B$ = {\it Kitchen Utensils} & & & & & & \\
\hline
{\it Fork} & 0.7 & 1 & 0.95 & $0^\circ$ & 1 & 0.05  \\
{\it Freezer} & 1 & 0.6 & 0.95 & $0^\circ $  & 1 & 0.05  \\
{\it Extractor Fan} & 1 & 0.4 & 0.9 & $0^\circ$ & 1 &  0.1 \\
{\it Carving Knife} & 0.7 & 1 & 1 & $0^\circ$ & 1 &  0 \\
{\it Cooking Stove} & 1 & 0.5 & 1 & $0^\circ$ & 1 &  0 \\
{\it Iron} & 1 & 0.3 & 0.95 & $0^\circ$  & 1 &  0.05 \\
{\it Chopping Board} & 0.45 & 1 & 0.95 & $0^\circ$ & 1 &  0.05 \\
{\it Television} & 0.95 & 0 & 0.85 & $0^\circ$ & 0.904761905 &  0.054762 \\
{\it Vacuum Cleaner} & 1 & 0 & 1 & $0^\circ$ & 1 &  0 \\
{\it Rolling Pin} & 0.45 & 1 & 1 & $0^\circ$ & 1 &  0 \\
{\it Frying Pan} & 0.7 & 1 & 0.95 & $0^\circ$ & 1 &  0.05 \\
\hline
$A$ = {\it Fruits} & & & & & & \\
$B$ = {\it Vegetables} & & & & & & \\
\hline
{\it Apple} & 1 & 0 & 1 & 0$^\circ$ & 1 & 0  \\
{\it Broccoli} & 0 & 0.8 & 1 & 0$^\circ$ & 0.666666667 & 0.333333 \\
{\it Raisin} & 1 & 0 & 0.9 & $0^\circ$ & 1 & 0.1 \\
{\it Coconut} & 0.7 & 0 & 1 & 0$^\circ$ & 0.538461538 & 0.461538  \\
{\it Mushroom} & 0 & 0.5 & 0.8 & 0$^\circ$ & 0.333333333 & 0.566667 \\
{\it Yam} & 0.43 & 0.67 & 1 & $0^\circ $ & 0.551042236 & 0.298958 \\
{\it Elderberry} & 1 & 0 & 0.55 & $0^\circ $ & 1 & 0.45 \\
\hline
\hline 
\end{longtable}
\normalsize
For {\it Wall Hangings}, {\it Wall Mirror} with respect to the concepts {\it House Furnishings} and {\it Furniture} both models, the one-particle model of Aerts (2007a), and the two particle model in this article, are comparable, and attain a value that is less than 0.1 of the experimental value, while {\it Park Bench} with respect to the same pair of concepts is modeled better (difference about 0.03) by the one quantum particle model than by the two identical particles model (difference about 0.1). {\it Camping}, {\it Autograph Hunting}, {\it Keep Fit} and {\it Rock Climbing} with respect to the concepts {\it Hobbies} and {\it Games} do better with the two identical particles model as compared to the one-particle model, while {\it Skating}, {\it Jogging} and {\it Noughts} with respect to the same pair of concepts do better with the one-particle model as compared to the two-particle model. {\it Gold Fish}, {\it Prize Bull}, {\it Siamese Cat} and {\it Chicken} with respect to the concepts {\it Pets} and {\it Farmyard Animals} do better in the two identical particles model, while {\it Camel}, {\it Guide Dog for the Blind}, {\it Ginger Tom-Cat} and {\it Cart Horse} with respect to the same pair of concepts do better in the one-particle model. 

For the item {\it Parsley} with respect to the concepts {\it Spices} and {\it Herbs}, the one-particle model does much better (0.002) as compared to the two identical particles model (0.22), but for the other items, {\it Chili Pepper}, {\it Sugar} and {\it Chires}, with respect to the same pair of concepts, both models are comparable. For the pair of concepts {\it Instruments} and {\it Tools}, for the considered items in Tables 2 and 3, i.e. {\it Magnetic Compass}, {\it Tuning Fork}, {\it Pen-Knife} and {\it Skate Board}, both models deliver comparable matches, except of course for the one item {\it Pliers}, where the two identical particles model delivers a perfect match with the experimental value. For the pair of concepts {\it Sportswear} and {\it Sports Equipments}, apart from the perfect matches for the items {\it Baseball Bat}, {\it Sailing Life Jacket} and {\it Tennis Racket}, a similar pattern exists; for some of the remaining items one of the quantum models gives a better match, while for the other items the other quantum model gives a better match. A similar pattern repeats itself for the pair of concepts {\it Household Appliances} and {\it Kitchen Utensils} and {\it Fruits} and {\it Vegetables}.

In conclusion, we can state that both models apply and that each one does extremely well for a substantial part of the items, delivering perfect matches with the experimental values. We have also seen that both models do moderately well for a substantial group of items, and that there is a non negligible group of items where they do not well at all. The models are mostly complementary, in the sense that where the one does extremely well, the other one fails, and vice versa. It is this state of affairs which was our inspiration to propose the quantum field theoretic model that we will elaborate in the next section. We will see that the quantum field theoretic model not only produces perfect matches for almost all items, but also introduces a cognitive interaction dynamics that allows very straightforward interpretation for the tested cognitive situations.

\section{Development of a Quantum Field Model}
Quantum Field Theory is amongst the most sophisticated of all physical theories. When working out the quantum model in Hilbert space presented in Aerts (2007a), we wondered if it would be necessary to have recourse to a field theory to resolve the encountered problem. But we also felt reluctant, most of all because we were very much aware of the complexity of quantum field theory from a mathematical point of view, but also from a conceptual point of view. 

However, in our search for proper models, including the one particle quantum model in Aerts (2007a) and the two identical particle quantum model introduced above, for example by establishing for which items and concept disjunctions the one-particle model worked well and for which items and concept disjunctions the two identical particles theory presented in this article worked well, we gradually recognized that a quantum field theoretic model would bring the solution. Moreover, and reflecting on the differences in dynamical interaction offered by the one particle model and the two identical particles model, we were much surprised to observe that complicated conceptual aspects of quantum field theoretic dynamics were much in line with our intuition of how things might have happened in the minds of the tested subjects in the Hampton (1988b) experiment, and with the many analogous experiments on concept combinations. It dawned upon us that the interpretation of quantum field theory for physics is much less paradoxical than it had seemed to be, and that it is even quite natural for cognition. This was an equally important aspect for us turning to the development of a field theoretic model as the growing conviction of being able to find perfect matches of predictions with experiments within such a field theoretic model for cognition.

There is another event that strengthens our convinction that we have found the right theory for the type of cognitive interactions taking place in the experiments of the style of Hampton (1988b), but most probably also a theory for other aspects of cognitive interaction, and even possibly a theory that can be developed into a general framework for overall cognitive interaction, and if so containing an important breakthrough for modeling in cognition. Once we had built the quantum field theory model, and after establishing that we could find perfect matches for almost all data in Hampton, we turned to experiments on conjunctions, which are much more numerous. For these experiments, the field theoretic model produces predictions matching the experimental results even better than for the disjunction. We present this quantum field theoretic modeling for the conjunction of concepts in Aerts (2007b).

Our belief in the field theoretic model was further strenghtened by the question: ``Why would cognitive interaction not be at least as complicated as, and plausibly even more complicated than elementary particle interaction, so why would it not need a theory at least equally complex and sophisticated as quantum field theory, which is the theory describing elementary particle interaction?"

\subsection{Fock Space and a Field Theoretic Dynamics for Cognitive Interaction}

The prime aspect of quantum field theory, which makes it different, and a generalization of quantum mechanics, is that states of a quantum field do not necessarily correspond to a fixed number of quantum particles. Hence, concretely, the state space of a quantum field contains states corresponding to different numbers of particles, and even superpositions of states corresponding to different numbers of particles. If the quantum field is in such a superposition, this means that the number of particles contained in this state is indeterminate. Technically, the set of states of a quantum field is a Fock space, which is the direct sum of tensor products of Hilbert spaces, where each Hilbert space is the state space of a one-particle quantum entity. 

Hence we need, for the situation we want to consider, a Fock space consisting of the direct sum of a one-particle Hilbert space, and a two identical particles Hilbert space, because these are the situations that are relevant to our problem. This means that we need the Fock space
\begin{equation}
{\cal F} = \compl^2 \oplus (\compl^2 \otimes \compl^2)
\end{equation}
which is a 6-dimensional complex vector space. Vectors $\left| AB \right\rangle $ and $\left| A'B' \right\rangle $ are the base vectors of the first  $\compl^2$ in ${\cal F}$ we need to consider and vectors  $\left| {e_1 } \right\rangle  \otimes \left| {e_1 } \right\rangle $, $\left| {e_2 } \right\rangle  \otimes \left| {e_2 } \right\rangle $, $\frac{1}{{\sqrt 2 }}(\left| {e_1 } \right\rangle  \otimes \left| {e_2 } \right\rangle  + \left| {e_2 } \right\rangle  \otimes \left| {e_1 } \right\rangle )$ and  $\frac{1}{{\sqrt 2 }}(\left| {e_1 } \right\rangle  \otimes \left| {e_2 } \right\rangle  - \left| {e_2 } \right\rangle  \otimes \left| {e_1 } \right\rangle )$ are the base vectors of $\compl^2 \otimes \compl^2$ in ${\cal F}$ we need to consider. Hence the base of the Fock space ${\cal F}$ we consider consists of the 6 vectors 
\begin{equation}
\left\{ \! {\left|  {AB}  \right\rangle \! , \!\left| {A'B'} \right\rangle \! , \! \left| {e_1 } \right\rangle \! \otimes \! \left| {e_1 } \right\rangle \! , \! \frac{1}{{\sqrt 2 }}(\left| {e_1 } \right\rangle \! \otimes \! \left| {e_2 } \right\rangle \! + \! \left| {e_2 } \right\rangle \! \otimes \! \left| {e_1 } \right\rangle )} , \! {\left| {e_2 } \right\rangle  \! \otimes \! \left| {e_2 } \right\rangle \! , \! \frac{1}{{\sqrt 2 }}(\left| {e_1 } \right\rangle \! \otimes \! \left| {e_2 } \right\rangle \! - \! \left| {e_2 } \right\rangle \! \otimes \! \left| {e_1 } \right\rangle )} \! \right\}
\end{equation}
Let us now turn directly to the modeling of the situation of an item $X$ and its weights with respect to the disjunction of two concepts  $A$ and  $B$. Each of the base vectors represents an archetypical situation that we have considered already. Hence:  $\left| AB \right\rangle $ represents the situation of the new `concept $A$ or $B$';  $\left| A'B' \right\rangle $ is the situation orthogonal to this. Both  $\left| AB \right\rangle $ and  $\left| A'B' \right\rangle $ are `one-item' situations, i.e. membership and non membership of one item  $X$ with respect to the new `concept $A$ or $B$' are considered. We refer to Aerts (2007a) for a detailed analysis of the one item situation.

The four other base vectors represent `two identical items' situations, i.e. membership and non membership of two identical items  $X$ are considered with respect to combinations of concepts $A$ and  $B$. More specifically,   $\left| {e_1 } \right\rangle  \otimes \left| {e_1 } \right\rangle $ represents membership of both identical items $X$ of both concepts $A$ and  $B$,  $\frac{1}{{\sqrt 2 }}(\left| {e_1 } \right\rangle  \otimes \left| {e_2 } \right\rangle  + \left| {e_2 } \right\rangle  \otimes \left| {e_1 } \right\rangle )$ represents membership of one of the identical items $X$ of one of the concept $A$ or $B$ and non membership of the other item $X$ of the other concept, and  $\left| {e_2 } \right\rangle  \otimes \left| {e_2 } \right\rangle $ represents non membership of both identical items  of both concepts $A$ and  $B$. What about the base vector  $\frac{1}{{\sqrt 2 }}(\left| {e_1 } \right\rangle  \otimes \left| {e_2 } \right\rangle  - \left| {e_2 } \right\rangle  \otimes \left| {e_1 } \right\rangle )$? This vector is the famous vacuum of quantum field theory. It represents the situation of zero items. Calculation of the weights indeed shows that for any vector representing item  $X$, the weight with respect to base vector  $\frac{1}{{\sqrt 2 }}(\left| {e_1 } \right\rangle  \otimes \left| {e_2 } \right\rangle  - \left| {e_2 } \right\rangle  \otimes \left| {e_1 } \right\rangle )$ is zero.

The 6 base vectors we consider are often denoted in a different way in quantum field theory, namely by indicating the number of quantum particles involved. Using this standard quantum field notation we have
\begin{eqnarray}
 \left| {00} \right\rangle  &=& \frac{1}{{\sqrt 2 }}(\left| {e_1 } \right\rangle  \otimes \left| {e_1 } \right\rangle  - \left| {e_2 } \right\rangle  \otimes \left| {e_1 } \right\rangle ) \\ 
 \left| {10} \right\rangle  &=& \left| {AB} \right\rangle ,\left| {01} \right\rangle  = \left| {A'B'} \right\rangle  \\ 
 \left| {20} \right\rangle  &=& \left| {e_1 } \right\rangle  \otimes \left| {e_1 } \right\rangle ,\left| {11} \right\rangle  = \frac{1}{{\sqrt 2 }}(\left| {e_1 } \right\rangle  \otimes \left| {e_1 } \right\rangle  + \left| {e_2 } \right\rangle  \otimes \left| {e_1 } \right\rangle ),\left| {02} \right\rangle  = \left| {e_2 } \right\rangle  \otimes \left| {e_2 } \right\rangle
\end{eqnarray}
where $\left| {00} \right\rangle $ is the vacuum, in our case the situation with no items present to be tested, $\left| {01} \right\rangle $ and $\left| {10} \right\rangle $ are the one-particle states, in our case the situations with one item  $X$ to be tested, hence  $\left| {01} \right\rangle $ represents the situation where the one item  $X$ is a member of the new `concept $A$ or $B$', and  $\left| {10} \right\rangle $ the situation where the one item  $X$ is not a member of the new `concept $A$ or $B$'. 

Vectors  $\left| {20} \right\rangle $,  $\left| {11} \right\rangle $ and $\left| {02} \right\rangle $ represent the two identical particles states. In our case, this means the situations where two identical items are involved, and  $\left| {20} \right\rangle $ is the situation of membership of both items with respect to concepts $A$ and  $B$, while  $\left| {11} \right\rangle $ is the situation of membership of one of the identical items with respect to one of the concepts  $A$ or  $B$ and non membership of the other item with respect to the other concept, and  $\left| {20} \right\rangle $ the situation of non membership of both items with respect to concepts  $A$ and  $B$.

The state of the item  $X$ is represented by means of a vector of the Fock space  ${\cal F}$; this means in general a linear combination of the base vectors. Taking into account the results in Aerts (2007a) with the `one-particle quantum model' and the results of the foregoing sections with respect to the `two identical particles quantum model', we propose the following vector for the representation of the state of item $X$ within the quantum field theoretic model
\begin{equation} \label{eq2.4}
\left| x \right\rangle  = ce^{i\gamma } \left| X \right\rangle  + c'e^{i\gamma '} \frac{1}{E}(\left| U \right\rangle  \otimes \left| V \right\rangle  + \left| V \right\rangle  \otimes \left| U \right\rangle )
\end{equation}
with
\begin{equation}
c^2  + c'^2  = 1
\end{equation}
where, see (65) and (66) of Aerts (2007a), we have
\begin{eqnarray}
 \left| X \right\rangle  &=& ae^{i\alpha } \left| A \right\rangle  + a'e^{i\alpha '} \left| {A'} \right\rangle  \\ 
  &=& be^{i\beta } \left| B \right\rangle  + b'e^{i\beta '} \left| {B'} \right\rangle 
\end{eqnarray}
and, see (\ref{eq1.26}) and (\ref{eq1.27}), we have
\begin{eqnarray}
 \left| U \right\rangle  &=& ae^{i\alpha } \left| {e_1 } \right\rangle  + a'e^{i\alpha '} \left| {e_2 } \right\rangle  \\ 
 \left| V \right\rangle  &=& be^{i\beta } \left| {e_1 } \right\rangle  + b'e^{i\beta '} \left| {e_2 } \right\rangle 
\end{eqnarray}
and, see (\ref{eq1.31}) and (77), (78), (87) and (173) of Aerts (2007a), we have
\begin{equation} \label{eq2.8}
E = \sqrt {2 + 2\left| {\left\langle {U}
 \mathrel{\left | {\vphantom {U V}}
 \right. \kern-\nulldelimiterspace}
 {V} \right\rangle } \right|^2 }  = \sqrt {2 + 2a^2 b^2  + 2a'^2 b'^2  + 4aa'bb'\cos (\beta  - \alpha  - \beta ' + \alpha ')} 
\end{equation}
\begin{equation} \label{eq2.9}
D = \sqrt {2 + 2\Re \left\langle {A}
 \mathrel{\left | {\vphantom {A B}}
 \right. \kern-\nulldelimiterspace}
 {B} \right\rangle }  = \sqrt {2 + 2ab\cos (\beta  - \alpha ) + 2a'b'\cos (\beta ' - \alpha ')} 
\end{equation}
\begin{eqnarray}
 \left| {AB} \right\rangle  &=& \frac{1}{D}(\left| A \right\rangle  + \left| B \right\rangle ) \\ 
 \left| {A'B'} \right\rangle  &=& \frac{1}{D}(\left| {A'} \right\rangle  + \left| {B'} \right\rangle ) 
\end{eqnarray}
Let us calculate the amplitudes corresponding to the different situations.

1) The amplitude corresponding to the item  $X$ that is a member of the new `concept  $A$ or $B$' is given by
\begin{equation} \label{eq2.11}
\left\langle {{AB}}
 \mathrel{\left | {\vphantom {{AB} x}}
 \right. \kern-\nulldelimiterspace}
 {x} \right\rangle  = ce^{i\gamma } \left\langle {{AB}}
 \mathrel{\left | {\vphantom {{AB} X}}
 \right. \kern-\nulldelimiterspace}
 {X} \right\rangle  = \frac{{ce^{i\gamma } }}{D}(ae^{i\alpha }  + be^{i\beta } )
\end{equation}

2) The amplitude corresponding to the item  $X$ that is not a member of the new `concept $A$ or $B$' is given by
\begin{equation} \label{eq2.12}
\left\langle {{A'B'}}
 \mathrel{\left | {\vphantom {{A'B'} x}}
 \right. \kern-\nulldelimiterspace}
 {x} \right\rangle  = ce^{i\gamma } \left\langle {{A'B'}}
 \mathrel{\left | {\vphantom {{A'B'} X}}
 \right. \kern-\nulldelimiterspace}
 {X} \right\rangle  = \frac{{ce^{i\gamma } }}{D}(a'e^{i\alpha '}  + b'e^{i\beta '} )
\end{equation}

3) The amplitude for one of two identical items  $X$ that is a member of concept $A$ and the other one of the two identical items  $X$ that is a member of concept $B$ is given by
\begin{eqnarray} 
 (\left\langle {e_1 } \right| \otimes \left\langle {e_1 } \right|)(\left| x \right\rangle ) &=& \frac{{c'e^{i\gamma '} }}{E}(\left\langle {e_1 } \right| \otimes \left\langle {e_1 } \right|)(\left| U \right\rangle  \otimes \left| V \right\rangle  + \left| V \right\rangle  \otimes \left| U \right\rangle ) \\ 
  &=& \frac{{c'e^{i\gamma '} }}{E}(\left\langle {{e_1 }}
 \mathrel{\left | {\vphantom {{e_1 } U}}
 \right. \kern-\nulldelimiterspace}
 {U} \right\rangle \left\langle {{e_1 }}
 \mathrel{\left | {\vphantom {{e_1 } V}}
 \right. \kern-\nulldelimiterspace}
 {V} \right\rangle  + \left\langle {{e_1 }}
 \mathrel{\left | {\vphantom {{e_1 } V}}
 \right. \kern-\nulldelimiterspace}
 {V} \right\rangle \left\langle {{e_1 }}
 \mathrel{\left | {\vphantom {{e_1 } U}}
 \right. \kern-\nulldelimiterspace}
 {U} \right\rangle ) \\ 
  &=& \frac{{2c'e^{i\gamma '} }}{E}\left\langle {{e_1 }}
 \mathrel{\left | {\vphantom {{e_1 } U}}
 \right. \kern-\nulldelimiterspace}
 {U} \right\rangle \left\langle {{e_1 }}
 \mathrel{\left | {\vphantom {{e_1 } V}}
 \right. \kern-\nulldelimiterspace}
 {V} \right\rangle  \\ \label{eq2.13}
  &=& \frac{{2c'e^{i\gamma '} }}{E}abe^{i(\alpha  + \beta )}
\end{eqnarray}
4) The amplitude for one of two identical items $X$ to be a member of one of the concepts $A$ or $B$ and for the other of the two identical items $X$ to be not a member of the other one of the concepts $A$ or $B$ is given by
\begin{eqnarray}
&& \!\!\!\! \frac{1}{{\sqrt 2 }}(\left\langle {e_1 } \right| \otimes \left\langle {e_2 } \right| + \left\langle {e_2 } \right| \otimes \left\langle {e_1 } \right|)(\left| x \right\rangle ) \\
&& \!\!\!\! = \frac{{c'e^{i\gamma '} }}{{\sqrt 2 E}}(\left\langle {e_1 } \right| \otimes \left\langle {e_2 } \right| + \left\langle {e_2 } \right| \otimes \left\langle {e_1 } \right|)(\left| U \right\rangle  \otimes \left| V \right\rangle  + \left| V \right\rangle  \otimes \left| U \right\rangle ) \\ 
&& \!\!\!\! = \! \frac{{c'e^{i\gamma '} }}{{\sqrt 2 E}}(\left\langle {{e_1 }}
 \mathrel{\left |  {\vphantom {{e_1 } U}}
 \right. \kern-\nulldelimiterspace}
 {U} \right\rangle \! \left\langle {{e_2 }}
 \mathrel{\left | {\vphantom {{e_2 } V}}
 \right. \kern-\nulldelimiterspace}
 {V} \right\rangle \! + \! \left\langle {{e_1 }}
 \mathrel{\left | {\vphantom {{e_1 } V}}
 \right. \kern-\nulldelimiterspace}
 {V} \right\rangle \! \left\langle {{e_2 }}
 \mathrel{\left | {\vphantom {{e_2 } U}}
 \right. \kern-\nulldelimiterspace}
 {U} \right\rangle \! + \! \left\langle {{e_2 }}
 \mathrel{\left | {\vphantom {{e_2 } U}}
 \right. \kern-\nulldelimiterspace}
 {U} \right\rangle \! \left\langle {{e_1 }}
 \mathrel{\left | {\vphantom {{e_1 } V}}
 \right. \kern-\nulldelimiterspace}
 {V} \right\rangle \! + \! \left\langle {{e_2 }}
 \mathrel{\left | {\vphantom {{e_2 } V}}
 \right. \kern-\nulldelimiterspace}
 {V} \right\rangle \! \left\langle {{e_1 }}
 \mathrel{\left | {\vphantom {{e_1 } U}}
 \right. \kern-\nulldelimiterspace}
 {U} \right\rangle ) \\ 
&& \!\!\!\! = \frac{{2c'e^{i\gamma '} }}{{\sqrt 2 E}}(\left\langle {{e_1 }}
 \mathrel{\left | {\vphantom {{e_1 } U}}
 \right. \kern-\nulldelimiterspace}
 {U} \right\rangle \left\langle {{e_2 }}
 \mathrel{\left | {\vphantom {{e_2 } V}}
 \right. \kern-\nulldelimiterspace}
 {V} \right\rangle  + \left\langle {{e_1 }}
 \mathrel{\left | {\vphantom {{e_1 } V}}
 \right. \kern-\nulldelimiterspace}
 {V} \right\rangle \left\langle {{e_2 }}
 \mathrel{\left | {\vphantom {{e_2 } U}}
 \right. \kern-\nulldelimiterspace}
 {U} \right\rangle ) \\ \label{eq2.14}
&& \!\!\!\! = \frac{{2c'e^{i\gamma '} }}{{\sqrt 2 E}}(ab'e^{i(\alpha  + \beta ')}  + a'be^{i(\alpha ' + \beta )} )
\end{eqnarray}

5) The amplitude for one of the identical items  $X$ not to be a member of concept $A$ and the other of the identical items $X$ not to be a member of the concept $B$ is given by
\begin{eqnarray}
 (\left\langle {e_2 } \right| \otimes \left\langle {e_2 } \right|)(\left| x \right\rangle ) &=& \frac{{c'e^{i\gamma '} }}{E}(\left\langle {e_2 } \right| \otimes \left\langle {e_2 } \right|)(\left| U \right\rangle  \otimes \left| V \right\rangle  + \left| V \right\rangle  \otimes \left| U \right\rangle ) \\ 
  &=& \frac{{c'e^{i\gamma '} }}{E}(\left\langle {{e_2 }}
 \mathrel{\left | {\vphantom {{e_2 } U}}
 \right. \kern-\nulldelimiterspace}
 {U} \right\rangle \left\langle {{e_2 }}
 \mathrel{\left | {\vphantom {{e_2 } V}}
 \right. \kern-\nulldelimiterspace}
 {V} \right\rangle  + \left\langle {{e_2 }}
 \mathrel{\left | {\vphantom {{e_2 } V}}
 \right. \kern-\nulldelimiterspace}
 {V} \right\rangle \left\langle {{e_2 }}
 \mathrel{\left | {\vphantom {{e_2 } U}}
 \right. \kern-\nulldelimiterspace}
 {U} \right\rangle ) \\ 
  &=& \frac{{2c'e^{i\gamma '} }}{E}\left\langle {{e_2 }}
 \mathrel{\left | {\vphantom {{e_2 } U}}
 \right. \kern-\nulldelimiterspace}
 {U} \right\rangle \left\langle {{e_2 }}
 \mathrel{\left | {\vphantom {{e_2 } V}}
 \right. \kern-\nulldelimiterspace}
 {V} \right\rangle  \\ \label{eq2.15}
  &=& \frac{{2c'e^{i\gamma '} }}{E}a'b'e^{i(\alpha ' + \beta ')}
\end{eqnarray}
If we know the amplitudes for these events, we can calculate the weights.

1) The weight for item  $X$ to be a member of the new `concept $A$ or $B$' is given by
\begin{eqnarray}
 \mu (AB) &=& \left| {\left\langle {{AB}}
 \mathrel{\left | {\vphantom {{AB} x}}
 \right. \kern-\nulldelimiterspace}
 {x} \right\rangle } \right|^2  \\ 
  &=& (\frac{{ce^{i\gamma } }}{D}(ae^{i\alpha }  + be^{i\beta } ))^*(\frac{{ce^{i\gamma } }}{D}(ae^{i\alpha }  + be^{i\beta } )) \\ 
  &=& (\frac{{ce^{ - i\gamma } }}{D}(ae^{ - i\alpha }  + be^{ - i\beta } ))(\frac{{ce^{i\gamma } }}{D}(ae^{i\alpha }  + be^{i\beta } )) \\ 
  &=& \frac{{c^2 }}{{D^2 }}(a^2  + b^2  + 2ab\cos (\beta  - \alpha ))
\end{eqnarray}

2) The weight for item  $X$ to not be a member of the new `concept $A$ or $B$' is given by
\begin{eqnarray}
 \mu (A'B') &=& \left| {\left\langle {{A'B'}}
 \mathrel{\left | {\vphantom {{A'B'} x}}
 \right. \kern-\nulldelimiterspace}
 {x} \right\rangle } \right|^2  \\ 
  &=& (\frac{{ce^{i\gamma } }}{D}(a'e^{i\alpha '}  + b'e^{i\beta '} ))^*(\frac{{ce^{i\gamma } }}{D}(a'e^{i\alpha '}  + b'e^{i\beta '} )) \\ 
  &=& (\frac{{ce^{ - i\gamma } }}{D}(a'e^{ - i\alpha '}  + b'e^{ - i\beta '} ))(\frac{{ce^{i\gamma } }}{D}(a'e^{i\alpha '}  + b'e^{i\beta '} )) \\ 
  &=& \frac{{c^2 }}{{D^2 }}(a'^2  + b'^2  + 2a'b'\cos (\beta ' - \alpha '))
\end{eqnarray}

3) The weight for one of two identical items  $X$ to be a member of concept $A$ and the other one of the two identical items  $X$ to be a member of concept $B$ is given by
\begin{eqnarray}
 \mu (A \wedge B) &=& \left| {(\left\langle {e_1 } \right| \otimes \left\langle {e_1 } \right|)(\left| x \right\rangle )} \right|^2  \\ 
  &=& (\frac{{2c'e^{i\gamma '} }}{E}abe^{i(\alpha  + \beta )} )^*(\frac{{2c'e^{i\gamma '} }}{E}abe^{i(\alpha  + \beta )} ) \\ 
  &=& \frac{{2c'e^{ - i\gamma '} }}{E}abe^{ - i(\alpha  + \beta )} \frac{{2c'e^{i\gamma '} }}{E}abe^{i(\alpha  + \beta )}  \\ 
  &=& \frac{{4c'^2 }}{{E^2 }}a^2 b^2
\end{eqnarray}

4) The weight for one of two identical items  $X$ to be a member of one of concepts $A$ or $B$ and for the other of the two identical items  $X$ to be not a member of the other one of concepts $A$ or $B$ is given by
\begin{eqnarray}
&& \mu ((A \wedge B') \vee (A' \wedge B)) = \left| {\frac{1}{{\sqrt 2 }}(\left\langle {e_1 } \right| \otimes \left\langle {e_2 } \right| + \left\langle {e_2 } \right| \otimes \left\langle {e_1 } \right|)(\left| x \right\rangle )} \right|^2  \\ 
&&  = (\frac{{2c'e^{i\gamma '} }}{{\sqrt 2 E}}(ab'e^{i(\alpha  + \beta ')}  + a'be^{i(\alpha ' + \beta )} ))^*(\frac{{2c'e^{i\gamma '} }}{{\sqrt 2 E}}(ab'e^{i(\alpha  + \beta ')}  + a'be^{i(\alpha ' + \beta )} )) \\ 
&&  = (\frac{{2c'e^{ - i\gamma '} }}{{\sqrt 2 E}}(ab'e^{ - i(\alpha  + \beta ')}  + a'be^{ - i(\alpha ' + \beta )} ))(\frac{{2c'e^{i\gamma '} }}{{\sqrt 2 E}}(ab'e^{i(\alpha  + \beta ')}  + a'be^{i(\alpha ' + \beta )} )) \\ 
&&  = \frac{{2c'^2 }}{{E^2 }}(a^2 b'^2  + a'^2 b^2  + 2aa'bb'\cos (\alpha  + \beta ' - \alpha ' - \beta )) 
\end{eqnarray}
5) The weight for one of the identical items  $X$ to be not a member of concept $A$ and also the other of the identical items  $X$ to be not a member of concept $B$ is given by
\begin{eqnarray}
 \mu (A' \wedge B') &=& \left| {(\left\langle {e_2 } \right| \otimes \left\langle {e_2 } \right|)(\left| x \right\rangle )} \right|^2  \\ 
  &=& (\frac{{2c'e^{i\gamma '} }}{E}a'b'e^{i(\alpha ' + \beta ')} )*)(\frac{{2c'e^{i\gamma '} }}{E}a'b'e^{i(\alpha ' + \beta ')} ) \\ 
  &=& \frac{{2c'e^{ - i\gamma '} }}{E}a'b'e^{ - i(\alpha ' + \beta ')} \frac{{2c'e^{i\gamma '} }}{E}a'b'e^{i(\alpha ' + \beta ')}  \\ 
  &=& \frac{{4c'^2 }}{{E^2 }}a'^2 b'^2 
\end{eqnarray}
We have now derived all equations needed to calculate the weight for the disjunction of two concepts within the quantum field theoretic model. For this, we need to sum three weights (1) weight $\mu(AB)$ for item  $X$ to be a member of the new `concept  $A$ or $B$', plus (2) weight  $\mu(A \wedge B)$ of one of the identical items $X$ to be a member of concept $A$ and the other one of the identical items $X$ to be a member of concept  $B$, plus (3) weight  $\mu((A \wedge B') \vee (A' \wedge B))$ for one of the two identical items $X$ to be a member of one of the concepts  $A$ or $B$ and the other of the identical items to be a member of the other concept. This gives
\begin{eqnarray}
 \mu (A\ {\rm{ or }}\ B) \!\! &=& \!\! \mu (AB) + \mu (A \wedge B) + \mu ((A \wedge B') \vee (A' \wedge B)) \\ 
\!\!\!\!  &=& \frac{{c^2 }}{{D^2 }}(a^2  + b^2  + 2ab\cos (\beta  - \alpha )) \nonumber \\ 
\!\!\!\!  && + \frac{{4c'^2 }}{{E^2 }}a^2 b^2  + \frac{{2c'^2 }}{{E^2 }}(a^2 b'^2  + a'^2 b^2  + 2aa'bb'\cos (\alpha  + \beta ' - \alpha ' - \beta )) \\ 
\!\!\!\!  &=& \frac{{c^2 }}{{D^2 }}(a^2  + b^2  + 2ab\cos (\beta  - \alpha )) \nonumber \\ 
\!\!\!\!  && + \frac{{2c'^2 }}{{E^2 }}(2a^2 b^2  + a^2 b'^2  + a'^2 b^2  + 2\sqrt {a^2 a'^2 b^2 b'^2 } \cos (\alpha  + \beta ' - \alpha ' - \beta )) \\ 
\!\!\!\!  &=& \frac{{c^2 }}{{D^2 }}(a^2  + b^2  + 2ab\cos (\beta  - \alpha )) \nonumber \\ 
\!\!\!\!  && \!\!\!\! \!\!\!\! \!\!\!\! \!\!\!\! \!\!\!\! \!\!\!\!+ \frac{{2c'^2 }}{{E^2 }}(a^2 (b^2  \! + \! b'^2 ) \! + \! (a^2 \! + \! a'^2 )b^2 \! + \! 2\sqrt {a^2 (1 \! - \! a^2 )b^2 (1 \! - \! b^2 )} \cos (\alpha \! + \! \beta ' \! - \! \alpha ' \! - \! \beta )) \\ 
&& \!\!\!\! \!\!\!\! \!\!\!\! \!\!\!\! \!\!\!\! \!\!\!\! = \frac{{c^2 }}{{D^2 }}(a^2  + b^2  + 2ab\cos (\beta  - \alpha )) \nonumber \\ 
&& \!\!\!\! \!\!\!\! \!\!\!\! \!\!\!\! \!\!\!\! \!\!\!\! + \frac{{2c'^2 }}{{E^2 }}(a^2  + b^2  + 2\sqrt {(a^2  - a^4 )(b^2  - b^4 )} \cos (\alpha  + \beta ' - \alpha ' - \beta ))
\end{eqnarray}
Substituting (\ref{eq2.8}) and (\ref{eq2.9}), we get
\begin{eqnarray}
 \mu (A\ {\rm{ or }}\ B) &=& \frac{{c^2 (a^2  + b^2  + 2ab\cos (\beta  - \alpha ))}}{{2 + 2ab\cos (\beta  - \alpha ) + 2a'b'\cos (\beta ' - \alpha ')}} \nonumber \\ 
&&  \!\!\!\! \!\!\!\! \!\!\!\! \!\!\!\!+ \frac{{2c'^2 (a^2  + b^2  + 2\sqrt {(a^2  - a^4 )(b^2  - b^4 )} \cos (\alpha  + \beta ' - \alpha ' - \beta ))}}{{2 + 2a^2 b^2  + 2a'^2 b'^2  + 4aa'bb'\cos (\beta  - \alpha  - \beta ' + \alpha ')}} \\ 
  &=& \frac{{c^2 (a^2  + b^2  + 2ab\cos (\beta  - \alpha ))}}{{2 + 2ab\cos (\beta  - \alpha ) + 2\sqrt {(1 - a^2 )(1 - b^2 )} \cos (\beta ' - \alpha ')}} \nonumber \\ 
&&  \!\!\!\! \!\!\!\! \!\!\!\! \!\!\!\! + \frac{{c'^2 (a^2 \! + b^2 \! + 2\sqrt {(a^2 \! - \! a^4 )(b^2 \! - \! b^4 )} \cos (\alpha \! + \beta ' \! - \alpha ' \! - \! \beta ))}}{{1 \! + \! a^2 b^2 \! + \! (1 \! - \! a^2 )(1 \! - \! b^2 ) \! + \! 2\sqrt {(a^2 \! - \! a^4 )(b^2 \! - \! b^4 )} \cos (\beta  \! - \! \alpha \! - \! \beta ' \! + \! \alpha ')}} \\ 
  &=& \frac{{c^2 (a^2  + b^2  + 2ab\cos (\beta  - \alpha ))}}{{2 + 2ab\cos (\beta  - \alpha ) + 2\sqrt {(1 - a^2 )(1 - b^2 )} \cos (\beta ' - \alpha ')}} \nonumber \\ 
&& \!\!\!\! \!\!\!\! \!\!\!\! \!\!\!\! + \frac{{c'^2 (a^2  + b^2  + 2\sqrt {(a^2  - a^4 )(b^2  - b^4 )} \cos (\alpha  + \beta ' - \alpha ' - \beta ))}}{{2 + 2a^2 b^2  - a^2  - b^2  + 2\sqrt {(a^2  - a^4 )(b^2  - b^4 )} \cos (\beta  - \alpha  - \beta ' + \alpha ')}} 
\end{eqnarray}
Taking into account that $a = \mu(A)$ and  $b = \mu(B)$, we have our final equation
\begin{eqnarray} \label{eq2.23}
\!\!\!\! \!\!\!\! \mu _{quant} (A\ {\rm{ or }}\ B) \!\! \!\! &=& \!\! \!\! \frac{{c^2 (\mu (A) \! + \! \mu (B) \! + \! 2\sqrt {\mu (A)\mu (B)} \cos (\beta \! - \! \alpha ))}}{{2 \! + \! 2\sqrt {\mu (A)\mu (B)} \cos (\beta \! - \! \alpha ) \! + \! 2\sqrt {(1 \! - \! \mu (A))(1 \! - \! \mu (B))} \cos (\beta ' \! - \! \alpha ')}} \nonumber \\ 
&& \!\!\!\! \!\!\!\! \!\!\!\! \!\!\!\! \!\!\!\! \!\!\!\! \!\!\!\! \!\!\!\! \!\!\!\! \!\!\!\! \!\!\!\! \!\!\!\! \!\!\!\! \!\!\!\! + \frac{{c'^2 (\mu (A) + \mu (B) + 2\sqrt {(\mu (A) \! - \! \mu (A)^2 )(\mu (B) \! - \! \mu (B)^2 )} \cos (\alpha \! + \! \beta ' \! - \! \alpha ' \! - \! \beta ))}}{{2 \! + \! 2\mu (A)\mu (B) \! - \! \mu (A) \! - \! \mu (B) \! + \! 2\sqrt {(\mu (A) \! - \! \mu (A)^2 )(\mu (B) \! - \! \mu (B)^2 )} \cos (\beta \!  - \! \alpha \!  - \! \beta ' \! + \! \alpha ')}} 
\end{eqnarray}
This is the quantum field theoretic equation that gives us weight $\mu_{quant}(A\ {\rm or}\ B)$ of an item $X$ with respect to the disjunction of concepts  $A$ or $B$ in function of weight $\mu(A)$ of this item  $X$ with respect to concept $A$ and weight $\mu(B)$ of item  $X$ with respect to concept  $B$. We will now investigate if this equation predicts the experimental results tested in Hampton (1988b). Before we do so in the next section we will consider some special cases of (\ref{eq2.23}).

i) The case where $\mu (A) = \mu (B) = 0$. Equation (\ref{eq2.23}) then gives
\begin{equation}
\mu _{quant} (A\ {\rm{ or }}\ B) = 0
\end{equation}

ii) The case where  $\mu (A) = \mu (B) = 1$. Equation (\ref{eq2.23}) then gives
\begin{equation}
\mu _{quant} (A\ {\rm{ or }}\ B) = \frac{{c^2 (2 + 2\cos (\beta  - \alpha ))}}{{2 + 2\cos (\beta  - \alpha )}} + \frac{{2c'^2 }}{2} = c^2  + c'^2  = 1
\end{equation}

iii) The case where $\mu (A) = 0$
and  $\mu (B) = 1$
or $\mu (A) = 1$
 and  $\mu (B) = 0$. Equation (\ref{eq2.23}) then gives
\begin{equation}
\mu _{quant} (A\ {\rm{ or }}\ B) = \frac{{c^2 }}{2} + \frac{{c'^2 }}{{2 - 1}} = \frac{{c^2 }}{2} + c'^2  = \frac{{c^2  + 2c'^2 }}{2} = \frac{{1 + c'^2 }}{2} = \frac{{2 - c^2 }}{2}
\end{equation}

In this case, $\mu _{quant} (A\ {\rm{ or }}\ B)$ only depends on the value of quantum parameter $c$. The square of this parameter represents the weight of the `one-item situation' as compared to the square of parameter   representing the weight of the `two identical items situation'. Hence, within the field theoretic model, if the weight of the `one-item situation' is 1, hence  $c^2 = 1$, we get  $\mu _{quant} (A\ {\rm{ or }}\ B) = 0.5$, indeed the medium between  $\mu(A) = 0$ and  $\mu(B) = 1$ (or  $\mu(A) = 1$ and  $\mu(B) = 0$). While if the weight of the `one-item situation' equals zero, hence  $c^2 = 0$, and hence the weight of the `two identical items situation' equals 1, meaning $c'^2 = 1$, we get  $\mu _{quant} (A\ {\rm{ or }}\ B) = 1$. Indeed, in this case it is sufficient for one of weights $\mu(A)$ or $\mu(B)$ to be equal to 1, to also have the weight of the disjunction  $\mu _{quant} (A\ {\rm{ or }}\ B)$ become equal to 1.

iv) The case where  $\mu(A) = 1$. Equation (\ref{eq2.23}) then gives us
\begin{equation}
\mu _{quant} (A\ {\rm{ or }}\ B) = \frac{{c^2 (1 + \mu (B) + 2\sqrt {\mu (B)} \cos (\beta  - \alpha ))}}{{2 + 2\sqrt {\mu (B)} \cos (\beta  - \alpha )}} + c'^2 
\end{equation}

v) The case where  $\mu(B) = 1$. Equation (\ref{eq2.23}) then gives us

\begin{equation}
\mu _{quant} (A\ {\rm{ or }}\ B) = \frac{{c^2 (1 + \mu (A) + 2\sqrt {\mu (A)} \cos (\beta  - \alpha ))}}{{2 + 2\sqrt {\mu (A)} \cos (\beta  - \alpha )}} + c'^2 
\end{equation}

vi) The case where  $\mu(A) = 0$. Equation (\ref{eq2.23}) then gives us

\begin{equation}
\mu _{quant} (A\ {\rm{ or }}\ B) = \frac{{c^2 \mu (B)}}{{2 + 2\sqrt {(1 - \mu (B))} \cos (\beta ' - \alpha ')}} + \frac{{c'^2 \mu (B)}}{{2 - \mu (B)}}
\end{equation}

vi) The case where  $\mu(B) = 0$. Equation (\ref{eq2.23}) then gives us

\begin{equation}
\mu _{quant} (A\ {\rm{ or }}\ B) = \frac{{c^2 \mu (A)}}{{2 + 2\sqrt {(1 - \mu (A))} \cos (\beta ' - \alpha ')}} + \frac{{c'^2 \mu (A)}}{{2 - \mu (A)}}
\end{equation}

\subsection{Predicting the Experimental Results by means of the Quantum Field Theoretic Model}

In this section we want to apply the equation that we derived within the quantum field theoretic model to make predictions for the experimental data collected in Hampton (1988b). In Table 4 below we present the results. If we consider equation (\ref{eq2.23}), we can see that for any item  $X$ there are three quantum parameters that can take different values, given weights   and  $\mu_{exp}(A)$ of item  $X$ with respect to concepts $A$ and  $B$. We already know quantum angles  $\beta - \alpha$ and  $\beta' - \alpha'$ from the foregoing models, i.e. the one-item model presented in Aerts (2007a), and the two identical items model of the foregoing sections. In this quantum field model, too, they are the fundamental quantum parameters at the origin of the quantum effect of interference. Their values are given in degrees, with  $90^\circ$ corresponding to a right angle.

In this quantum field model there is an additional parameter that did not appear in the foregoing models, and we denoted it by  $c^2$. It represents the weight of the `one-item model' within the quantum field model, and hence  represents the weight of the `two identical items' model within the quantum field model. More concretely, if  $c^2 = 1$, and hence  $c'^2 = 0$, this means that item $X$ interacts as `one item' with concepts $A$ and  $B$, and more specifically as `one item' with the new concept  $A$ or $B$, and it is this interaction that gives rise to weight  $\mu _{quant} (A\ {\rm{ or }}\ B)$ of item  $X$ with respect to the disjunction, as encountered in equation (\ref{eq2.23}). If  $c^2 = 0$ and hence  $c'^2 = 1$, this means that item   interacts under the form of `two identical items  $X$' with concepts $A$ and $B$ and it is this interaction that gives rise to weight $\mu _{quant} (A\ {\rm{ or }}\ B)$ of item  $X$ with respect to the disjunction, as encountered in equation (\ref{eq2.23}). 

In the general case, where $c^2$ and $c'^2$ are both different from 0 and 1, state  $\left| x \right\rangle $ given in equation (\ref{eq2.4}) is a genuine superposition state of  $\left| X \right\rangle $, the `one-item state' of item  $X$, and  $\frac{1}{E}(\left| U \right\rangle  \otimes \left| V \right\rangle  + \left| V \right\rangle  \otimes \left| U \right\rangle )$, the `two identical items state' of item  $X$, as can be seen from equation (\ref{eq2.4}). Hence this state  $\left| x \right\rangle $ represents a situation where the subject, when asked to estimate the membership weight of item  $X$ with respect to the disjunction of the two concepts $A$ and  $B$, does something `in between' or `mixed', but actually it is better to say 'superposed', because `in between' and `mixed' would be classical ways of describing what happens. More concretely, the subject partly makes this estimate by considering the disjunction of concept  $A$ and concept  $B$ as a new `concept $A$ or $B$', estimating the membership of the item with respect to this new `concept $A$ or $B$', but partly the subject also does something else. He or she considers the disjunction not as a new concept, but relates the membership weight of item  with respect to concept  $A$ or concept  $B$ to the membership weights of item $X$ with respect to concept  $A$ and the membership weight of item $X$ with respect to concept  $B$ and taking into account the way in which the `or' relates these three membership weights. This way consists of considering two identical items  $X$ and their membership weights with respect to both concepts $A$ and  $B$, and then making the `yes' decision for membership, if this `yes' decision for membership is made for one of the two concepts. This is the reason why this part is well modeled by the `two-identical particle quantum model'.

Before we comment in detail the content of Table 4, we introduce a graphical representation of the situation. Consider the item {\it Mantelpiece} for the pair of concepts {\it House Furnishings} and {\it Furniture} and their disjunction {\it House Furnishings or Furniture}. In Table 4 below we have presented the data for a perfect match with experimental results, with  $c^2 = 0.2865$, and the quantum angles $\beta  - \alpha  = {\rm{71}}{\rm{.79797}}^\circ $ and $\beta ' - \alpha ' = {\rm{0}}^\circ $.

In Figure 3, also below, we have represented the graph of $\mu _{quant} (A{\rm{ or }}B)$ of equation (\ref{eq2.23}) in function of  $\beta - \alpha$ and $\beta' - \alpha'$ for the item {\it Mantelpiece}, where we have chosen  $c^2 = 0.2865$, and hence $\mu(A) = 0.4$ and  $\mu(B) = 0.8$.
\begin{figure}[h]
\centerline {\includegraphics[width=16cm]{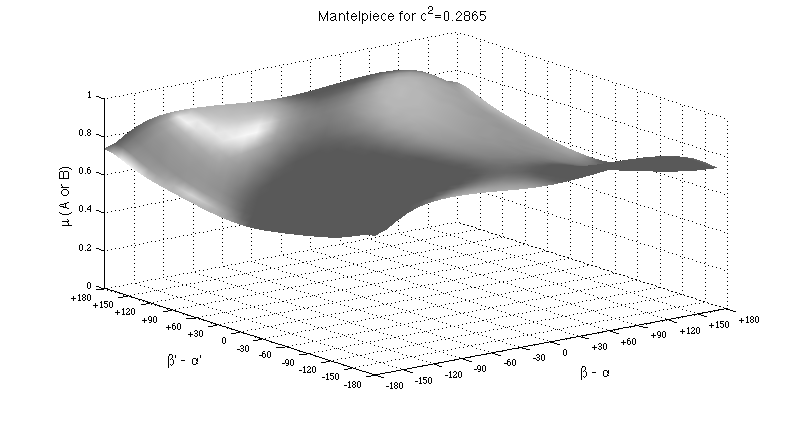}}
\caption{A graphical representation of the membership weight predicted by the quantum field model of the item {\it Mantelpiece} with respect to the disjunction of the concepts {\it House Furnishings} and {\it Furniture} for the value of $c^2 = 0.2865$ and in function of the quantum angles $\beta - \alpha$ and  $\beta' - \alpha'$. This value of  $c^2$ produces a volume under the surface equal to the volume under the horizontal plane, i.e. it is the optimum of all  $c^2$ values, yielding perfect matches between theory and experiment.}
\end{figure}
On the two horizontal axes in Figure 3 the values of the quantum angles $\beta - \alpha$ and $\beta' - \alpha'$ are represented, and on the vertical axis the values of  $\mu_{quant}(A\ {\rm or}\ B)$ are represented. The surface visible in Figure 3 represents the points  $(\beta - \alpha, \beta' - \alpha', \mu_{quant}(A\ {\rm or}\ B)$ for the item {\it Mantelpiece} with respect to the pair of concepts {\it House Furnishings} and {\it Furniture} and their disjunction for a value of  $c^2 = 0.2865$, where  $\beta - \alpha$,  $\beta' - \alpha'$ and $\mu_{quant}(A\ {\rm or}\ B)$ satisfy equation (\ref{eq2.23}). The experimental value in Hampton (1988b) of the weight of the disjunction membership for {\it Mantelpiece} is given by  $\mu_{quant}(A\ {\rm or}\ B) = 0.75$.

This means that for each value of the couple $(\beta - \alpha, \beta' - \alpha')$ in the horizontal plane of Figure 3 where the value of $\mu_{quant}(A\ {\rm or}\ B)$ reaches 0.75, we have a perfect match of the value   predicted by the quantum field model with the experimental value $\mu_{exp}(A\ {\rm or}\ B) = 0.75$. In Figure 4 we have represented the constant value of 0.75 by a horizontal plane at a height of 0.75 parallel to the horizontal ground plane. Where this horizontal plane cuts the surface, we have points with quantum angle values that yield perfect matches between the values predicted by the quantum field model and the experimental values. In Figure 4, we have indicated one of these points -- the one we have chosen in Table 4 -- by means of a dot and its coordinate lines. It is point ($71.79797^\circ$,  $0^\circ$, 0.75), containing the values in Table 4 for the item {\it Mantelpiece}.
\begin{figure}[h]
\centerline {\includegraphics[width=16cm]{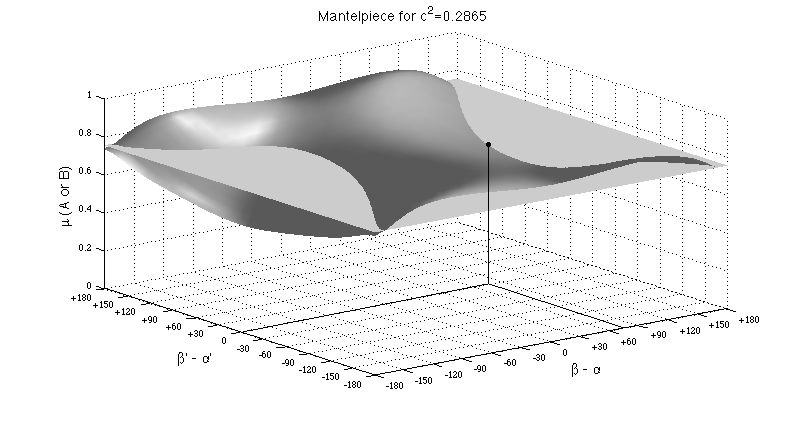}}
\caption{The surface is a graphical representation of the membership weight predicted by the quantum field model of the item {\it Mantelpiece} with respect to the disjunction of the concepts {\it House Furnishings} and {\it Furniture} for  $c^2 = 0.2865$ and in function of the quantum angles $\beta - \alpha$ and  $\beta' - \alpha'$. The horizontal plane is a graphical representation of the membership weight measured in Hampton (1988b) for the item {\it Mantelpiece} with respect to the disjunction of the concepts {\it House Furnishings} and {\it Furniture}. The intersection of the surface with the plane gives the points with perfect matches between theory and experiment. The black dot represents the values chosen in Table 4 for {\it Mantelpiece}.}
\end{figure}
We can easily see in Figure 4 that there are a great many other points that produce perfect matches between theory and experiment, namely all the points where the horizontal plane cuts the surface. Both Figure 3 and Figure 4 represent the item {\it Mantelpiece} for  $c^2 = 0.2865$. Let us explain why we choose  $c^2 = 0.2865$. Figure 5 represents the situations of the item {\it Mantelpiece} for four other values of  $c^2$, more specifically for $c^2 = 0$,  $c^2 = 0.07$,  $c^2 = 0.4$, and  $c^2 = 1$. We have also added the horizontal plane at value 0.75, as in Figure 4.
\begin{figure}[h]
\centerline {\includegraphics[width=16cm]{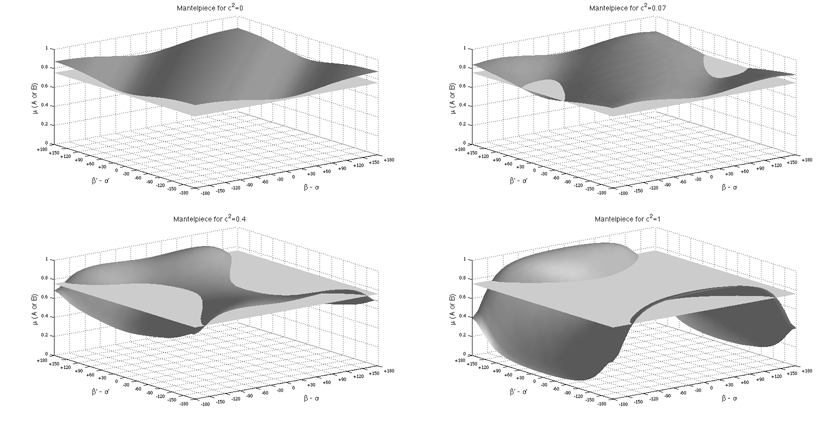}}
\caption{The situation of Figure 4 for different values of  $c^2$. For  $c^2 = 0$, there are no values of the quantum angles that give rise to perfect matches between the predictions of the quantum field model and the experimental results of Hampton (1988b), because the intersection between the surface and the horizontal plane is empty. For the other values of $c^2$ represented in the figures, i.e.  $c^2 = 0.07$,  $c^2 = 0.4$ and  $c^2 = 1$, different values of the quantum angles exist giving rise to perfect matches between the predictions of the quantum field model and the experimental results of Hampton (1988b), namely the values corresponding to the points of the intersection of the surface with the horizontal plane.}
\end{figure}

In Figure 5 we can see that for $c^2 = 0$ there are no points in common between the surface and the horizontal plane, which means that for $c^2 = 0$ there is no perfect match between the value predicted by the quantum field model and the experimental value. For $c^ = 0.07$ the surface already cuts the horizontal plane at a height of 0.75, so that points exist that give rise to perfect matches between theory and experiment for this value of  $c^2$. The same applies to  $c^2 = 0.4$ and  $c^2 = 1$.

However, we have chosen $c^2 = 0.2865$ in Table 4, because this value results in a surface that is generally closest to the horizontal plane at a height of 0.75. We have expressed this by comparing the volume contained under the surface with the volume contained under the horizontal plane at a height of 0.75, and choosing the value of $c^2$ such that these two volumes are as close as possible or even equal in size. The size of the volume under the surface is given by
\begin{equation}
\int_{ - \pi }^{ + \pi } {d(\beta  - \alpha )\int_{ - \pi }^{ + \pi } {d(\beta ' - \alpha ')\mu _{quant} (A\ {\rm{ or }}\ B)} } 
\end{equation}
where $\mu _{quant} (A\ {\rm{ or }}\ B)$ is given by equation (\ref{eq2.23}). The size of the volume under the horizontal plane at 0.75 is given by
\begin{equation}
0.75(2\pi )^2  = {\rm{29}}{\rm{.6088}}
\end{equation}
For $c^2 = 0.2865$ we have
\begin{equation}
\int_{ - \pi }^{ + \pi } {d(\beta  - \alpha )\int_{ - \pi }^{ + \pi } {d(\beta ' - \alpha ')\mu _{quant} (A\ {\rm{ or }}\ B)} }  = 29.6088
\end{equation}
which shows that for this value of  $c^2$ the two volumes are equal in size.

Let us work out one more example in detail. Consider the item {\it Keep Fit} in relation to the concepts {\it Hobbies} and {\it Games} and their disjunction {\it Hobbies or Games}. For the experimental membership weight for the disjunction of this item, Hampton (1988b) measured   $\mu _{\exp } (A\ {\rm{ or }}\ B) = 0.95$. In Figure 6 we have represented the situations as described by the quantum field model for four different values of  $c^2$. More specifically, for $c^2 = 0$ we can see in Figure 6 that no perfect match is possible between the quantum field membership weight $\mu _{quant } (A\ {\rm{ or }}\ B)$ and the experimental membership weight  $\mu _{\exp } (A\ {\rm{ or }}\ B)$. Indeed, the surface does not cut the horizontal plane at 0.95. 
\begin{figure}[h]
\centerline {\includegraphics[width=16cm]{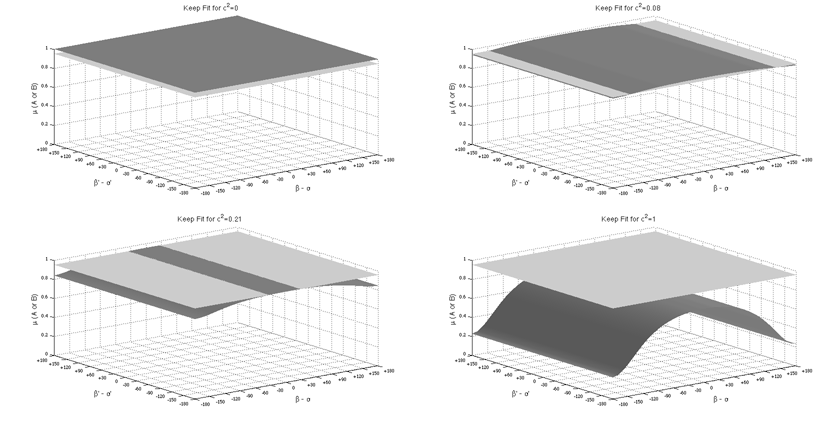}}
\caption{Graphical representation as in Figure 5 of the item {\it Keep Fit} with respect to the disjunction of the concepts {\it Sportswear} and {\it Sports Equipment}. In this case no perfect matches between theory and experiment exist for values $c^2 = 0$ and  $c^2 = 1$. Hence a perfect match between theory and experiment can be obtained only by applying the genuine quantum field model for modeling, i.e. for values of  $c^2$ different from 0 and 1.}
\end{figure}
For $c^2 = 0.08$ the surface cuts the plane along two lines. Hence for this value of $c^2$ all the points of the lines give rise to values with perfect matches between theory and experiment. Also for $c^2 = 0.21$ the surface cuts the plane along two lines, so that for this value of  $c^2$ these points give rise to values with perfect matches between theory and experiment as well. For  $c^2 = 1$ the surface does not cut the horizontal plane, so that for this value no perfect match between theory and experiment is possible.

The volume contained under the horizontal plane at a height of 0.95 is given by
\begin{equation}
0.95(2\pi )^2  = 37.5045
\end{equation}
For $c^2 = 0.1195$ we find that the volume under the surface is given by
\begin{equation}
\int_{ - \pi }^{ + \pi } {d(\beta  - \alpha )\int_{ - \pi }^{ + \pi } {d(\beta ' - \alpha ')\mu _{quant} (A\ {\rm{ or }}\ B)} }  = 37.5045
\end{equation}
which is the reason why we have chosen  $c^2 = 0.1193$ for the item {\it Keep Fit} in Table 4. Figure 7 provides a graphical representation of this situation.
\begin{figure}[h]
\centerline {\includegraphics[width=16cm]{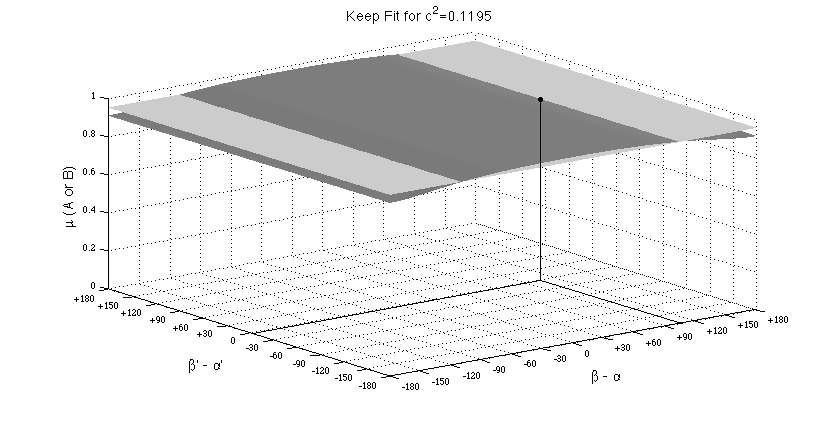}}
\caption{Graphical representation of the item {\it Keep Fit} with respect to the disjunction of the concepts {\it Sportswear} and {\it Sports Equipment} for  $c^2 = 0.1195$ and in function of the quantum angles $\beta - \alpha$ and  $\beta' - \alpha'$. This value of  $c^2$ produces a volume under the surface equal to the volume under the horizontal plane, i.e. it is the optimum of all  $c^2$ values, giving rise to perfect matches between theory and experiment.}
\end{figure}

\noindent
For the items {\it Window Seat}, {\it Painting}, {\it Light Fixture}, {\it Kitchen Count}, {\it Bath Tub}, {\it Rug}, {\it Space Rack}, {\it Bar}, {\it Lamp}, {\it Wall Mirror} and {\it Sculpture}, the values of $c^2$ are likewise chosen in such a way as to achieve that the volumes contained under the corresponding horizontal plane and the surface differ as little as possible in size. All these items require a value of $c^2$ different from 0 and different from 1, which means that for matches between theory and experiment to be perfect the genuine quantum field model should be used in these situation. For the items {\it Desk Chair}, {\it Ashtray}, {\it Door Bell}, {\it Hammock}, {\it Refrigerator}, {\it Waste Paper Basket} and {\it Sink Unit}, the difference in size between the two volumes is least if  $c^ = 1$. This means that for these items a description with a perfect match between theory and experiment can be reached by applying the `one-item quantum model'.

For the concepts {\it House Furnishings} and {\it Furniture} we have two of 24 items that yield no perfect match, not even if we use the quantum field model. These are the items {\it Wall Hangings} and {\it Park Bench}. For {\it Wall Hangings} the best approximation between theory and experiment is reached for  $c^2 = 0$, where the difference is 0.02, while for {\it Park Bench} the best approximation between theory and experiment is reached for  $c^2 = 1$, where the difference is 0.03. For the item {\it Shelves} only $c^2 = 0$ produces a perfect match between theory and experiment. Hence for {\it Shelves} subjects did not consider {\it House Furnishings} or {\it Furniture} to be a new `concept {\it House Furnishings or Furniture}' and evaluated the membership weight of {\it Shelves} with respect to this new concept. Rather, in this case, subjects must have reasoned along the following lines: ``Since with a weight equal to 1, {\it Shelves} are a member of {\it Furniture}, they are also, with a weight equal to 1, a member of {\it House Furnishings or Furniture}."

There is one item, namely {\it Wall Mirror}, for which no perfect match exists if $c^2 = 0$ or  $c^2 = 1$, which means that the intrinsic nature of the quantum field model is needed to obtain a perfect match between theory and experiment for this item. Hence the state $\left| x \right\rangle $ of equation (\ref{eq2.4}) for {\it Wall Mirror} must be a genuine superposition state of the `one-item model' and the `two identical items model'. This means that subjects partly decided to take into consideration the new `concept {\it House Furnishings or Furniture}', and estimated the membership weights of {\it Wall Mirror} with respect to this new concept. This is the reason why subjects decided that, although the item has membership weight equal to 1 with respect to the concept {\it Furniture}, the membership weight with respect to the new `concept {\it House Furnishings or Furniture}' is less than 1, namely 0.95. However, 0.95 is too close to 1 for the `one-item model' alone to be able to model the situation with a perfect match. That is why we need the full quantum field model. Interpreting this, we find that subjects also partly considered the situation neglecting the new concept {\it House Furnishings or Furniture} and deciding that since {\it Wall Mirror} has a membership equal to 1 with respect to {\it Furniture}, it also has a membership weight equal to 1 for {\it House Furnishings or Furniture}. The fact that the experimental value is not 1 but 0.95, indicates that this represents only part of the reasoning of the subjects, the other part being that the subjects took into consideration the new `concept {\it House Furnishings or Furniture}.

\tiny
\setlongtables 
\begin{longtable}{|l|l|l|l|l|l|l|l|l|} 
\caption{The list of items of Hampton's (1988b) experiment on the disjunction of concepts modeled by the quantum field theoretic model. $\mu_{exp}(A)$, $\mu_{exp}(B)$ and $\mu _{\exp } (A{\rm{\ or\ }}B)$ are the membership weights of concepts $A$, $B$ and the disjunction $A$ or $B$, respectively, for the considered item, as measured in Hampton (1988b). $\beta - \alpha$, $\beta' - \alpha'$ and $c^2$ are the quantum parameters, which need to be chosen for the predicted quantum weights $\mu_{quant}(A\ {\rm or}\ B)$ to be equal to the experimental weights $\mu _{\exp } (A{\rm{\ or\ }}B)$.} \\
\hline 
 & $\mu_{exp}(A)$ & $\mu_{exp}(B)$ & $\mu _{\exp } (A{\rm{\ or\ }}B)$ & $\beta - \alpha$ & $\beta' - \alpha'$ & $c^2$ & $\mu_{quant}(AB)$ & $|\mu_{exp} - \mu_{quant}| $ \\ 
\endfirsthead 
\hline 
 & $\mu_{exp}(A)$ & $\mu_{exp}(B)$ & $\mu _{\exp } (A{\rm{\ or\ }}B)$ & $\beta - \alpha$ & $\beta' - \alpha'$ & $c^2$ & $\mu_{quant}(AB)$ & $|\mu_{exp} - \mu_{quant}| $ \\
\hline \hline
\endhead 
\hline \hline  
$A$ = {\it House Furnishings}  &  &  &  & & & & &\\
$B$ = {\it Furniture} &  &  &  & & & & &\\
\hline
{\it Mantelpiece} & 0.4 & 0.8 & 0.75 & 71.79797$^\circ$ & 0$^\circ$ & 0.2865 & 0.75 & 0 \\
{\it Window Seat} & 0.9 & 0.9 & 0.8 & 97.871824$^\circ$ & 0$^\circ$ & 0.9754 & 0.8 & 0 \\
{\it Painting} & 0.5 & 0.9 & 0.85 & 80.04715$^\circ$ & 0$^\circ$ & 0.2621 & 0.85 & 0 \\
{\it Light Fixture} & 0.4 & 0.8 & 0.775 & 77.33493$^\circ$ & 0$^\circ$ & 0.1933 & 0.775 & 0 \\
{\it Kitchen Count} & 0.55 & 0.8 & 0.625 & 63.89759$^\circ$ & 0$^\circ$ & 0.9386 & 0.625 & 0 \\
{\it Bath Tub} & 0.7 & 0.5 & 0.75 & 80.54885$^\circ$ & 0$^\circ$ & 0.1752 & 0.75 & 0 \\
{\it Deck Chair} & 0.3 & 0.1 & 0.35 & 0$^\circ$ & 97.740547$^\circ$ & 1 & 0.35 & 0 \\
{\it Shelves} & 0.4 & 1 & 1 & arbitrary & arbitrary & 0 & 1 & 0 \\
{\it Rug} & 0.6 & 0.9 & 0.95 & 0$^\circ$ & 86.0354$^\circ$ & 0.0001 & 0.95 & 0 \\
{\it Bed} & 1 & 1 & 1 & arbitrary & arbitrary & 0 & 1 & 0 \\
{\it Wall Hangings} & 0.4 & 0.9 & 0.95 & 0$^\circ$ & 0$^\circ$ & 0 & 0.92998583 & 0.020014 \\
{\it Space Rack} & 0.5 & 0.7 & 0.65 & 51.61357$^\circ$ & 0$^\circ$ & 0.614 & 0.65 & 0 \\
{\it Ashtray} & 0.7 & 0.3 & 0.25 & 113.207338$^\circ$ & 0$^\circ$ & 1 & 0.25 & 0 \\
{\it Bar} & 0.6 & 0.35 & 0.55 & 53.10437$^\circ$ & 0$^\circ$ & 0.541 & 0.55 & 0 \\
{\it Lamp} & 0.7 & 1 & 0.9 & 122.740905$^\circ$ & arbitrary & 0.365 & 0.9 & 0 \\
{\it Wall Mirror} & 0.6 & 1 & 0.95 & 118.36495$^\circ$ & arbitrary & 0.158 & 0.95 & 0 \\
{\it Door Bell} & 0.1 & 0.5 & 0.55 & 0$^\circ$ & 113.882924$^\circ$ & 1 & 0.55 & 0 \\
{\it Hammock} & 0.5 & 0.2 & 0.35 & 0$^\circ$ & 21.78679$^\circ$ & 1 & 0.35 & 0 \\
{\it Desk} & 1 & 1 & 1 & arbitrary & arbitrary & arbitrary & 1 & 0 \\
{\it Refrigerator} & 0.7 & 0.9 & 0.575 & 111.824212$^\circ$ & 0$^\circ$ & 1 & 0.575 & 0 \\
{\it Park Bench} & 0 & 0.3 & 0.05 & arbitrary & 0$^\circ$ & 1 & 0.08166999 & 0.03166999 \\
{\it Waste Paper Basket} & 0.5 & 1 & 0.6 & 122.02776$^\circ$ & arbitrary & 1 & 0.6 & 0 \\
{\it Sculpture} & 0.4 & 0.8 & 0.8 & 141.46342$^\circ$ & 90$^\circ$ & 0.1 & 0.8 & 0 \\
{\it Sink Unit} & 0.6 & 0.9 & 0.6 & 95.857926$^\circ$ & 0$^\circ$ & 1 & 0.6 & 0 \\
\hline
$A$ = {\it Hobbies} & & & & & & & & \\
$B$ = {\it Games} & & & & & & & & \\
\hline
{\it Gardening} & 1 & 0 & 1 & arbitrary & arbitrary & 0 & 1 & 0 \\
{\it Theatre-Going} & 1 & 0 & 1 & arbitrary & arbitrary & 0 & 1 & 0 \\
{\it Archery} & 1 & 0.9 & 0.95 & 136.137045$^\circ$ & arbitrary & 0 & 1 & 0 \\
{\it Monopoly} & 0.7 & 1 & 1 & arbitrary & arbitrary & 0 & 1 & 0 \\
{\it Tennis} & 1 & 1 & 1 & arbitrary & arbitrary & arbitrary & 1 & 0 \\
{\it Bowling} & 1 & 1 & 1 & arbitrary & arbitrary & arbitrary & 1 & 0 \\
{\it Fishing} & 1 & 0.6 & 1 & arbitrary & arbitrary & 0 & 1 & 0  \\
{\it Washing Dishes} & 0.1 & 0 & 0.15 & arbitrary & 136.116765$^\circ$ & 0.9231 & 0.15 & 0  \\
{\it Eating Ice-Cream Cones} & 0.2 & 0 & 0.1 & arbitrary & 0$^\circ$ & 0.1905044 & 0.1 & 0  \\
{\it Camping} & 1 & 0.1 & 0.9 & 99.35442$^\circ$ & arbitrary & 0.2108 & 0.9 & 0  \\
{\it Skating} & 1 & 0.5 & 0.9 & 114.50157$^\circ$ & arbitrary & 0.2827 & 0.9 & 0  \\
{\it Judo} & 1 & 0.7 & 0.8 & 122.722587$^\circ$ & arbitrary & 0.7303 & 0.8 & 0  \\
{\it Guitar Playing} & 1 & 0.1 & 1 & arbitrary & arbitrary & 0 & 1 & 0  \\
{\it Autograph Hunting} & 1 & 0.2 & 0.9 & 103.65817$^\circ$ & arbitrary & 0.2236 & 0.9 & 0  \\
{\it Discus Throwing} & 1 & 0.75 & 0.7 & 132.343613$^\circ$ & arbitrary & 1 & 0.7 & 0  \\
{\it Jogging} & 1 & 0.4 & 0.9 & 110.84949$^\circ$ & arbitrary & 0.2583 & 0.9 & 0  \\
{\it Keep Fit} & 1 & 0.3 & 0.95 & 107.36806$^\circ$ & arbitrary & 0.1195 & 0.95 & 0  \\
{\it Noughts} & 0.5 & 1 & 0.9 & 114.50157$^\circ$ & arbitrary & 0.2827 & 0.9 & 0  \\
{\it Karate} & 1 & 0.7 & 0.8 & 122.722587$^\circ$ & arbitrary & 0.7303 & 0.8 & 0  \\
{\it Bridge} & 1 & 1 & 1 & arbitrary & arbitrary & arbitrary & 1 & 0  \\
{\it Rock Climbing} & 1 & 0.2 & 0.95 & 103.76366$^\circ$ & arbitrary & 0.1117 & 0.95 & 0  \\
{\it Beer Drinking} & 0.8 & 0.2 & 0.575 & 34.62136$^\circ$ & 0$^\circ$ & 0.7001 & 0.575 & 0  \\
{\it Stamp Collecting} & 1 & 0.1 & 1 & arbitrary & arbitrary & 0 & 1 & 0  \\
{\it Wrestling} & 0.9 & 0.6 & 0.625 & 90$^\circ$ & 0$^\circ$ & 1 & 0.625 & 0 \\
\hline
$A$ = {\it Pets} & & & & & & & & \\
$B$ = {\it Farmyard Animals} & & & & & & & & \\
\hline
{\it Goldfish} & 1 & 0 & 0.95 & arbitrary & arbitrary & 0.1 & 0.95 & 0 \\
{\it Robin} & 0.1 & 0.1 & 0.1 & 96.37937$^\circ$ & 0$^\circ$ & 0 & 0.1 & 0  \\
{\it Blue-Tit} & 0.1 & 0.1 & 0.1 & 96.37937$^\circ$ & 0$^\circ$ & 0 & 0.1 & 0  \\
{\it Collie Dog} & 1 & 0.7 & 1 & arbitrary & arbitrary & 0 & 1 & 0 \\
{\it Camel} & 0.4 & 0 & 0.1 & arbitrary & 0$^\circ$ & 1 & 0.11270167 & 0.01270167  \\
{\it Squirrel} & 0.2 & 0.1 & 0.1 & 125.685335$^\circ$ & 0$^\circ$ & 0 & 0.1 & 0  \\
{\it Guide Dog for the Blind} & 0.7 & 0 & 0.9 & arbitrary & 180$^\circ$ & 1 & 0.77386128 & 0.126139  \\
{\it Spider} & 0.5 & 0.35 & 0.55 & 96.68877$^\circ$ & 0$^\circ$ & 0 & 0.55 & 0  \\
{\it Homing Pig} & 0.9 & 0.1 & 0.8 & 67.96923$^\circ$ & 0$^\circ$ & 0.132 & 0.8 & 0  \\
{\it Monkey} & 0.5 & 0 & 0.25 & arbitrary & 0$^\circ$ & 0.4459029 & 0.25 & 0  \\
{\it Circus Horse} & 0.4 & 0 & 0.3 & arbitrary & 118.037828$^\circ$ & 0.755 & 0.3 & 0  \\
{\it Prize Bull} & 0.1 & 1 & 0.9 & 99.35442$^\circ$ & arbitrary & 0.2108 & 0.9 & 0  \\
{\it Rat} & 0.5 & 0.7 & 0.4 & 97.29629$^\circ$ & 0$^\circ$ & 1 & 0.4 & 0  \\
{\it Badger} & 0 & 0.25 & 0.1 & arbitrary & 0$^\circ$ & 0.5648772 & 0.1 & 0  \\
{\it Siamese Cat} & 1 & 0.1 & 0.95 & 99.1892$^\circ$ & arbitrary & 0.1055 & 0.95 & 0  \\
{\it Race Horse} & 0.6 & 0.25 & 0.65 & 51.469775$^\circ$ & 0$^\circ$ & 0 & 0.65 & 0  \\
{\it Fox} & 0.1 & 0.3 & 0.2 & 108.007533$^\circ$ & 0$^\circ$ & 0 & 0.2 & 0  \\
{\it Donkey} & 0.5 & 0.9 & 0.7 & 77.339965$^\circ$ & 0$^\circ$ & 0.7422 & 0.7 & 0  \\
{\it Field Mouse} & 0.1 & 0.7 & 0.4 & 0$^\circ$ & 40.202965$^\circ$ & 1 & 0.4 & 0  \\
{\it Ginger Tom-Cat} & 1 & 0.8 & 0.95 & 128.190115$^\circ$ & arbitrary & 0.2235 & 0.95 & 0  \\
{\it Husky in Sledream} & 0.4 & 0 & 0.425 & arbitrary & 133.115215$^\circ$ & 1 & 0.425 & 0  \\
{\it Cart Horse} & 0.4 & 1 & 0.85 & 110.87858$^\circ$ & arbitrary & 0.3873 & 0.85 & 0  \\
{\it Chicken} & 0.3 & 1 & 0.95 & 107.36806$^\circ$ & arbitrary & 0.1195 & 0.95 & 0  \\
{\it Doberman Guard Dog} & 0.6 & 0.85 & 0.8 & 75.95264$^\circ$ & 0$^\circ$ & 0.4335 & 0.8 & 0  \\
\hline
$A$ = {\it Spices} & & & & & & & & \\
$B$ = {\it Herbs} & & & & & & & & \\
\hline
{\it Molasses} & 0.4 & 0.05 & 0.425 & 0$^\circ$ & 111.708232$^\circ$ & 1 & 0.425 & 0 \\
{\it Salt} & 0.75 & 0.1 & 0.6 & 61.511235$^\circ$ & 0$^\circ$ & 0.2486 & 0.6 & 0 \\
{\it Peppermint} & 0.45 & 0.6 & 0.6 & 58.55685$^\circ$ & 0$^\circ$ & 0.509 & 0.6 & 0 \\
{\it Curry} & 0.9 & 0.4 & 0.75 & 73.448515$^\circ$ & 0$^\circ$ & 0.5 & 0.75 & 0 \\
{\it Oregano} & 0.7 & 1 & 0.875 & 122.716483$^\circ$ & arbitrary & 0.4565 & 0.875 & 0 \\
{\it MSG} & 0.15 & 0.1 & 0.425 & 0$^\circ$ & 128.141267$^\circ$ & 1 & 0.425 & 0 \\
{\it Chili Pepper} & 1 & 0.6 & 0.95 & 118.36495$^\circ$ & arbitrary & 0.158 & 0.95 & 0 \\
{\it Mustard} & 1 & 0.8 & 0.85 & 128.18468$^\circ$ & arbitrary & 0.6706 & 0.85 & 0 \\
{\it Mint} & 0.6 & 0.8 & 0.925 & 90$^\circ$ & 180$^\circ$ & 0.3505096 & 0.925 & 0 \\
{\it Cinnamon} & 1 & 0.4 & 1 & arbitrary & arbitrary & 0 & 1 & 0 \\
{\it Parsley} & 0.5 & 0.9 & 0.95 & 0$^\circ$ & 180$^\circ$ & 1 & 0.947213595 & 0.002786 \\
{\it Saccharin} & 0.1 & 0 & 0.15 & arbitrary & 136.116765$^\circ$ & 0.9231 & 0.15 & 0 \\
{\it Poppyseeds} & 0.4 & 0.4 & 0.4 & 0$^\circ$ & 0$^\circ$ & 1 & 0.4 & 0 \\
{\it Pepper} & 0.9 & 0.6 & 0.95 & 86.0985$^\circ$ & 0$^\circ$ & 0 & 0.95 & 0 \\
{\it Turmeric} & 0.7 & 0.45 & 0.675 & 65.31911$^\circ$ & 0$^\circ$ & 0.4123 & 0.675 & 0 \\
{\it Sugar} & 0 & 0 & 0.2 & arbitrary & arbitrary & arbitrary & 0 & 0.2 \\
{\it Vinegar} & 0.1 & 0 & 0.35 & arbitrary & 154.6230665$^\circ$ & 1 & 0.35 & 0 \\
{\it Sesame Seeds} & 0.35 & 0.4 & 0.625 & 167.32$^\circ$ & 180$^\circ$ & 0.051493 & 0.625 & 0 \\
{\it Lemon Juice} & 0.1 & 0 & 0.15 & arbitrary & 136.116765$^\circ$ & 0.9231 & 0.15 & 0 \\
{\it Chocolate} & 0 & 0 & 0 & arbitrary & arbitrary & arbitrary & 0 & 0 \\
{\it Horseradish} & 0.2 & 0.4 & 0.625 & 0$^\circ$ & 130.528224$^\circ$ & 1 & 0.7 & 0 \\
{\it Vanilla} & 0.6 & 0 & 0.275 & arbitrary & 81.735685$^\circ$ & 1 & 0.275 & 0 \\
{\it Chires} & 0.6 & 1 & 0.95 & 118.36495$^\circ$ & arbitrary & 0.158 & 0.95 & 0 \\
{\it Root Ginger} & 0.7 & 0.15 & 0.675 & 50.25749$^\circ$ & 0$^\circ$ & 0 & 0.675 & 0 \\
\hline
$A$ = {\it Instruments} & & & & & & & & \\
$B$ = {\it Tools} & & & & & & & & \\
\hline
{\it Broom} & 0.1 & 0.7 & 0.6 & 87.91571$^\circ$ & 0$^\circ$ & 0 & 0.6  & 0 \\
{\it Magnetic Compass} & 0.9 & 0.5 & 1 & 0$^\circ$ & 180$^\circ$ & 1 & 0.947213595  & 0.052786 \\
{\it Tuning Fork} & 0.9 & 0.6 & 1 & 0$^\circ$ & 180$^\circ$ & 1 & 0.967423461  & 0.032577 \\
{\it Pen-Knife} & 0.65 & 1 & 0.95 & 120.44306$^\circ$ & arbitrary & 0.169 & 0.95  & 0 \\
{\it Rubber Band} & 0.25 & 0.5 & 0.25 & 83.918343$^\circ$ & 0$^\circ$ & 1 & 0.25  & 0 \\
{\it Stapler} & 0.85 & 0.8 & 0.85 & 84.2862$^\circ$ & 0$^\circ$ & 0.4755 & 0.85  & 0 \\
{\it Skate Board} & 0.1 & 0 & 0 & arbitrary & 0$^\circ$ & 1 & 0.025658351  & 0.025658 \\
{\it Scissors} & 0.85 & 1 & 0.9 & 131.642835$^\circ$ & arbitrary & 0.5165 & 0.9  & 0 \\
{\it Pencil Eraser} & 0.4 & 0.7 & 0.45 & 71.796025$^\circ$ & 0$^\circ$ & 1 & 0.45  & 0 \\
{\it Tin Opener} & 0.9 & 0.9 & 0.95 & 97.78503$^\circ$ & 0$^\circ$ & 1 & 0.45  & 0 \\
{\it Bicycle Pump} & 1 & 0.9 & 0.7 & 151.451209$^\circ$ & arbitrary & 1 & 0.7  & 0 \\
{\it Scalpel} & 0.8 & 1 & 0.925 & 128.16295$^\circ$ & arbitrary & 1 & 0.7  & 0 \\
{\it Computer} & 0.6 & 0.8 & 0.6 & 75.43202$^\circ$ & 0$^\circ$ & 1 & 0.6  & 0 \\
{\it Paper Clip} & 0.3 & 0.7 & 0.6 & 56.91307$^\circ$ & 0$^\circ$ & 0.4746 & 0.6  & 0 \\
{\it Paint Brush} & 0.65 & 0.9 & 0.95 & 90.901$^\circ$ & 0$^\circ$ & 0.217 & 0.95  & 0 \\
{\it Step Ladder} & 0.2 & 0.9 & 0.85 & 83.70581$^\circ$ & 0$^\circ$ & 0.06 & 0.85  & 0 \\
{\it Door Key} & 0.3 & 0.1 & 0.95 & 0$^\circ$ & 169.469637$^\circ$ & 1 & 0.95  & 0 \\
{\it Measuring Calipers} & 0.9 & 1 & 0.9 & 136.128333$^\circ$ & arbitrary & 0.6322 & 0.9  & 0 \\
{\it Toothbrush} & 0.4 & 0.4 & 0.5 & 98.40885$^\circ$ & 0$^\circ$ & 0.012 & 0.5  & 0 \\
{\it Sellotape} & 0.1 & 0.2 & 0.325 & 0$^\circ$ & 106.763935$^\circ$ & 1 & 0.325  & 0 \\
{\it Goggles} & 0.2 & 0.3 & 0.15 & 145.56166$^\circ$ & 0$^\circ$ & 0 & 0.15  & 0 \\
{\it Spoon} & 0.65 & 0.9 & 0.7 & 83.93378$^\circ$ & 0$^\circ$ & 0.9115 & 0.7  & 0 \\
{\it Pliers} & 0.8 & 1 & 1 & arbitrary & arbitrary & 0 & 1  & 0 \\
{\it Computer} & 0.6 & 0.8 & 0.6 & 97.18770632$^\circ$ & 67.36029726$^\circ$ & 0.6 & 0  & 0 \\
{\it Meat Thermometer} & 0.75 & 0.8 & 0.9 & 81.12534$^\circ$ & 0$^\circ$ & 0.15 & 0.9  & 0 \\
\hline
$A$ = {\it Sportswear} & & & & & & & & \\
$B$ = {\it Sports Equipment} & & & & & & & & \\
\hline
{\it American Foot} & 1 & 1 & 1 & arbitrary & arbitrary & arbitrary & 1 & 0  \\
{\it Referee's Whistle} & 0.6 & 0.2 & 0.45 & 0$^\circ$ & 47.85689$^\circ$ & 0.9085 & 0.45 & 0  \\
{\it Circus Clowns} & 0 & 0 & 0.1 & arbitrary & arbitrary & arbitrary & 0 & 0.1  \\
{\it Backpack} & 0.6 & 0.5 & 0.6 & 52.23231$^\circ$ & 0$^\circ$ & 0.6325 & 0.6 & 0 \\
{\it Diving Mask} & 1 & 1 & 0.95 & arbitrary & arbitrary & arbitrary & 1 & 0.05  \\
{\it Frisbee} & 0.3 & 1 & 0.85 & 107.36806$^\circ$ & arbitrary & 0.3585 & 0.85 & 0  \\
{\it Sunglasses} & 0.4 & 0.2 & 0.1 & 180$^\circ$ & 0$^\circ$ & 0.4708025 & 0.1 & 0  \\
{\it Suntan Lotion} & 0 & 0 & 0.1 & arbitrary & arbitrary & arbitrary & 0 & 0.1  \\
{\it Gymnasium} & 0 & 0.9 & 0.825 & arbitrary & arbitrary & 0 & 0.818181818 & 0.006818  \\
{\it Motorcycle Helmet} & 0.7 & 0.9 & 0.75 & 86.184155$^\circ$ & 0$^\circ$ & 0.7975 & 0.75 & 0  \\
{\it Rubber Flipper} & 1 & 1 & 1 & arbitrary & arbitrary & arbitrary & 1 & 0  \\
{\it Wrist Sweat} & 1 & 1 & 0.95 & arbitrary & arbitrary & arbitrary & 1 &0.05  \\
{\it Golf Ball} & 0.1 & 1 & 1 & arbitrary & arbitrary & 0 & 1 & 0  \\
{\it Cheerleaders} & 0.3 & 0.4 & 0.45 & 91.624316$^\circ$ & 0$^\circ$ & 0 & 0.45 & 0  \\
{\it Lineman's Flag} & 0.1 & 1 & 0.75 & 99.35442$^\circ$ & arbitrary & 0.527 & 0.75 & 0  \\
{\it Underwater} & 1 & 0.65 & 0.6 & 134.242395$^\circ$ & arbitrary & 1 & 0.6 & 0  \\
{\it Baseball Bat} & 0.2 & 1 & 1 & arbitrary & arbitrary & 0 & 1 & 0  \\
{\it Bathing Costume} & 1 & 0.8 & 0.8 & 128.169742$^\circ$ & arbitrary & 0.8945 & 0.8 & 0  \\
{\it Sailing Life Jacket} & 1 & 0.8 & 1 & arbitrary & arbitrary & 0 & 1 & 0  \\
{\it Ballet Shoes} & 0.7 & 0.6 & 0.6 & 54.70783$^\circ$ & 0$^\circ$ & 0.9827 & 0.6 & 0  \\
{\it Hoola Hoop} & 0.1 & 0.6 & 0.5 & 86.0985$^\circ$ & 0$^\circ$ & 0 & 0.5 & 0  \\
{\it Running Shoes} & 1 & 1 & 1 & arbitrary & arbitrary & arbitrary & 1 & 0  \\
{\it Cricket Pitch} & 0 & 0.5 & 0.525 & arbitrary & arbitrary & 0 & 1 & 0  \\
{\it Tennis Racket} & 0.2 & 1 & 1 & arbitrary & arbitrary & 0 & 1 & 0  \\
\hline
$A$ = {\it Household Appliances} & & & & & & & & \\
$B$ = {\it Kitchen Utensils} & & & & & & & & \\
\hline
{\it Fork} & 0.7 & 1 & 0.95 & 122.7409$^\circ$ & arbitrary & 0.1825 & 0.95 & 0  \\
{\it Apron} & 0.3 & 0.4 & 0.5 & 71.83197$^\circ$ & 0$^\circ$ & 0 & 0.5 & 0  \\
{\it Hat Stand} & 0.45 & 0 & 0.3 & arbitrary & 116.35571$^\circ$ & 0.2145 & 0.3 & 0 \\
{\it Freezer} & 1 & 0.6 & 0.95 & 118.36495$^\circ$ & arbitrary & 0.158 & 0.95 & 0  \\
{\it Extractor Fan} & 1 & 0.4 & 0.9 & 110.84949$^\circ$ & arbitrary & 0.6325 & 0.6 & 0  \\
{\it Cake Tin} & 0.4 & 0.7 & 0.95 & 118.36497$^\circ$ & 0$^\circ$ & 0.6325 & 0.6 & 0  \\
{\it Carving Knife} & 0.6 & 0.5 & 0.6 & arbitrary & arbitrary & 0 & 1 & 0  \\
{\it Cooking Stove} & 1 & 0.5 & 1 & arbitrary & arbitrary & 0 & 1 & 0  \\
{\it Iron} & 1 & 0.3 & 0.95 & 107.36806$^\circ$ & arbitrary & 0.1195 & 0.95 & 0  \\
{\it Food Processor} & 1 & 1 & 1 & arbitrary & arbitrary & arbitrary & 1 & 0  \\
{\it Chopping Board} & 0.45 & 1 & 0.95 & 112.67467$^\circ$ & arbitrary & 0.1348 & 0.95 & 0  \\
{\it Television} & 0.95 & 0 & 0.85 & arbitrary & 96.51724$^\circ$ & 0.1312 & 0.85 & 0  \\
{\it Vacuum Cleaner} & 1 & 0 & 1 & arbitrary & arbitrary & 0 & 1 & 0  \\
{\it Rubbish Bin} & 0.5 & 0.5 & 0.8 & 179.99999$^\circ$ & 180$^\circ$ & 0.1942029 & 0.8 & 0  \\
{\it Vegetable Rack} & 0.4 & 0.7 & 0.7 & 79.67065$^\circ$ & 0$^\circ$ & 0.194 & 0.7 & 0  \\
{\it Broom} & 0.55 & 0.4 & 0.625 & 95.88754$^\circ$ & 0$^\circ$ & 0 & 0.625 & 0  \\
{\it Rolling Pin} & 0.45 & 1 & 1 & arbitrary & arbitrary & 0 & 0.625 & 0  \\
{\it Table Mat} & 0.25 & 0.4 & 0.325 & 0$^\circ$ & 11.74284$^\circ$ & 1 & 0.325 & 0  \\
{\it Whisk} & 1 & 1 & 1 & arbitrary & arbitrary & 0 & 1 & 0  \\
{\it Blender} & 1 & 1 & 1 & arbitrary & arbitrary & arbitrary & 1 & 0  \\
{\it Electric Toothbrush} & 0.8 & 0 & 0.55 & arbitrary & 103.651897$^\circ$ & 0.5316 & 0.55 & 0  \\
{\it Frying Pan} & 0.7 & 1 & 0.95 & 122.76533$^\circ$ & arbitrary & 0.1824 & 0.95 & 0  \\
{\it Toaster} & 1 & 1 & 1 & arbitrary & arbitrary & arbitrary & 1 & 0  \\
{\it Spatula} & 0.55 & 0.9 & 0.95 & 29.4211$^\circ$ & 0$^\circ$ & 0 & 0.95 & 0  \\
\hline
$A$ = {\it Fruits} & & & & & & & \\
$B$ = {\it Vegetables} & & & & & & & \\
\hline
{\it Apple} & 1 & 0 & 1 & arbitrary & arbitrary & 0 & 1 & 0  \\
{\it Parsley} & 0 & 0.2 & 0.45 & arbitrary & 180$^\circ$ & 0.5 & 0.05 & 0  \\
{\it Olive} & 0.5 & 0.1 & 0.8 & 70.2$^\circ$ & 180$^\circ$ & 0.77615234 & 0.8 & 0  \\
{\it Chili Pepper} & 0.05 & 0.5 & 0.5 & 0$^\circ$ & 180$^\circ$ & 1 & 0.5 & 0  \\
{\it Broccoli} & 0 & 0.8 & 1 & 0$^\circ$ & 180$^\circ$ & 1 & 0.7236068 & 0.2763932  \\
{\it Root Ginger} & 0 & 0.3 & 0.55 & arbitrary & 150.372432$^\circ$ & 1 & 0.55 & 0  \\
{\it Pumpkin} & 0.7 & 0.8 & 0.925 & 93.1273$^\circ$ & 0$^\circ$ & 0 & 0.925 & 0  \\
{\it Raisin} & 1 & 0 & 0.9 & arbitrary & arbitrary & 0.2 & 0.9 & 0  \\
{\it Acorn} & 0.35 & 0 & 0.4 & arbitrary & 134.242395$^\circ$ & 1 & 0.4 & 0  \\
{\it Mustard} & 0 & 0.2 & 0.175 & arbitrary & 128.170223$^\circ$ & 0.568 & 0.175 & 0  \\
{\it Rice} & 0 & 0.4 & 0.325 & arbitrary & 119.771108$^\circ$ & 1 & 0.325 & 0  \\
{\it Tomato} & 0.7 & 0.7 & 1 & 179.999998$^\circ$ & 180$^\circ$ & 0.31034484 & 1 & 0  \\
{\it Coconut} & 0.7 & 0 & 1 & arbitrary & 180$^\circ$ & 1 & 0.77386128 & 0.22613872  \\
{\it Mushroom} & 0 & 0.5 & 0.8 & arbitrary & 180$^\circ$ & 1 & 0.83355339 & 0.05355339  \\
{\it Wheat} & 0 & 0.1 & 0.2 & arbitrary & 142.238756$^\circ$ & 1 & 0.2 & 0  \\
{\it Green Pepper} & 0.3 & 0.6 & 0.8 & 111.332$^\circ$ & 180$^\circ$ & 0.51942375 & 0.8 & 0  \\
{\it Watercress} & 0 & 0.6 & 0.8 & arbitrary & 180$^\circ$ & 0.95813878 & 0.8 & 0  \\
{\it Peanut} & 0.3 & 0.1 & 0.4 & 0$^\circ$ & 107.614671$^\circ$ & 1 & 0.4 & 0  \\
{\it Black Pepper} & 0.15 & 0.2 & 0.225 & 0$^\circ$ & 54.20827$^\circ$ & 0.3823 & 0.225 & 0  \\
{\it Garlic} & 0.1 & 0.2 & 0.5 & 0$^\circ$ & 131.169683$^\circ$ & 1 & 0.5 & 0  \\
{\it Yam} & 0.45 & 0.65 & 0.85 & 124$^\circ$ & 180$^\circ$ & 0.39637625 & 0.85 & 0  \\
{\it Elderberry} & 1 & 0 & 0.55 & arbitrary & arbitrary & 0.4 & 0.55 & 0  \\
{\it Almond} & 0.2 & 0.1 & 0.425 & 0$^\circ$ & 122.484807$^\circ$ & 1 & 0.425 & 0  \\
{\it Lentils} & 0 & 0.6 & 0.525 & arbitrary & 180$^\circ$ & 0.24874756 & 0.525 & 0  \\
 \\
\hline
\hline 
\end{longtable}
\normalsize
For the concepts {\it Hobbies} and {\it Games} and their disjunction {\it Hobbies} or {\it Games}, the quantum field theoretic model gives rise to perfect matches for all 24 items tested in Hampton (1988b). In choosing the values for  $c^2$, we proceeded in the same way as explained in the foregoing paragraphs, namely to match the sizes of the volumes in the graphical representations. That is why we chose  $c^2 = 0$ for five items, {\it Gardening}, {\it Theatre-Going}, {\it Monopoly}, {\it Fishing} and {\it Guitar Playing}, and  $c^2 = 1$ for one item, {\it Discus Throwing}. For the items {\it Archery}, {\it Washing Dishes}, {\it Eating Ice-Cream Cones}, {\it Camping}, {\it Skating}, {\it Judo}, {\it Autograph Hunting}, {\it Jogging}, {\it Keep Fit}, {\it Noughts}, {\it Karate}, {\it Rock Climbing} and {\it Beer Drinking} we chose  different from 0 and different from 1. Hence these items are all represented by a vector that is a genuine superposition of the `one item' situation and the `two identical items' situation.

For the items {\it Camping}, {\it Skating}, {\it Autograph Hunting}, {\it Jogging}, {\it Keep Fit}, {\it Noughts} and {\it Rock Climbing} only a solution giving rise to a perfect match between theory and experiment exists for $c^2$ different from 0 and different from 1. Remember that for the concepts {\it House Furnishings} and {\it Furniture} there was only one item, namely the item {\it Wall Mirror}, for which no solution giving rise to a perfect match between theory and experiment existed for $c^2 = 0$ or  $c^2 = 1$. Hence for the concepts {\it Hobbies} and {\it Games} we have seven such items.

For the concepts {\it Pets} and {\it Farmyard Animals} the quantum field theoretic model delivers a perfect match for 22 of the 24 items. For seven items, {\it Robin}, {\it Blue-Tit}, {\it Collie Dog}, {\it Squirrel}, {\it Spider}, {\it Race Horse} and {\it Fox} we have $c^2 = 0$, while for three items, {\it Rat}, {\it Field Mouse} and {\it Husky} in {\it Sledream} we have  $c^2 = 1$. For twelve items, {\it Goldfish}, {\it Homing Pig}, {\it Monkey}, {\it Prize Bull}, {\it Badger}, {\it Siamese Cat}, {\it Donkey}, {\it Ginger Tom-Cat}, {\it Cart Horse}, {\it Chicken} and {\it Dobberman Guard Dog} we have  $c^2$ different from 0 and different from 1, and hence these items need a genuine quantum field theoretic model to produce a perfect match with the experiment.

For two items, {\it Camel} and {\it Guide Dog for the Blind}, we cannot produce a perfect match with the experiment. For {\it Camel} the match is close (difference of 0.01270167), comparable to the other problem items we found for the concepts {\it House Furnishings} and {\it Furniture}, namely {\it Wall Hangings} and {\it Park Bench} (differences of 0.020014 and 0.03166999, respectively). For {\it Guide Dog for the Blind}, the difference is in the order of ten times bigger (0.126139), and hence we do not have a close match for this item.

There are six items, {\it Goldfish}, {\it Prize Bull}, {\it Siamese Cat}, {\it Ginger Tom-Cat}, {\it Cart Horse} and {\it Chicken}, for which a perfect match of theory with experiment does not exist if $c^2 = 0$ or  $c^2 = 1$, which means that only within the quantum field model a solution for a perfect match exists.

For the concepts {\it Spices} and {\it Herbs} we have again 22 of the 24 items with perfect matches with experiment for the predictions of the field theoretic model. There are seven items with  $c^2 = 1$, so that they fit into a `one-item situation' with estimation with respect to the new `concept {\it Spices or Herbs}'. There are three items with  $c^2 = 0$, which therefore fits into a `two identical items situation' without consideration of the new `concept {\it Spices or Herbs}'.

There are twelve items with  $c^2$ different from 1 and different from 0, and hence needing genuine quantum field modeling, with partly a consideration of the new `concept {\it Spices or Herbs}'. We have two items where the quantum field model does not produce a perfect match with the experiment; these are {\it Parsley} (difference of 0.002786), with a very close match, and {\it Sugar} (difference of 0.2), with no close match. There are two items, {\it Chili Pepper} and {\it Chires} for which a perfect match of theory with experiment does not exist for values of $c^2 = 0$ or  $c^2 = 1$, which means that only within the quantum field model a solution for a perfect match exists.

For the concepts {\it Instruments} and {\it Tools} we get 21 perfect matches with experiments out of the 24 tested items, and three items for which the quantum field theoretic model does not produce a perfect match. These are {\it Magnetic Compass}, {\it Tuning Fork} and {\it Skate Board}. All three of them, however, get a close match, the differences being 0.052786, 0.032577 and 0.025658, respectively. There are six items, {\it Rubber Band}, {\it Pencil Eraser}, {\it Bicycle Pump}, {\it Computer}, {\it Door Key} and {\it Sellotape}, that are modeled with  $c^2 = 1$, hence considering the new `concept {\it Instruments or Tools}' and estimating the membership value with respect to this new concept. Three items, {\it Broom}, {\it Goggles} and {\it Pliers}, are modeled with  $c^2 = 0$, so within the `two identical items' situation and neglecting the formation of a new `concept {\it Instruments or Tools}'.

For twelve items, {\it Pen-Knife}, {\it Stapler}, {\it Scissors}, {\it Tin Opener}, {\it Scalpel}, {\it Paper Clip}, {\it Paint Brush}, {\it Step Ladder}, {\it Measuring Calipers}, {\it Toothbrush}, {\it Spoon} and {\it Meat Thermometer}, the values of  $c^2$ are different from 1 and 0, requiring genuine quantum field modeling. There is only one item, namely {\it Pen-Knife}, for which a perfect match between theory and experiment does not exist for $c^2 = 0$ or  $c^2 = 1$, which means that only within the quantum field model a solution for a perfect match exists.

For the next pair of concepts, {\it Sportswear} and {\it Sports Equipment}, we have 19 of the 24 items where the quantum field theoretic model gives rise to a perfect match, the remaining five items not yielding a perfect match. Of these five items, one, {\it Gymnasium}, is a very close match (difference of 0.006818), two, {\it Diving Mask} and {\it Wrist Sweat}, are close matches, both with a difference of 0.05, and two, {\it Circus Clown} and {\it Suntan Lotion}, are not close matches, both with a difference of 0.1. For two items, {\it Underwater} and {\it Cricket Pitch}, we have  $c^2 = 1$ as the best solution for a perfect match, so that they fit within the `one-item situation' with {\it Sportswear} or {\it Sports Equipment} as a new concept. For seven items, {\it Gymnasium}, {\it Golf Ball}, {\it Cheerleaders}, {\it Baseball Bat}, {\it Sailing Life Jacket}, {\it Hoola Hoop} and {\it Tennis Racket}, we have  $c^2 = 0$, so that they fit within the `two identical items situation', neglecting {\it Sportswear or Sports Equipment} as a new concept.

For eight items, Referee's {\it Whistle}, {\it Backpack}, {\it Frisbee}, {\it Sunglasses}, {\it Motorcycle Helmet}, {\it Lineman's Flag}, {\it Bathing Costume} and {\it Ballet Shoes}, we have  different from 1 and different from 0, which means that they need a genuine quantum field description with superposition between the `one-item situation' and the `two identical items situation'. There is one item, namely Lineman's Flag, for which a perfect match of theory with experiment does not exist for $c^2 = 0$ or  $c^2 = 1$, which means that only within the quantum field model a solution for a perfect match exists.

For the next pair of concepts, {\it Household Appliances} and {\it Kitchen Utensils}, we get a perfect match for all 24 items. We have one item, {\it Table Mat}, with  $c^2 = 1$, so that we have a `one-item situation' modeling with respect to the new concept {\it Household Appliances or Kitchen Utensils}. There are seven items, {\it Apron}, {\it Carving Knife}, {\it Cooking Stove}, {\it Vacuum Cleaner}, {\it Broom}, {\it Rolling Pin} and {\it Spatula}, with  $c^2 = 0$, involving a `two identical situations' modeling neglecting the new concept {\it Household Appliances or Kitchen Utensils}.

For twelve items, {\it Fork}, {\it Hat Stand}, {\it Freezer}, {\it Extractor Fan}, {\it Cake Tin}, {\it Iron}, {\it Chopping Board}, {\it Television}, {\it Rubbish Bin}, {\it Vegetable Rack}, {\it Electric Toothbrush} and {\it Frying Pan}, we have $c^2$ different from 1 and different from 0, requiring genuine quantum field modeling with superposition between the `one-item situation' and the `two identical items situation'. There are seven items, {\it Fork}, {\it Freezer}, {\it Extractor Fan}, {\it Iron}, {\it Chopping Board}, {\it Television} and {\it Frying Pan}, for which a perfect match between theory and experiment does not exist for $c^2 = 0$ or  $c^2 = 1$, which means that only within the quantum field model a solution for a perfect match exists.

This brings us to the last pair of concepts, {\it Fruits} and {\it Vegetables}. Here we have 21 of the 24 items with a perfect match, and three items for which the quantum field model does not produce a perfect match. One of them, {\it Mushroom}, produces a close match (difference of 0.05355339), and two, {\it Broccoli} and {\it Coconut}, produce no close matches (differences of 0.2763932 and 0.22613872, respectively). For nine items, {\it Parsley}, {\it Chili Pepper}, {\it Root Ginger}, {\it Acorn}, {\it Rice}, {\it Wheat}, {\it Peanut}, {\it Garlic} and {\it Almond}, we have  $c^2 = 1$, so that they fit in the `one-item situation' with consideration of the new `concept {\it Fruits or Vegetables}'. For two items, {\it Apple} and {\it Pumpkin}, we have  $c^2 = 0$, so that this item fits within the `two identical items situation', neglecting the new concept {\it Fruits or Vegetables}.

For ten items, {\it Olive}, {\it Raisin}, {\it Mustard}, {\it Tomato}, {\it Green Pepper}, {\it Watercress}, {\it Black Pepper}, {\it Yam}, {\it Elderberry} and {\it Lentils}, we have $c^2$ different from 1 and different from 0, which means that these items need a genuine quantum field theoretic modeling, and a situation which is a superposition between the `one-item' and `two identical items' situations. There are two items, {\it Raisin} and {\it Elderberry}, for which a perfect match between theory and experiment does not exist if  $c^2 = 0$ or  $c^2 = 1$, which means that only within the quantum field model a solution for a perfect match exists.

\subsection{Calculation of the Vector Representing the Items in the Canonical Base of Fock Space}

We now have all the necessary data and values to calculate the vector expressed in the canonical base of Fock space representing the states of the considered items. Hence, let us calculate  $\left| x \right\rangle $ expressed in function of the canonical base $\left\{ {\left| {00} \right\rangle ,\left| {10} \right\rangle ,\left| {01} \right\rangle ,\left| {20} \right\rangle ,\left| {11} \right\rangle ,\left| {02} \right\rangle } \right\}$ of Fock space. From (\ref{eq2.11}), (\ref{eq2.12}), (\ref{eq2.13}), (\ref{eq2.14}) and (\ref{eq2.15}) it follows that
\begin{eqnarray}
 \left\langle {{00}}
 \mathrel{\left | {\vphantom {{00} x}}
 \right. \kern-\nulldelimiterspace}
 {x} \right\rangle  &=& 0 \\ 
 \left\langle {{10}}
 \mathrel{\left | {\vphantom {{10} x}}
 \right. \kern-\nulldelimiterspace}
 {x} \right\rangle  &=& \frac{{ce^{i\gamma } }}{D}(ae^{i\alpha }  + be^{i\beta } ) \\ 
 \left\langle {{01}}
 \mathrel{\left | {\vphantom {{01} x}}
 \right. \kern-\nulldelimiterspace}
 {x} \right\rangle  &=& \frac{{ce^{i\gamma } }}{D}(a'e^{i\alpha '}  + b'e^{i\beta '} ) \\ 
 \left\langle {{20}}
 \mathrel{\left | {\vphantom {{20} x}}
 \right. \kern-\nulldelimiterspace}
 {x} \right\rangle  &=& \frac{{2c'e^{i\gamma '} }}{E}abe^{i(\alpha  + \beta )}  \\ 
 \left\langle {{11}}
 \mathrel{\left | {\vphantom {{11} x}}
 \right. \kern-\nulldelimiterspace}
 {x} \right\rangle  &=& \frac{{2c'e^{i\gamma '} }}{{\sqrt 2 E}}(ab'e^{i(\alpha  + \beta ')}  + a'be^{i(\alpha ' + \beta )} ) \\ 
 \left\langle {{02}}
 \mathrel{\left | {\vphantom {{02} x}}
 \right. \kern-\nulldelimiterspace}
 {x} \right\rangle  &=& \frac{{2c'e^{i\gamma '} }}{E}a'b'e^{i(\alpha ' + \beta ')} 
\end{eqnarray}
which gives us the following expression for $\left| x \right\rangle $ in function of the canonical base of Fock space
\begin{eqnarray}
 \left| x \right\rangle  &=& \frac{{ce^{i\gamma } }}{D}(ae^{i\alpha }  + be^{i\beta } )\left| {10} \right\rangle  + \frac{{ce^{i\gamma } }}{D}(a'e^{i\alpha '}  + b'e^{i\beta '} )\left| {01} \right\rangle  + \frac{{2c'e^{i\gamma '} }}{E}abe^{i(\alpha  + \beta )} \left| {20} \right\rangle  \\ 
  && + \frac{{2c'e^{i\gamma '} }}{{\sqrt 2 E}}(ab'e^{i(\alpha  + \beta ')}  + a'be^{i(\alpha ' + \beta )} )\left| {11} \right\rangle  + \frac{{2c'e^{i\gamma '} }}{E}a'b'e^{i(\alpha ' + \beta ')} \left| {02} \right\rangle
\end{eqnarray}
Let us calculate this vector explicitly for the item {\it Mantelpiece} with respect to the pair of concepts {\it House Furnishings} and {\it Furniture}. For Mantelpiece we have  $a^2 = \mu(A) = 0.4$, and hence  $a = \sqrt{\mu(A)} = 0.63245553$,  $a'^2 = 1 - \mu(A) = 0.6$, and hence  $a' = \sqrt{1 - \mu(A)} = 0.77459667$,  $b^2 = \mu(B) = 0.8$, and hence  $b = \sqrt{\mu(B)} = 0.89442719$, and  $b'^2 = 1 - \mu(B) = 0.2$, and hence  $b' = \sqrt{1 - \mu(B)} = 0.44721360$. We have $c^2 = 0.2865$ and  $c'^2 = 1 - 0.2865 = 0.7135$, which gives $c = 0.53525695$ and  $c' = 0.84468929$. We have $\beta - \alpha = 0^\circ$ and  $\beta' - \alpha' = 71.79797^\circ$. Let us choose  $\alpha = \gamma = 0^\circ$,  $\alpha' = \gamma' = 0^\circ$,  $\beta = 0^\circ$ and  $\beta' = 71.79797^\circ$. Using (\ref{eq2.8}) and (\ref{eq2.9}), we get  $E = 1.76772338$ and  $D = 1.82969564$. This gives us
\begin{eqnarray}
 \left\langle {{10}}
 \mathrel{\left | {\vphantom {{10} x}}
 \right. \kern-\nulldelimiterspace}
 {x} \right\rangle  &=& \frac{{ce^{i\gamma } }}{D}(ae^{i\alpha }  + be^{i\beta } ){\rm{ = 0}}{\rm{.44667242}} \\ 
 \left\langle {{01}}
 \mathrel{\left | {\vphantom {{01} x}}
 \right. \kern-\nulldelimiterspace}
 {x} \right\rangle  &=& \frac{{ce^{i\gamma } }}{D}(a'e^{i\alpha '}  + b'e^{i\beta '} ) = {\rm{0}}{\rm{.26746592 + 0}}{\rm{.12428085i}} \\ 
 \left\langle {{20}}
 \mathrel{\left | {\vphantom {{20} x}}
 \right. \kern-\nulldelimiterspace}
 {x} \right\rangle  &=& \frac{{2c'e^{i\gamma '} }}{E}abe^{i(\alpha  + \beta )}  = {\rm{0}}{\rm{.54061447}} \\ 
 \left\langle {{11}}
 \mathrel{\left | {\vphantom {{11} x}}
 \right. \kern-\nulldelimiterspace}
 {x} \right\rangle  &=& \frac{{2c'e^{i\gamma '} }}{{\sqrt 2 E}}(ab'e^{i(\alpha  + \beta ')}  + a'be^{i(\alpha ' + \beta )} ) = {\rm{0}}{\rm{.52789077 + 0}}{\rm{.18157182i}} \\ 
 \left\langle {{02}}
 \mathrel{\left | {\vphantom {{02} x}}
 \right. \kern-\nulldelimiterspace}
 {x} \right\rangle  &=& \frac{{2c'e^{i\gamma '} }}{E}a'b'e^{i(\alpha ' + \beta ')}  = {\rm{0}}{\rm{.10341193 + 0}}{\rm{.31449164i}}
\end{eqnarray}
Hence we have
\begin{eqnarray}
 \left| x \right\rangle _{Mantelpiece}  &=& {\rm{0}}{\rm{.44667242}}\left| {{\rm{10}}} \right\rangle  + ({\rm{0}}{\rm{.26746592 + 0}}{\rm{.12428085i)}}\left| {{\rm{01}}} \right\rangle \nonumber \\
&& + {\rm{0}}{\rm{.54061447}}\left| {{\rm{20}}} \right\rangle  + ({\rm{0}}{\rm{.52789077 + 0}}{\rm{.18157182i)}}\left| {{\rm{11}}} \right\rangle  \nonumber \\
&& + ({\rm{0}}{\rm{.10341193 + 0}}{\rm{.31449161i)}}\left| {{\rm{02}}} \right\rangle 
\end{eqnarray}
This is the vector representing the item {\it Mantelpiece} in the 6-dimensional complex Hilbert space which is the Fock space of the field theoretic model that we introduced. Observe that some of the components of the vector are complex numbers.

\section{Conclusion}
We introduced a quantum field model for the description of the disjunction of two concepts that is able to predict with very great accuracy the outcomes of experiments that measure the membership weight of items with respect to the disjunction of two concepts in function of the membership weight of these items with respect to each of the concepts apart. The predictions are made using formula (\ref{eq2.23}), where the membership weight $\mu_{quant}(A\ {\rm or}\ B)$ of an item with respect to the disjunction of concept $A$ and concept $B$ is given in function of the membership weights $\mu(A)$ and $\mu(B)$ of this item with respect to concept $A$ and concept  $B$. The angles  $\beta - \alpha$ and  $\beta' - \alpha'$ are the quantum angles characterizing the specific item and introduce the quantum effect of interference giving rise to the exact deviations from a classical interpretation of the disjunction as measured in an experiment performed by Hampton (1988b).

These deviations, which Hampton (1988b) called overextension and underextension, are similar to the well-known deviations from a classical interpretation measured in the case of the conjunction of two concepts, an effect often referred to as the guppy effect and a situation often referred to as the pet-fish problem. Hence, according to our quantum field model, the effects of overextension and underextension are due to cognitive interference and can be modeled to a very great accuracy by our quantum field theoretic model. Next to these very accurate predictions obtained by the quantum field model, the quantum field model also introduces a dynamics for situations encountered in the experiment.

The quantum field model that we introduced consists of two sub models. One of the sub models is a `one-item model' and the other sub model is a `two identical items model'. The complete quantum field model consists of a superposition of these two sub models. The sub model which is the `one-item model' describes the following aspect: the item in question is compared with a new concept whose vector state is mathematically the superposition of the vector states representing the two considered concepts. This new concept is the concept  $A$ or  $B$. The other sub model, the `two identical items model', describes the following aspect: the item in question is compared with concept  $A$ and with concept  $B$ and a decision about membership weight in relation with the disjunction of  $A$ and  $B$ is taken depending on the decision about membership weight in relation to concept $A$ and concept  $B$ apart, and making use of a quantum version of the `or', hence `yes' for membership of the disjunction follows from `yes' for membership of one of the concepts  $A$ or  $B$.

In this article we only treat the situation of the disjunction of concepts. The theory's full power shows when we use it also for the modeling of the situation of the conjunction of concepts (Aerts 2007b) and for the elaboration of a fundamental mechanism for the formation of concepts (Aerts 2007c).

The content of this article is part of an ongoing research interest in applying quantum structures to domains of science different from the micro-world with applications to economics (Schaden, 2002; Baaquie, 2004; Haven, 2005; Bagarello, 2006), operations research and management sciences (Bordley, 1998; Bordley \& Kadane, 1999; Mogiliansky, 2006), psychology and cognition (Aerts \& Aerts, 1994; Grossberg, 2000; Gabora \& Aerts, 2002a,b; Aerts \& Gabora, 2005a,b; Busemeyer, Wang \& Townsend, 2006; Aerts, 2007a,b), game theory (Eisert, Wilkens \& Lewenstein, 1999; Piotrowski \& Sladkowski, 2003), and language and artificial intelligence (Widdows, 2003, 2006; Widdows \& Peters, 2003; Aerts \& Czachor, 2004; Van Reisbergen, 2004; Aerts, Czachor \& D'Hooghe, 2005, Bruza \& Cole, 2005). 

\section*{References}
\begin{description}
\item Aerts, D. (2007a). Quantum interference and superposition in cognition: Development of a theory for the disjunction of concepts. Manuscript submitted for publication.

\item Aerts, D. (2007b). A quantum field model for the conjunction of concepts. Manuscript submitted for publication.

\item Aerts, D. (2007c). A fundamental model for general concept formation. Manuscript submitted for publication. 

\item Aerts, D., \& Aerts, S. (1994). Applications of quantum statistics in psychological studies of decision processes. {\it Foundations of Science, 1}, 85-97. Reprinted in B. Van Fraassen (1997), (Eds.), {\it Topics in the Foundation of Statistics} (pp. 111-122). Dordrecht: Kluwer Academic.

\item Aerts, D., \& Czachor, M. (2004). Quantum aspects of semantic analysis and symbolic artificial intelligence. {\it Journal of Physics A, Mathematical and Theoretical, 37}, L123-L132. 

\item Aerts, D., Czachor, M., \& D'Hooghe, B. (2006). Towards a quantum evolutionary scheme: violating bell's inequalities in language. In N. Gontier, J. P. Van Bendegem, \& D. Aerts (Eds.), {\it Evolutionary Epistemology, Language and Culture - A Non Adaptationist Systems Theoretical Approach}. Dordrecht: Springer.

\item Aerts, D., \& Gabora, L. (2005a). A theory of concepts and their combinations II: A Hilbert space representation. {\it Kybernetes, 34}, 192-221.

\item Aerts, D., \& Gabora, L. (2005b). A theory of concepts and their combinations I: The structure of the sets of contexts and properties. {\it Kybernetes, 34}, 167-191.

\item Baaquie, B. E. (2004). {\it Quantum Finance: Path Integrals and Hamiltonians for Options and Interest Rates}. Cambridge UK: Cambridge University Press.

\item Bagarello, F. (2006). An operatorial approach to stock markets. {\it Journal of Physics A, 39}, 6823-6840. 

\item Bordley, R. F. (1998). Quantum mechanical and human violations of compound probability principles: Toward a generalized Heisenberg uncertainty principle. {\it Operations Research, 46}, 923-926. 

\item Bordley, R. F., \& Kadane, J. B. (1999). Experiment dependent priors in psychology and physics. {\it Theory and Decision, 47}, 213-227. 

\item Bruza, P. D. and Cole, R. J. (2005). Quantum logic of semantic space: An exploratory investigation of context effects in practical reasoning. In S. Artemov, H. Barringer, A. S. d'Avila Garcez, L.C. Lamb, J. Woods (Eds.) {\it We Will Show Them: Essays in Honour of Dov Gabbay}. College Publications.

\item Busemeyer, J. R., Wang, Z., \& Townsend, J. T. (2006). Quantum dynamics of human decision making. {\it Journal of Mathematical Psychology, 50}, 220-241. 

\item Chater, N., Lyon, K., \& Meyers, T.  (1990).  Why are conjunctive categories extended?  {\it Journal of Experimental Psychology: Learning, Memory \& Cognition, 16}, 497-508. 

\item Dirac, P. A. M. (1958). {\it Quantum mechanics}, 4th ed. London: Oxford University Press.

\item Eisert, J., Wilkens, M., \& Lewenstein, M. (1999). Quantum games and quantum strategies. {\it Physical Review Letters, 83}, 3077-3080. 

\item Feynman, R. P. (1985). {\it QED: The Strange Theory of Light and Matter}. Princeton, New Jersey: Princeton University Press. 

\item Feynman, R. P., Leighton, R. B., \& Sands, M. (1966). {\it The Feynman Lectures on Physics III: Quantum Mechanics}. Reading, Massachusetts: Addison-Wesley. 

\item Gabora, L., \& Aerts, D. (2002a). Contextualizing concepts. In {\it Proceedings of the 15th International FLAIRS Conference. Special track: Categorization and Concept Representation: Models and Implications}, Pensacola Florida, May 14-17, American Association for Artificial Intelligence (pp. 148-152). 

\item Gabora, L., \& Aerts, D. (2002b). Contextualizing concepts using a mathematical generalization of the quantum formalism. {\it Journal of Experimental and Theoretical Artificial Intelligence, 14}, 327-358. Preprint at http://arXiv.org/abs/quant-ph/0205161 

\item Grossberg, S. (2000). The complementary brain: Unifying brain dynamics and modularity. {\it Trends in Cognitive Science, 4}, 233-246.

\item Hampton, J. A. (1987). Inheritance of attributes in natural concept conjunctions. {\it Memory \& Cognition, 15}, 55-71. 

\item Hampton, J. A. (1988a). Overextension of conjunctive concepts: Evidence for a unitary model for concept typicality and class inclusion. {\it Journal of Experimental Psychology: Learning, Memory, and Cognition, 14}, 12-32.

\item Hampton, J. A. (1988b). Disjunction of natural concepts. {\it Memory \& Cognition, 16}, 579-591.

\item Hampton, J. A. (1991).  The combination of prototype concepts.  In P. Schwanenflugel (Ed.), {\it The Psychology of Word Meanings}. Hillsdale, NJ: Erlbaum.

\item Hampton, J. A. (1993).  Prototype models of concept representation.  In I. Van Mechelen, J. Hampton, R. S. Michalski, \& P. Theuns (Eds.), {\it Categories and Concepts: Theoretical Views and Inductive Data Analysis} (pp. 67-95).  London, UK: Academic Press.

\item Hampton, J.A. (1996). Conjunctions of visually-based categories: overextension and compensation. {\it Journal of Experimental Psychology: Learning, Memory and Cognition, 22}, 378-396.

\item Hampton, J. A. (1997a). Conceptual combination: Conjunction and negation of natural concepts. {\it Memory \& Cognition, 25}, 888-909.

\item Hampton, J. A. (1997b). Conceptual combination. In K. Lamberts \& D. Shanks (Eds.), {\it Knowledge, Concepts, and Categories} (pp. 133-159). Hove: Psychology Press. 

\item Haven, E. (2005). Pilot-wave theory and financial option pricing. {\it International Journal of Theoretical Physics, 44}, 1957-1962.

\item Komatsu, L. K. (1992).  Recent views of conceptual structure.  {\it Psychological Bulletin, 112}, 500-526. 

\item Kunda, Z., Miller, D. T., \& Claire, T. (1990).  Combining social concepts: The role of causal reasoning.  {\it Cognitive Science, 14}, 551-577. 

\item Mogiliansky, A. L., Zamir, S., \& Zwirn, H. (2006). Type indeterminacy: A model of the KT(Kahneman-Tversky)-man [preprint]. http://lanl.arxiv.org/abs/physics/0604166

\item Osherson , D. N. \& Smith, E. E. (1981).  On the adequacy of prototype theory as a theory of concepts.  {\it Cognition, 9}, 35-58. 

\item Osherson, D. N. \& Smith, E. E. (1982).  Gradedness and conceptual combination.  {\it Cognition, 12}, 299-318. 

\item Piotrowski, E. W., \& Sladkowski, J. (2003). An invitation to quantum game theory. {\it International Journal of Theoretical Physics, 42}, 1089.

\item Rips, L J. (1995). The current status of research on concept combination. {\it Mind \& Language, 10}, 72-104. 

\item Rosch, E. (1973a). Natural categories. {\it Cognitive Psychology, 4}, 328.

\item Rosch, E. (1973b). On the internal structure of perceptual and semantic categories. In T. E. Moore (Ed.), {\it Cognitive Development and the Acquisition of Language}. New York: Academic Press. 

\item Rosch, E. (1975). Cognitive representations of semantic categories. {\it Journal of Experimental Psychology: General, 104}, 192-232. 

\item Rosch, E. (1978). Principles of categorization. In E. Rosch \& B. Lloyd (Eds.), {\it Cognition and Categorization} (pp. 133-159). Hillsdale, NJ: Lawrence Erlbaum. 

\item Rosch, E. (1983). Prototype classification and logical classification: The two systems. In E. K. Scholnick (Ed.), {\it New Trends in Conceptual Representation: Challenges to Piaget's Theory?} (pp. 133-159). Hillsdale, NJ: Lawrence Erlbaum. 

\item Schaden, M. (2002). Quantum finance: A quantum approach to stock price fluctuations. {\it Physica A, 316}, 511. 

\item Smith, E. E., \& Medin, D. L. (1981). {\it Categories and Concepts}. Cambridge, MA.: Harvard University Press.

\item Smith, E.  E., \& Osherson, D. N. (1984).  Conceptual combination with prototype concepts.  {\it Cognitive Science, 8}, 357-361. 

\item Smith, E. E., Osherson, D. N., Rips, L. J., \& Keane, M. (1988).  Combining prototypes: A selective modification model.  {\it Cognitive Science, 12}, 485-527. 

\item Springer, K., \& Murphy, G. L. (1992).  Feature availability in conceptual combination.  {\it Psychological Science, 3}, 111-117. 

\item Storms, G., De Boeck, P., Van Mechelen, I., \& Geeraerts, D. (1993). Dominance and non-commutativity effects in concept conjunctions: Extensional or intensional basis? {\it Memory \& Cognition, 21}, 752-762.

\item Storms, G., De Boeck, P., Van Mechelen, I. \& Ruts, W. (1996). The dominance effect in concept conjunctions: Generality and interaction aspects. {\it Journal of Experimental Psychology: Learning, Memory \& Cognition, 22}, 1-15.

\item Storms, G., de Boeck, P., Hampton , J.A., \& van Mechelen, I. (1999). Predicting conjunction typicalities by component typicalities. {\it Psychonomic Bulletin and Review, 6}, 677-684. 

\item Van Reisbergen, K. (2004). {\it The Geometry of Information Retrieval}. Cambridge UK: Cambridge University Press. 

\item Widdows, D. (2003). Orthogonal negation in vector spaces for modelling word-meanings and document retrieval. In {\it Proceedings of the 41st Annual Meeting of the Association for Computational Linguistics} (pp. 136-143). Sapporo, Japan, July 7-12. 

\item Widdows, D. (2006). {\it Geometry and Meaning}. CSLI Publications: University of Chicago Press. 

\item Widdows, D., \& Peters, S. (2003). Word vectors and quantum logic: Experiments with negation and disjunction. In {\it Mathematics of Language 8} (pp. 141-154). Indiana: Bloomington. 

\item Zadeh, L. (1965). Fuzzy sets. {\it Information \& Control, 8}, 338-353.

\end{description}
\end{document}